\documentclass{masterthesis}

\title{Deep-Spying: Spying using Smartwatch and Deep Learning}

\author{Tony Beltramelli}
\supervisor{Sebastian Risi, Ph.D.}

\university{IT University of Copenhagen}
\location{Copenhagen, Denmark}

\degree{A thesis submitted in fulfillment of the requirements
for the degree of Master of Science.}

\date{December 2015}

\hypersetup{
  pdfinfo={
    Title={Deep-Spying: Spying using Smartwatch and Deep Learning},
    Author={Tony Beltramelli},
    Keywords={Security, Side-Channel Attack, Keystroke Inference, Motion Sensors, Deep Learning, Recurrent Neural Network, Wearable Computing}
  }
}

\begin{document}

\maketitle
\makefrontmatter

\chapter{Introduction}

This chapter will first introduce the reader to the problem being addressed in this research and the related implications. Finally, the methodology employed to provide a practical proof-of-concept system will be shortly described.

\section{Problem Statement}
\label{sec:problemstatement}

The keyboard is one of the oldest human-computer interface and still one of the most common devices to input information into various types of machines. Some of this information can be sensitive and highly valuable, such as passwords, PINs, social security numbers, and credit card numbers. Related works (detailed in Chapter \ref{ch:relatedwork}) have shown that the data from the motion sensors of a smartphone can be used to infer keystrokes entered on its touchscreen \cite{cai2011touchlogger, xu2012taplogger, owusu2012accessory}. Other research has proved that the motion sensors from a smartphone standing on a flat surface can be used to infer the keystrokes typed on a nearby physical computer keyboard \cite{marquardt2011sp}. Moreover, recently published works have demonstrated that smartwatches motion sensors could be exploited to infer keystrokes on both virtual and physical keyboards \cite{wang2015mole, maiti2015smart, liu2015good}.

The security of \emph{Wearable Wristband and Armband Devices (WADs)} such as smartwatches and fitness trackers is of great concern not directly because of the device itself being exploitable\footnote{A smartwatch is not dramatically different from a smartphone from a technological point of view, the sensors are the same and the features are largely similar. It is even reasonable to assume that a smartwatch is much more limited because of power and performance limitations.} but because of the very nature of wearable devices being wearable. A smartwatch is indeed potentially worn for an extended period such as the whole day, offering a pervasive attack surface to cyber-criminals.

The implications are therefore significant: exploiting motion sensors for keystrokes inference can happen continuously as long as a WAD is worn. More dramatically, the whole technological ecosystem of the user is compromised each time a WAD is worn. One can indeed extrapolate that keystrokes inference attack is possible on any keypad used by a person while she is wearing a WAD. For example, the virtual keyboard of a tablet computer, a smartphone touchscreen, the physical keyboard of a laptop computer, the keypad of an electronic building access system, the keypad of a hotel room safe or even the keyboard of a bank ATM. Eavesdropping on WAD sensors can thus have implications reaching far beyond a simple privacy leakage and have the potential to cause important damages.

Moreover, recent advances in Artificial Intelligence and notably \emph{Deep Learning} have allowed algorithms to solve problems with impressive performance sometimes even surpassing human experts. Deep neural networks are able to process robustly noisy real-world data and can automatically learn features from raw data. These powerful models have successfully been applied to complex tasks in the fields of Computer Vision, Natural Language Processing and Speech Recognition \cite{lecun2015deep, schmidhuber2015deep}. Deep Learning has however comparatively been used poorly to process time series data such as motion sensors \cite{langkvist2014review}. These state-of-the-art scientific tools used to require advanced knowledge to be implemented and applied successfully. However, their remarkable qualities lead to the development of various Open-Source projects \cite{caffe, torch, keras, lasagne, pybrain, deeplearning4j, convnetjs, brainstorm, tensorflow} making them available and free to use by anyone. Even though this offers great possibilities to everyone from businesses to hobbyists, it potentially puts Deep Learning in the toolbox of cyber-criminals.

To the best of our knowledge, keystroke inference in related works has only been performed using shallow models requiring manual feature extraction and carefully engineered signal processing pipelines. The use of a deep neural network approach can drastically cut down the number of technical steps towards successful keystroke inference. Thus making the attack a more plausible threat against users of WAD.\footnote{A bug in Android was discovered while working on this research project and was thought to lead to a vulnerability. In fact, the bug leads applications targeting Android Wear to grant some permissions without them being explicitly defined in the manifest file \cite{permissionsbug}. A responsible disclosure process was thus initiated with Google to fix the issue. After further investigations, the problem turned out to be a bug in the Android SDK with no serious security implications towards users.}

Two research goals can thus be formulated:

\begin{itemize}
    \item Assess the practicality of motion-based keystroke inference attack using wearable wristband/armband technology.
    \item Assess the practicality of motion-based keystroke inference attack using deep neural network models.
\end{itemize}

\section{Methodology}

The research questions enunciated in Section \ref{sec:problemstatement} are answered in a practical manner. First, a system is implemented to collect, process, analyze motion sensors data and perform experimental indirect passive attacks such as keylogging and touchlogging.
Second, experiments are conducted to collect data in a deployment environment. Finally, the results are interpreted and discussed.

Challenges that will be addressed in particular are:

\begin{enumerate}
\item Data Acquisition: The system should allow sensor recording at specific keystroke intervals in the continuous data stream.

\item Data Pre-processing: For comparison purpose in different contexts, data obtained from sensor outputs need to be reduced to a meaningful synthesis to limit the effect of noise. 

\item Feature Extraction: For comparison purpose with traditional methods, successful classification traditionally relies on carefully chosen discriminant features.

\item Classification: The artificial neural network should be trained as optimally as possible in order to avoid over-fitting and improve generalization.

\item Evaluation: The quality of the classifier must be quantified to assess its performance in practice with exotic data as input.
\end{enumerate}
\chapter{Related Work}
\label{ch:relatedwork}

\begin{figure}[H]
    \centering
    \includegraphics[width=.5\linewidth]{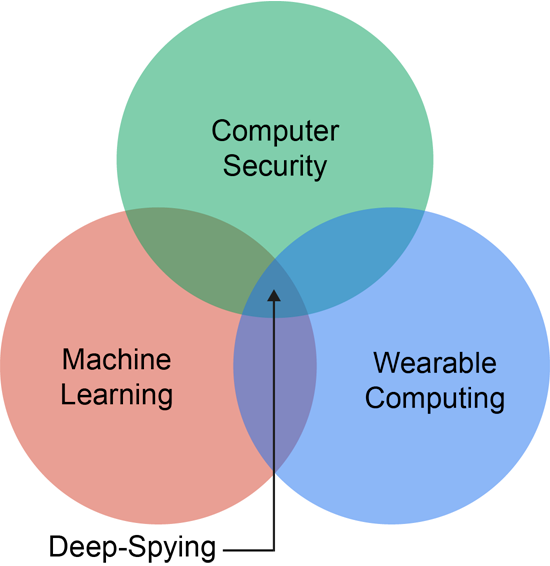}
    \caption{Related research fields.}
    \label{fig:thesistopic}
\end{figure}

This chapter's goal is twofold. First to define the key concepts of the theoretical and technical background on which this work is based. This research project is highly multidisciplinary and established at the intersection of various research fields (as illustrated in Figure \ref{fig:thesistopic}). The second goal is thus to review and reflect on the previous relevant studies and the current state-of-the-art in the related fields. The core focus is the security of wearable technologies and relies primarily on machine learning methods for data analysis.

\section{Key Concepts}

\subsection{Computer Security}

\noindent
\textbf{Passive Attack:}
A passive attack is characterized by an attacker eavesdropping on a communication channel. In such an attack, the attacker does not attempt to break the system or alter the transmitted data (i.e. active attack). Instead, the attacker is monitoring the exchanged packets to gain information about the target (e.g. the users, the system, the communicating entities) \cite{buttyan2007security}.

\noindent
\textbf{Side-channel Attack:}
A side-channel attack is defined by an attacker using side-channel information to obtain insights about a system. Side-channel information can be gained from data leaked at the physical layer (e.g. timing information, power consumption, electromagnetic emanations, sound) \cite{chen2010side}.

\noindent
\textbf{Keylogging and Touchlogging:}
Keylogging is the action of recording the keys entered on a keyboard by a user. Similarly, touchlogging is the action of recording the buttons pressed on a touch screen, or the coordinates of the touch events allowing the inference of the keys virtually touched by the user \cite{damopoulos2013keyloggers}. Keystroke inference refers to techniques used to perform keylogging or touchlogging from side-channels.

\subsection{Wearable Computing}

\noindent
\textbf{Wearable Technology:}
Wearable technologies are envisioned to be small and portable computers integrated into clothes or worn continuously (e.g. glasses, armband, wristband). These devices are designed to be extensively mobile and operate in environments that may have limited computing infrastructure support \cite{krumm2009ubiquitous}.

\noindent
\textbf{Motion Sensors:}
Modern mobile and wearable devices usually come with built-in motion sensors measuring the movements of the device. Analyzing the output data of such sensors allow the estimation of specific types of motion that the device undergoes such as translation, tilt, shake, rotation, or swing.
The typical motion sensors available in standard devices are listed in Table \ref{tab:motionsensors}. Software-based sensors usually derive their data from hardware-based sensors, namely the accelerometers (one for each axis x, y, and z), and the gyroscope \cite{motionsensorsandroid, motionsensorsapple, sensortypesandroid, al2013best}.

\begingroup
\renewcommand{\arraystretch}{1.5}
\begin{table}[ht]
    \centering
    \begin{tabular}{| l | p{5cm} | l |} \hline
        \textbf{Type} & \textbf{Description} & \textbf{Implementation} \\ \hline
Accelerometer & Acceleration force along three directions: lateral (x axis), longitudinal (y axis), and vertical (z axis) (including gravity). & Hardware \\ \hline
Gravity & Direction and magnitude of gravity along the x, y, and z axes. & Software \\ \hline
Linear Accelerometer & Acceleration force along the x, y, and z axes (excluding gravity). & Software \\ \hline
Gyroscope & Orientation change along three angles: pitch (x axis), roll (y axis), and azimuth (z axis). & Hardware \\ \hline
Rotation Vector & Orientation around the x, y, and z axes. & Software \\ \hline
    \end{tabular}
    \caption{Typical motion sensors on mobile and wearable devices.}
    \label{tab:motionsensors}
\end{table}
\endgroup

\subsection{Machine Learning}

\noindent
\textbf{Classification:}
In the field of machine learning, classification is the process of assigning categories to data. The classifier is a function assigning labels to data by building a statistical model (i.e. data structure) based on a training set containing example data with known associated classes. This process is known as supervised learning since the statistical model needs to be trained with expert-annotated data to classify subsequently unseen samples \cite{bishop2006pattern, han2011data}.

\noindent
\textbf{Feature Extraction:}
Building a relevant statistical model is only possible if the information used is describing the problem in a meaningful way. The term \emph{Feature Extraction} is used to refer to methods allowing the selection of such valuable information in a raw dataset. This dimensionality reduction process consists of building feature vectors of discrete length allowing to reduce the volume of data to process and improving its quality. \emph{Feature Engineering} is the process of manual feature selection from experts. \emph{Unsupervised Feature Learning} is a process where the model selects features automatically through training \cite{bishop2006pattern, schmidhuber2015deep}.

\section{Background: Artificial Neural Network}
\label{sec:backgroundann}

\begin{figure}[ht]
    \centering
    \includegraphics[width=.55\linewidth]{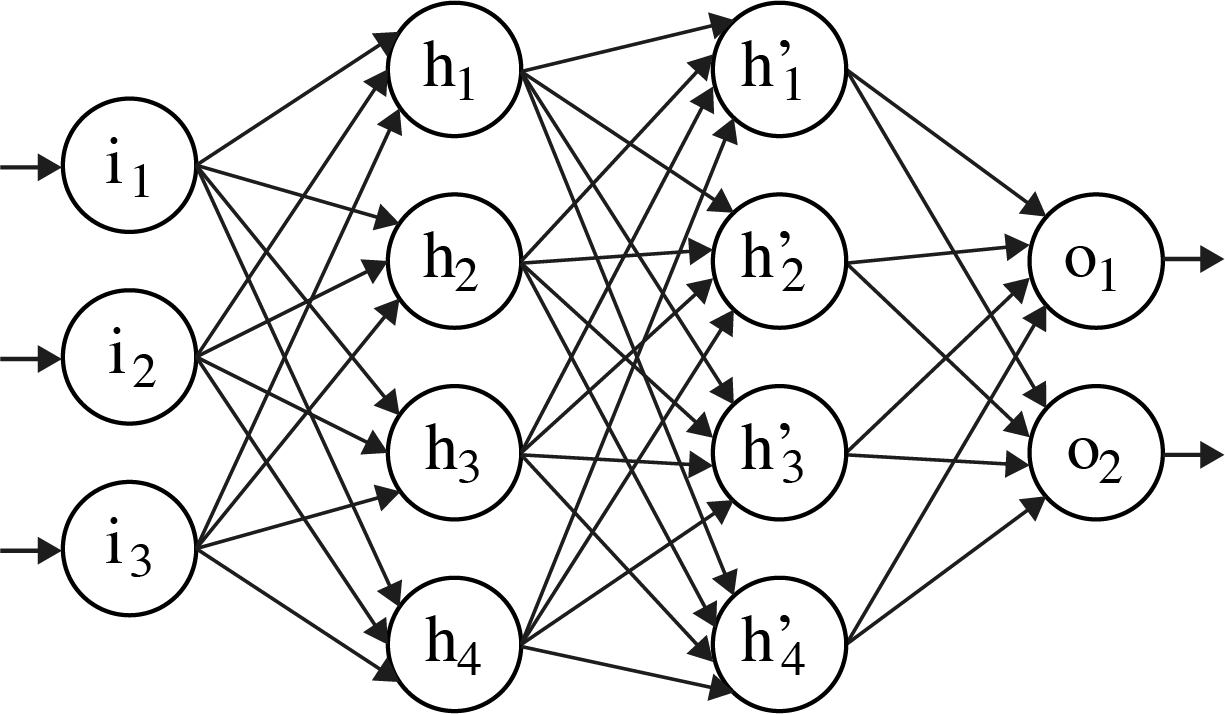}
    \caption{A feedforward neural network with $3$ input neurons, $2$ hidden layers $h$ and $h'$ containing 4 neurons each, and 2 output neurons.}
    \label{fig:feedforward_net}
\end{figure}

\emph{Artificial Neural Network (ANN)} is a class of biologically-inspired statistical model consisting of a set of connected nodes (i.e. neurons) where each connection (i.e. synapses) has a weight associated with it \cite{bishop2006pattern, haykin2004comprehensive}. A typical ANN consists of an input layer with a number of input neurons equal to the length of the feature vector. Its output layer can be built with a variable number of neurons depending on the task at hand (e.g. equal to the number of classes for classification, two output neurons for binary classification, one output neuron for regression). ANNs are usually additionally composed of hidden layers each containing a variable number of neurons as depicted in Figure \ref{fig:feedforward_net}. 

\begingroup
\renewcommand{\arraystretch}{1.5}
\begin{table}[ht]
    \centering
    \begin{tabular}{| c | c | c |} \hline
        \textbf{Function} & \textbf{Definition} & \textbf{Derivative} \\ \hline
        \begin{tabular}{@{}c@{}} Logistic Function \\ (Sigmoid) \end{tabular} & $\begin{aligned}\phi(x) = \frac{1}{1 + e^{-x}}\end{aligned}$ &
        $\begin{aligned}\frac{\partial \phi(x)}{\partial x} = \phi(x)(1 - \phi(x))\end{aligned}$ \\ \hline
        
        \begin{tabular}{@{}c@{}} Hyperbolic Tangent \\ (Tanh) \end{tabular} & $\begin{aligned}\phi(x) = \frac{e^{x} - e^{-x}}{e^{x} + e^{-x}}\end{aligned}$ &
        $\begin{aligned}\frac{\partial \phi(x)}{\partial x} = 1 - \phi(x)^2\end{aligned}$ \\ \hline
        
        \begin{tabular}{@{}c@{}} Rectified Linear Unit \\ (ReLU) \end{tabular} & $\begin{aligned}\phi(x) = \max(0, x)\end{aligned}$ &
        $\begin{aligned}\frac{\partial \phi(x)}{\partial x} = \begin{cases}
        0 & \text{if $x \leq 0$,} \\
        1 & \text{if $x > 0$;}
        \end{cases}
        \end{aligned}$ \\ \hline
        
        \begin{tabular}{@{}c@{}c@{}} Normalized \\ Exponential \\ (Softmax) \end{tabular} & $\begin{aligned}\phi(x_i) = \frac{e^{x_i}}{\sum\limits^{n}_{j=1}e^{x_j}}\end{aligned}$ &
        \begin{tabular}{@{}c@{}}
            $\begin{aligned} \frac{\partial \phi(x_i)}{\partial x_j} = \phi(x_i)(\delta_{ij} - \phi(x_j)) \end{aligned},$ \\
            
            $\delta_{ij} = 
            \begin{cases}
            0 & \text{if $i \neq j$,} \\
            1 & \text{if $i = j$;}
            \end{cases}$
        \end{tabular}
        \\ \hline
    \end{tabular}
    \caption{Popular activation functions used in ANNs.}
    \label{tab:activationfunction}
\end{table}
\endgroup

The network is activated by feeding its input layer with a feature vector that will be mapped to an output vector thanks to the network internal structure. The neurons map inputs to outputs by using a predefined activation function (examples listed in Table \ref{tab:activationfunction}). The output value of a given neuron $i$ can be computed as follows:

\begin{equation} \label{eq:ouput}
    y_{i} = \phi (x_{i}) = \phi \left(\sum^{n}_{j=1}W_{ij}y_{j}\right)
\end{equation}

\begin{figure}[H]
    \centering
    \includegraphics[width=.3\linewidth]{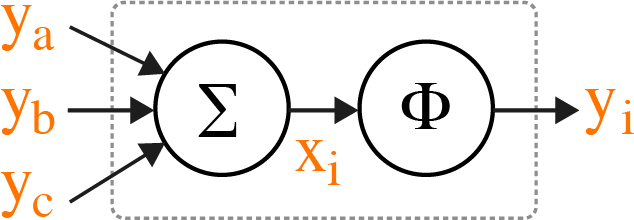}
    \caption{A standard feedforward cell $i$ with three other neurons $a$, $b$, and $c$ connected to its input.}
    \label{fig:feedforward_cell}
\end{figure}

With $\phi$ the activation function of the neuron, $n$ the number of neurons connected to neuron $i$, $W$ the weight associated with the connection between two neurons, and $y$ the output value. The input $x$ to a neuron is thus the weighted sum of outputs from connected neurons as illustrated in Figure \ref{fig:feedforward_cell}.

An ANN is trained by adjusting its weights until the correct output vector is generated from a given input so as to minimize the global error. The terms \emph{Feedforward Neural Network (FNN)} and \emph{Vanilla Neural Network} are used to refer to the most basic ANN architecture where the neurons are connected forward in an acyclic way. That is, the activation flow is unidirectional from the input layer to the output layer.

\subsection{Recurrent Neural Network}

\begin{figure}[H]
    \centering
    \includegraphics[width=.3\linewidth]{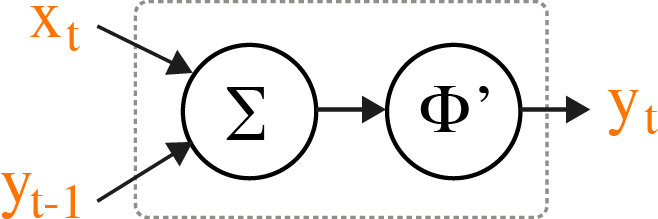}
    \caption{A standard RNN cell (see text for details).}
    \label{fig:recurrent_cell}
\end{figure}

\emph{Recurrent Neural Network (RNN)} is a class of ANN able to model relationships between sequential data-points. This type of model have gain great interest for application in context where data with time relation have to be processed. The network can indeed associate data in series by using a feedback loop allowing information to persist over time. That is, when an input vector is fed into the network, it will produce an output vector from the new input vector but also according to the vector previously seen as illustrated in Figure \ref{fig:recurrent_cell}. This memory state makes RNNs particularly suitable for processing sequential data such as time series, sound, video or text.

\begin{figure}[H]
    \begin{subfigure}{.5\textwidth}
        \centering
        \includegraphics[width=.21\linewidth]{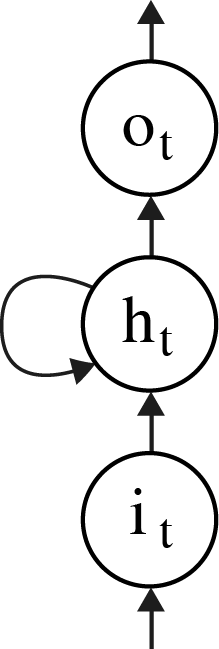}
        \caption{RNN with one hidden recurrent unit.}
    \end{subfigure}
    \begin{subfigure}{.5\textwidth}
        \centering
        \includegraphics[width=.7\linewidth]{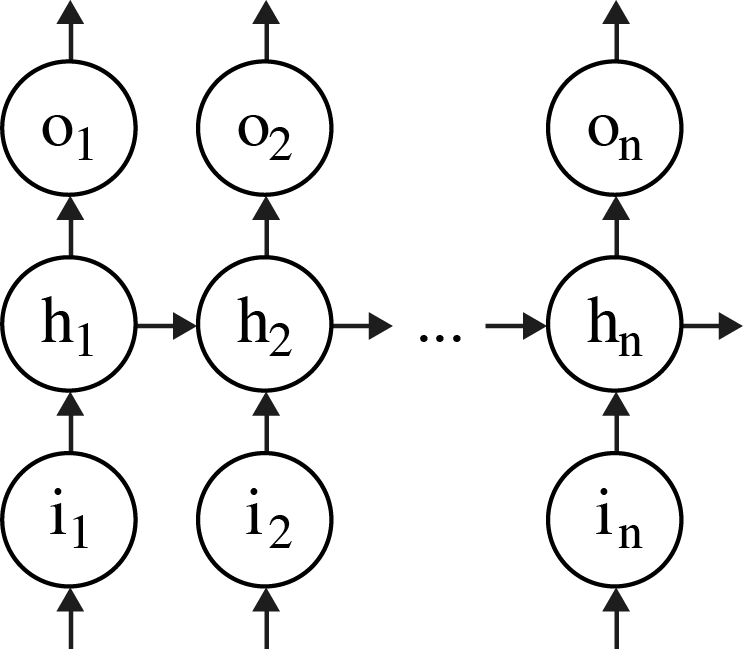}
        \caption{Unfolded RNN.}
    \end{subfigure}
    \caption{A RNN can be seen as an unfolded deep FNN. The depth corresponds to the length $n$ of the input sequence.}
    \label{fig:rnn}
\end{figure}

Mathematically, the output of an RNN cell can be expressed as follows:

\begin{equation}
    y_t = \phi'(W_{xi}x_{t} + W_{yi}y_{t-1})
\end{equation}

with $x_t$ the new input vector at time $t$, $y_{t-1}$ the previously produced output vector, and the activation function $\phi'$ the \emph{Hyperbolic Tangent (Tanh)}.
As shown in Figure \ref{fig:rnn}, RNNs can be seen as unfolded deep FNNs where each layer is connected to its past instance. It is thus possible to use RNN to map one input to one output, one input to many outputs, many inputs to one output, or many inputs to many outputs.

Despite the interesting properties of RNN, Bengio et al. \cite{bengio1994learning} have shown that standard RNNs are in practice unable to learn long-term dependencies in contexts where information need to be connected over long time intervals. In fact, training RNN with gradient descent methods such as \emph{Backpropagation} (details in Section \ref{ssec:backprop}) lead to gradually vanishing gradient because of nested activation functions. In the case of RNN where the depth can be both layer-related and time-related, this leads the network to be unable to associate information separated over long periods because the error cannot be preserved over such intervals.

\subsection{Long Short-Term Memory}

Hochreiter and Schmidhuber \cite{hochreiter1997long} overcame the limitations of standard RNN by introducing a new architecture termed \emph{Long Short-Term Memory (LSTM)} which allows the association of input with memories remote in time by preserving the backpropagated error through time and layers. While many LSTM implementation variants have been proposed \cite{greff2015lstm, jozefowicz2015empirical}, the following detailed LSTM cell use a forget gate \cite{gers2000learning} with no bias for simplicity reasons.

\begin{figure}[H]
    \centering
    \includegraphics[width=.7\linewidth]{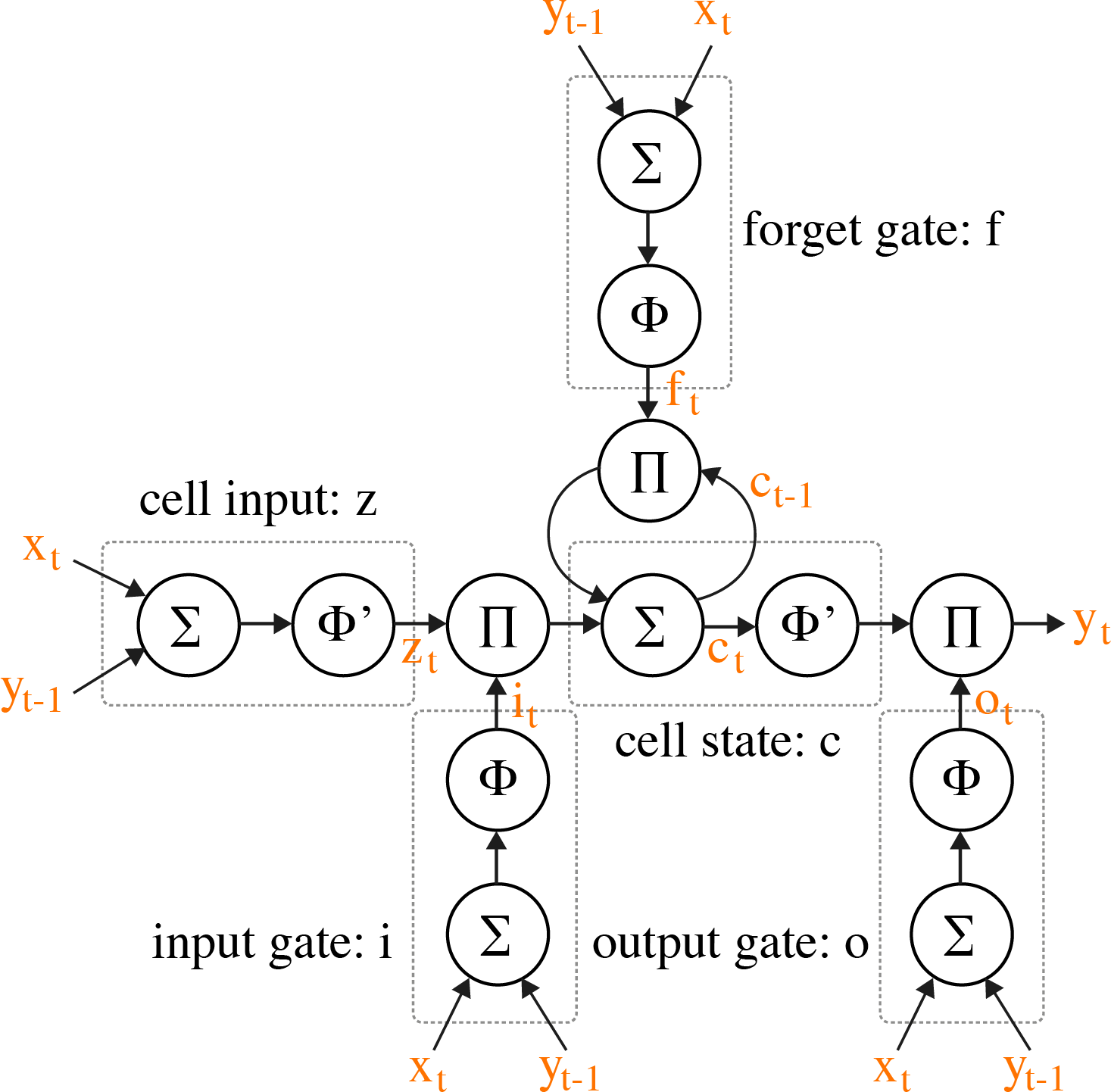}
    \caption{LSTM memory cell (see text for details).}
    \label{fig:lstm_cell}
\end{figure}

As depicted in Figure \ref{fig:lstm_cell}, an LSTM memory cell consists of many different gates that can learn to store, read, write and erase information. Weights are associated with the connection between the different gates and are updated during training. The cell state $c$ learns to memorize information by connecting one of its output to its inputs as traditional RNN cells. The input gate $i$ is used to control the error flow on the inputs of cell state $c$ to avoid input weight conflicts that occur in traditional RNN because the same weight has to be used for both storing certain inputs and ignoring others. The output gate $o$ controls the error flow from the outputs of the cell state $c$ to prevent output weight conflicts that happen in standard RNN because the same weight has to be used for both retrieving information and not retrieving others. The LSTM memory block can thus use $i$ to decide when to write information in $c$ and use $o$ to decide when to read information from $c$ \cite{hochreiter1997long}. Additionally, a forget gate $f$ is used to reset memory and, as a result, help the network process continuous sequences or sequences that are not segmented with precise starting and ending time \cite{gers2000learning}. The different gates outputs can be computed as follows:

\begin{equation}
    i_t = \phi(W_{xi}x_{t} + W_{yi}y_{t-1})
\end{equation}
\begin{equation}
    f_t = \phi(W_{xf}x_{t} + W_{yf}y_{t-1})
\end{equation}
\begin{equation}
    o_t = \phi(W_{xo}x_{t} + W_{yo}y_{t-1})
\end{equation}
\begin{equation}
    z_t = \phi'(W_{xz}x_{t} + W_{yz}y_{t-1})
\end{equation}
\begin{equation}
    c_t = f_{t}c_{t-1} + i_{t}z_{t}
\end{equation}
\begin{equation}
    y_t = o_{t}\phi'(c_{t})
\end{equation}

With $W$ the weight, $x_t$ the new input vector, $y_{t-1}$ the previously produced output vector, $c_{t-1}$ the previously produced cell state's output, $\phi$ the \emph{Logistic Function (Sigmoid)}, and $\phi'$ the \emph{Hyperbolic Tangent (Tanh)}.

One popular LSTM architecture alternative was proposed by Gers et al. \cite{gers2000recurrent} to allow recurrent networks to distinguish between sequences of variable length. This implementation adds peephole connections to pass the previous cell state to the input gate and the forget gate and the current cell state to the output gate. The gates outputs can thus be computed as follows:

\begin{equation}
    i_t = \phi(W_{xi}x_{t} + W_{yi}y_{t-1} + W_{ci}c_{t-1})
\end{equation}
\begin{equation}
    f_t = \phi(W_{xf}x_{t} + W_{yf}y_{t-1} + W_{cf}c_{t-1})
\end{equation}
\begin{equation}
    o_t = \phi(W_{xo}x_{t} + W_{yo}y_{t-1} + W_{co}c_{t})
\end{equation}

\subsection{Backpropagation}
 \label{ssec:backprop}

\emph{Backpropagation} is a popular algorithm designed to train ANNs using a gradient descent method to minimize the network prediction error \cite{rumelhart1985learning}. The network error is computed from a loss function (e.g. Mean Squared Error, Root Mean Squared Error, Quadratic Loss, Minkowski-R Error) by comparing the predicted vector with the expected vector. The error is propagated backward from the output layer to the input layer to adjust the weights along the way. Since each weight is responsible for a portion of the error, they are updated using the chain rule (the reader is invited to refer to Appendix \ref{ch:appendixbackprop} for additional theory details) to compute the partial derivative with respect to each weight such that:

\begin{equation}
    \frac{\partial E}{\partial W_{ij}} = e_{i}y_{j}
\end{equation}

With $E$ the output error to minimize, $W$ the weight to update, $y$ the output value, and $e$ the local error. The local error $e$ is computed differently depending on the neuron type. If the neuron belongs to the output layer, the error is proportional to the difference between the predicted value and the expected output. Otherwise, the error of hidden neurons is proportional to the weighted sum of errors from connected neurons \cite{bishop2006pattern, han2011data, haykin2004comprehensive}. That is, the error of a neuron $i$ is computed as follows:

\begin{equation}
    e_{i} =
    \begin{cases}
        \begin{aligned}
            \frac{\partial \phi(x_i)}{\partial x_{i}} (T_i - y_i)
        \end{aligned} &
        \text{if $i \in$ output layer,}
        \\
        \begin{aligned}
            \frac{\partial \phi(x_i)}{\partial x_{i}} \left(\sum\limits^{n}_{j=1}W_{ij} e_{j}\right)
        \end{aligned} &
        \text{otherwise;}
    \end{cases}
\end{equation}

With $\frac{\partial \phi(x_i)}{\partial x_{i}}$ the derivative of the activation function (see Table \ref{tab:activationfunction}), $x$ the input to the neuron (computed from Equation \ref{eq:ouput}), $T$ the target expected output\footnote{For regression tasks, the expected output vector usually consist of one or more continuous values. For classification tasks, the targeted output vector usually consists of binary values. For example for three classes, class $a$: $\langle 1, 0, 0 \rangle$, class $b$: $\langle 0, 1, 0 \rangle$, and class $c$: $\langle 0, 0, 1 \rangle$.}, and $y$ the predicted output. Since the error is being backpropagated, $j$ refer to neurons connected to $i$ in the next higher layer. The gradient thus represent the change to all the weights with regard to the change in the global output error. The weights can finally be updated such that:

\begin{equation}
    W_{ij} = W_{ij} - \eta\,e_{i}y_{j}
\end{equation}

With $\eta$ the learning rate, a constant value usually chosen in the range $(0.0, 1.0)$ used to tune the training algorithm by determining how much the weights are updated at each training iteration. A high learning rate can quicken the training process by doing big training steps but can prevent the global minima from being reached if too big. A low learning rate allows precise steps towards the solution but can lead to convergence in local minima if too small.

Many methods have been developed to further improve and optimize ANN training (e.g. Adaptive Learning Rate, Bias, Weight Decay). Moreover, different variants to the Backpropagation algorithm have been implemented to increase the performance of the algorithm or adapt the technique to different ANN architectures. A significant alternative is \emph{Backpropagation Through Time} \cite{werbos1990backpropagation} used to train RNNs.

\section{Literature Review}

The goal of this section is to investigate and understand the methods and techniques used in previous studies from relevant similar research fields. The objective being to compare alternative approaches, analyze their respective advantages and disadvantages, and inform and justify decisions made during the whole research process.

    \subsection{Motion-Based Keystroke Inference Attack}
    
Side-channel keystroke inference have been extensively explored in various studies. Traditionally by exploiting characteristics of physical keyboards such as electromagnetic waves \cite{vuagnoux2009compromising}, sound \cite{asonov2004keyboard, zhuang2009keyboard, berger2006dictionary}, and timing \cite{song2001timing, foo2010timing}. However, such side-channels are ineffective to exploit virtual keyboard, albeit sound have been successfully exploited on smartphones \cite{schlegel2011soundcomber}.

Studies have shown the great potential of recovering sound, music, voice conversations, and even typing by simply observing slight vibrations in the environment produced by physical events \cite{davis2014visual, barisani2009sniffing}. Such investigations have shown that motion invisible to the human eye can convey a surprisingly significant amount of information, establishing motion as a pertinent source of valuable data and thus a reliable side-channel. Although the camera can be used to detect motion, we are here interested in motion sensors because they are currently available on the majority of WAD, which is not the case for cameras.

Using motion as a side-channel imply that only movement dynamics will be used to attempt to infer information about the system. In our case, inferring keystrokes entered on a physical or virtual keyboard by a user. This attack works based on the observation that device motion during a keystroke is correlated to and consistent with keys typed by the user. In the case of such attack on a smartphone device, it is reasonable to assume that the motion of the device, while the user is typing, is affected by multiple factors. For example, the device dimension, the screen orientation, the sensor chips specifications, the keyboard layout, the user habits, and the relative user position and motion. This assumption leads many security researchers to question the practicality of such an attack. However, studies \cite{cai2012practicality, aviv2012practicality} have shown that motion-based keystroke inference attack remains effective and practical despite the obvious assumptions that the previously enunciated factors might alter the robustness and the accuracy of the inference. Motion is thus established as a significant side-channel allowing the leakage of sensitive information.
    
        \subsubsection{Keylogging}
    
Marquardt et al. \cite{marquardt2011sp} have shown that the motion sensors output from a smartphone standing on a flat surface can be used to infer keystrokes typed on a nearby physical computer keyboard standing on the same surface. Their attack scenario is based on two observations. First, that access to the accelerometer data was not protected by any mobile operating system, thus allowing any installed application to monitor the accelerometer events. Secondly, that many users place their smartphones on the desk next to their computer keyboard when they are working. In their experiment setup, an iPhone device is collecting the accelerometer data to send them to a remote server where data processing and classification is performed. They demonstrated the ability to recover up to $80\%$ of typed content by matching abstracted words with candidate dictionaries after classification. This research, therefore, shows the great potential of successfully inferring keystrokes from subtle motions such as small vibrations.
    
        \subsubsection{Touchlogging}
        \label{sssec:touchlogging}

Related works \cite{cai2011touchlogger, xu2012taplogger, owusu2012accessory, miluzzo2012tapprints} have shown that the data from the motion sensors of a smartphone can be used to infer keystrokes entered on its touchscreen. Cai et al. \cite{cai2011touchlogger} demonstrated that a malicious Android application can infer as much as $70\%$ of the keystrokes entered on a number-only virtual keyboard on an Android device. For their attack to work, the user, however, need to grant the application access to the motion sensors at install time. Cai et al. believed this assumption is not unrealistic considering that most users will not treat motion data as sensitive as camera or microphone for example.

In their paper, Owusu et al. \cite{owusu2012accessory} proposed a system that reads accelerometer data to extract 6-character passwords on an Android device. Their experiment consists of a QWERTY virtual keyboard used to perform the keystroke reconstruction attack. Additionally, a data collection screen is used to collect ground truth from acceleration measurements matching key presses at specific screen regions. They managed to break $59$ of $99$ passwords using only the accelerometer data.

Xu et al. \cite{xu2012taplogger} introduced a keystroke inference attack by using a Trojan application running on the Android platform. First, the host application is used to train the system when the user is interacting with it. Finally, keystroke inference is performed by the Trojan when the user enters sensitive information into the device (e.g. password of screen lock, numbers entered during a phone call). They were able to infer the majority of keys entered by the users in various experiment configurations.

Miluzzo et al. \cite{miluzzo2012tapprints} have demonstrated that the motion sensors built-in smartphones and tablets could be used to infer keystrokes entered on a complete 26-letters keyboard with an accuracy reaching as much as $90\%$. In their work, they have also shown that combining both the accelerometer and the gyroscope can leverage the accuracy of the classification. Their approach combines the results of multiple shallow classifiers to improve the prediction quality.

    \subsection{Classification of Motion Sensors Signal}
    \label{sssec:relatedworkclassifymotion}
    
Inferring keystrokes require the attacker to be able to associate measured raw sensor data with specific labels corresponding to the keys entered (i.e. the attacker needs to classify the motion signals). Motion sensors signal classification have been explored extensively in studies involving activity recognition \cite{bao2004activity, ravi2005activity, kwapisz2011activity} in the field of Pervasive Computing and gesture recognition \cite{pylvanainen2005accelerometer, schlomer2008gesture, liu2009uwave, wu2009gesture} in the area of Human-Computer Interaction. In both fields, the use of accelerometer sensors is historically studied more deeply although some studies explored sensors fusion to increase robustness \cite{shoaib2014fusion, lukowicz2002wearnet}. While approaches used in these fields can be borrowed, an important difference exists. In fact, the motion duration and the motion amplitude of a keystroke are respectively much shorter and lower than the motion caused by a gesture (e.g. finger swipe, hand waving) or an activity (e.g. running, sitting). Since classification is an approach to recognizing patterns in the signal, one can argue that patterns can emerge more significantly in long signals with high amplitude. Therefore making keystroke motions hard to classify.

        \subsubsection{Classifier Model}

According to Tanwani \cite{tanwani2009guidelines}, the classification accuracy of a given algorithm is largely dependent on the nature of the dataset rather than the classification algorithm itself. In fact, the choice of a classifier model varies greatly in the related studies with no major advantage of one model over one another; confirming that classification quality relies mostly on the feature extraction strategy and the quality of the training set (i.e. ground truth). Successful classification is traditionally only possible after suitable pre-processing to clean the signal and judicious feature extraction to select meaningful information from the data to represent the motion event in a relevant way. That is, good feature vectors contain features distinctive between motion signals from different keystrokes and consistent among motion signals caused by the same keystroke.

As mentioned by Cai et al. the smaller the required training set, the easier the attack. In their implementation \cite{cai2012practicality}, they showed that the inference accuracy level stabilizes when the training set reaches a certain size (i.e. 12 for alphabet-only keyboard and 8 for number-only keyboard). It is also important to note that the choice of the motion sensor can affect the quality of the classification. In fact, studies have shown that the gyroscope is a better side-channel than the accelerometer for keystroke inference \cite{al2013best, cai2012practicality, miluzzo2012tapprints}.

        \subsubsection{Data Analysis and Feature Extraction}

Processing data from motion sensors such as accelerometers and gyroscopes bring additional challenges. In fact, hardware specification such as the motion sensor sampling rate is not fixed and is much lower than the sampling rates of acoustic and electromagnetic sensors used in related eavesdropping attacks \cite{marquardt2011sp, cai2012practicality, al2013best}. In consequences, deciphering individual keystroke becomes difficult because the sensors return new values as sensor events only when these values are different from those reported in the previous event.

Cai et al. \cite{cai2012practicality} proposed a pre-processing solution allowing them to use signal analysis methods. The approach, termed de-jittering, consists of normalizing the sensors sampling rate to obtain a constant interval sampling. In their work, they build feature vectors from time domain data such as the duration of the motion data segment, the peaks time difference, the number of spikes, the peaks interval, the attenuation rate, and the vertex angle between peaks.
In another work, Cai et al. \cite{cai2011touchlogger} experimented with the patterns produced by the motion during keystrokes. They first identified the starting and ending time of keystrokes by calculating the Peak-to-Average Power Ratio of the pitch angle and the roll angle. Then they observed that when these angles values are plotted, distinctive lobes appear on the pattern with some interesting properties. In fact, they noticed that lobes directions are similar for same keys while the angles of the lobes vary for different keys. Based on this 2D representation, they built three pairs of features consisting of geometric metrics such as the angle between the direction axis of the upper lobe and the lower lobe with the x-axis, the angles of the two dominating edges, and finally the average width of both the upper and lower lobe.

Owusu et al. \cite{owusu2012accessory} solved the sampling rate problem by using an approach involving linear interpolation to create consistent sampling intervals throughout the recorded accelerometer data. They extracted the individual motion signals from each keypress using Root Mean Square anomalies for spike detection. In their work, they used a set of $46$ features consisting of $44$ acceleration stream information (i.e. min, max, Root Mean Square, number of local peaks, number of local crests, etc.) and two meta-features (i.e. total time and window size). A wrapper \cite{kohavi1997wrappers} was then used for feature subset selection to maximize the accuracy of the prediction.

In their work, Marquardt et al. \cite{marquardt2011sp} used a $100 ms$ long time window as Asonov et al. \cite{asonov2004keyboard} to extract features from the signal. They overcame the sampling rate problem by using a combination of domains to build the feature vectors. That is, time domain features (i.e. Root Mean Square, skewness, variance, kurtosis), spectral domain features (i.e. Fast Fourier Transform), and cepstrum features (i.e. Mel-frequency Cepstral Coefficients). For their attack to work, Marquardt et al. introduced an approach creating word profiles. First, each word in a training dictionary is broken down into its constituent characters and character-pairs. Secondly, they defined keypress events as pairs of successive events and their relation to each other. That is, the horizontal location (i.e. left or right) of each keypress and the distance between consecutive keypresses (i.e. near or far). Finally, words can be represented by profiles corresponding to the successive events leading to this word being entered. The feature vectors are then used to train two different classifiers, one for classifying left or right and one for classifying near or far. These labels are then used to specify the word profiles. During the logging phase, a word matching module is in charge of determining the actual text entered on the keyboard by scoring the predicted profiles against each word in a dictionary.

Xu et al. \cite{xu2012taplogger} selected features from the signal in a time window bounded by the touch-down and the touch-up events triggered when the user interact with the touchscreen. They used three features to determine if the touch event occurred on the left side or on the right side of the screen (i.e. roll angle variations) and three features to determine if the touch event occurred on the top or the bottom of the screen (i.e. pitch angle variations).

\subsection{Additional Mentions}

While \emph{Deep-Spying} was implemented, related studies worth mentioning were published. Wang et al. \cite{wang2015mole} developed a system to perform keylogging on a laptop keyboard using the motion sensors of a smartwatch. Their approach requires statistical features extraction to train a classifier, and non-trivial techniques such as point cloud fitting and Bayesian Inference.

Using the linear accelerometer built-in a smartwatch, Maiti et al. \cite{maiti2015smart} created a system to infer keystrokes entered by a user on a number-only keyboard displayed on a smartphone touchscreen. They managed to reach an accuracy of $90\%$ thanks to a solution relying on shallow classifiers trained using supervised learning with engineered feature vectors containing $54$ values.

In their paper, Liu et al. \cite{liu2015good} focused on performing keylogging with a smartwatch on both a number-only physical keypad and a standard QWERTY computer keyboard. They used the outputs of an accelerometer to train classification algorithms and acoustic emanations recorded from the microphone built-in the smartwatch to help identify keystrokes and segment the signal appropriately. The described approach depends on time domain features (i.e. displacement vectors) and spectral domain features (i.e. Fast Fourier Transform).

\chapter{Attack Description}
\label{ch:attack}

This chapter will provide an overview of three envisioned attack scenarios where an attacker (Eve) uses a smartphone wirelessly paired with a WAD worn by the victim (Alice) to perform a motion-based keystroke inference attack. All scenarios are similar regarding attacker goal, threat model, and methods employed. They only differ on the type of keypad being eavesdropped.

\section{Attacker Goals}

The attacker goal is to eavesdrop on the keys entered by the victim on a virtual or a physical keyboard. Eve could monitor Alice's keystrokes for various reasons.\footnote{Gaining access to Alice's private information could be used to impersonate Alice, to steal money or to get access to password-protected content for example.} Generally speaking, Eve might want to: 

\begin{itemize}
    \item Steal passwords and other credentials.
    \item Steal sensitive information such as PINs, social security numbers, and credit card numbers. 
    \item Direct spying by eavesdropping on messages typed.
\end{itemize}

\section{Threat Model}

Pairing a WAD with a smartphone is a risky task and can be a vector of choice for an attacker to gain access to the WAD. In fact, device pairing has been extensively studied in the field of Computer Security \cite{stajano2000resurrecting, mccune2005seeing, balfanz2002talking, castelluccia2005shake, goodrich2006loud} and has proven to be challenging. The establishment of a key between two devices in the presence of an active adversary remains a hard problem and existing solutions usually require trade-offs. With many users unaware of the risks, it is not unrealistic to consider the possibility of a WAD being paired with an attacker's smartphone. Bluetooth technology is currently the preferred medium used to pair smartwatches and fitness trackers with smartphones \cite{androidwear, applewatch, jawbone}. This communication channel being wireless further increases the risk of device pairing.\footnote{Wireless Networks offer by definition more opportunities to eavesdroppers than traditional wired networks because of their very nature of wireless transmission. The data is transmitted using radio waves through air, allowing anyone with a suitable receiver to collect and decode signals exchanged between two parties. The range of Bluetooth technology is application specific: Core Specification \cite{bluetoothspecifications} mandates a minimum range of $10$ meters while the signal can still be transmitted up to $100$ meters. External antennas can also potentially be used by an attacker to receive the signal from a further away location.}

Once the WAD is paired with the attacker's device, applications can be installed wirelessly on the WAD. Because the security risks of motion sensors are not well understood and often underestimated, current smartwatch operating systems (i.e. Android Wear 5.1 Lollipop, Apple Watch watchOS 2) does not require any user permissions for an application to use the motion sensors. Additionally, applications can run as Services in the background without displaying any GUI. Alice would, therefore, be unaware that an unknown application installed on her WAD by Eve is monitoring her motions.

\section{Attack Scenarios}

It is important to note that it is assumed that the victim is wearing the WAD on the wrist of the preferred hand used to interact with keyboards. In fact, in our attack scenario it would be harder, if not impossible, for the attacker to infer keystrokes if the victim is right-handed but is wearing the WAD on the left hand for example. The use of an exploit or the deployment of malware to compromise the WAD is not in the scope of this research work. Instead, it is assumed that the attacker has already compromised the WAD and can eavesdrop on the sensors output. The following attack scenarios will be studied to assess their feasibility.

\begin{figure}[H]
    \centering
    \includegraphics[width=.5\linewidth]{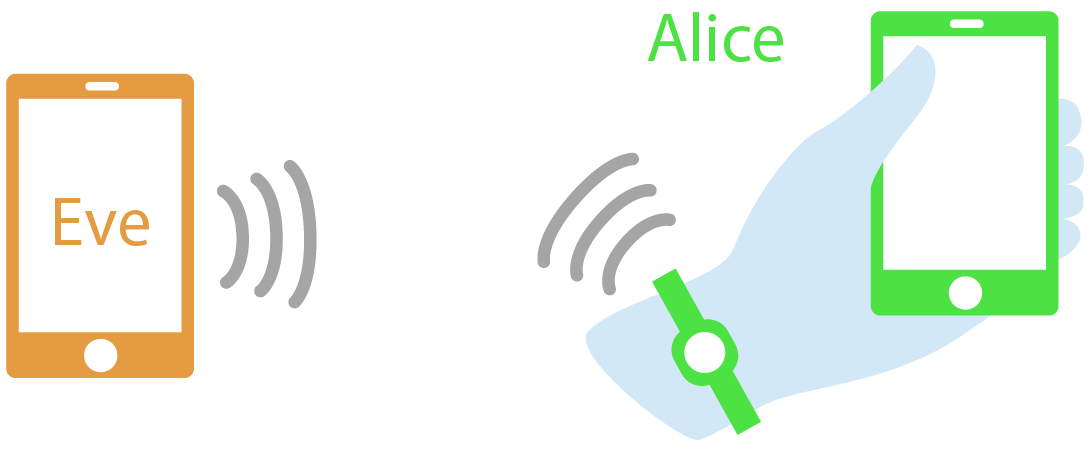}
    \caption{Smartphone attack overview.}
    \label{fig:attack_phone}
\end{figure}

\noindent
\textbf{Touchlogging Attack on Smartphone Touchscreen:}
Alice is typing her PIN code on the touchscreen virtual keyboard of her smartphone and Eve is trying to infer the keystrokes from the motion sensors signal of the compromised WAD as illustrated in Figure \ref{fig:attack_phone}.

\begin{figure}[H]
    \centering
    \includegraphics[width=.5\linewidth]{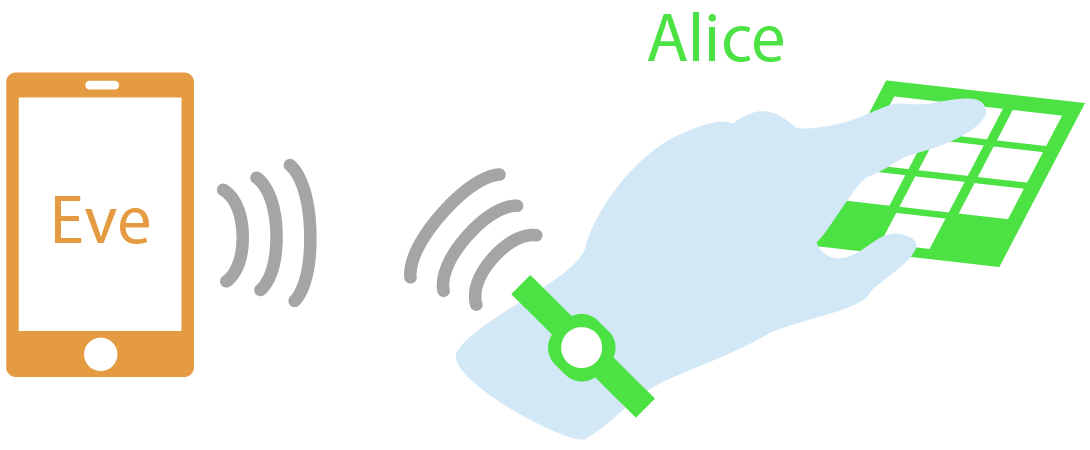}
    \caption{ATM attack overview.}
    \label{fig:attack_atm}
\end{figure}

\noindent
\textbf{Keylogging Attack on ATM-like Keypad:}
Alice is entering her credit card password on an ATM-style physical keypad and Eve is trying to infer the keystrokes from the motion sensors signal of the compromised WAD as shown in Figure \ref{fig:attack_atm}.

\noindent
\textbf{From Touchlogging Training to Keylogging Attack:}
This attack scenario is based on two assumptions. First, that it would be difficult, for example, for an attacker to gain access to an ATM to train his algorithm with labeled data. However, getting access to, or installing malware on the victim's smartphone is more realistic. The second assumption is that the motion of the user's hand while typing on a touchscreen is extrapolated to be similar to the motion when she is typing on a physical keypad. Conceding that the typing surfaces are oriented with equivalent angles and that Alice is typing using the same technique on both user interfaces. That is, if the thumb is preferred to enter keystrokes on touchscreens, the thumb should also be used to enter keys on physical keypads. This for the resulting motions to potentially display similar patterns. This attack scenario will thus explore the practicality of a machine learning algorithm trained for touchlogging to be used as such to perform keylogging attacks targeting unfamiliar devices.
\chapter{System}
\label{ch:system}

This chapter's purpose is twofold. First, to introduce the reader to the system architecture, its different components, and their relationships. Second, to describe each component respective purpose and the methods employed for their implementation.

\section{System Architecture}

The system should take WAD sensor data as input and infer keystrokes as output. The main architectural model adopted is \emph{Client-Server} because of its flexibility. In fact, this distributed system paradigm allows client machines with limited computational resources (e.g. mobile devices, wearable computers) to delegate heavy computations to more powerful machines such as a networked server. A server host provides services to the different clients connected to it while the clients initiate communication sessions with the server that await incoming requests.

\begin{figure}[ht]
    \centering
    \includegraphics[width=.7\linewidth]{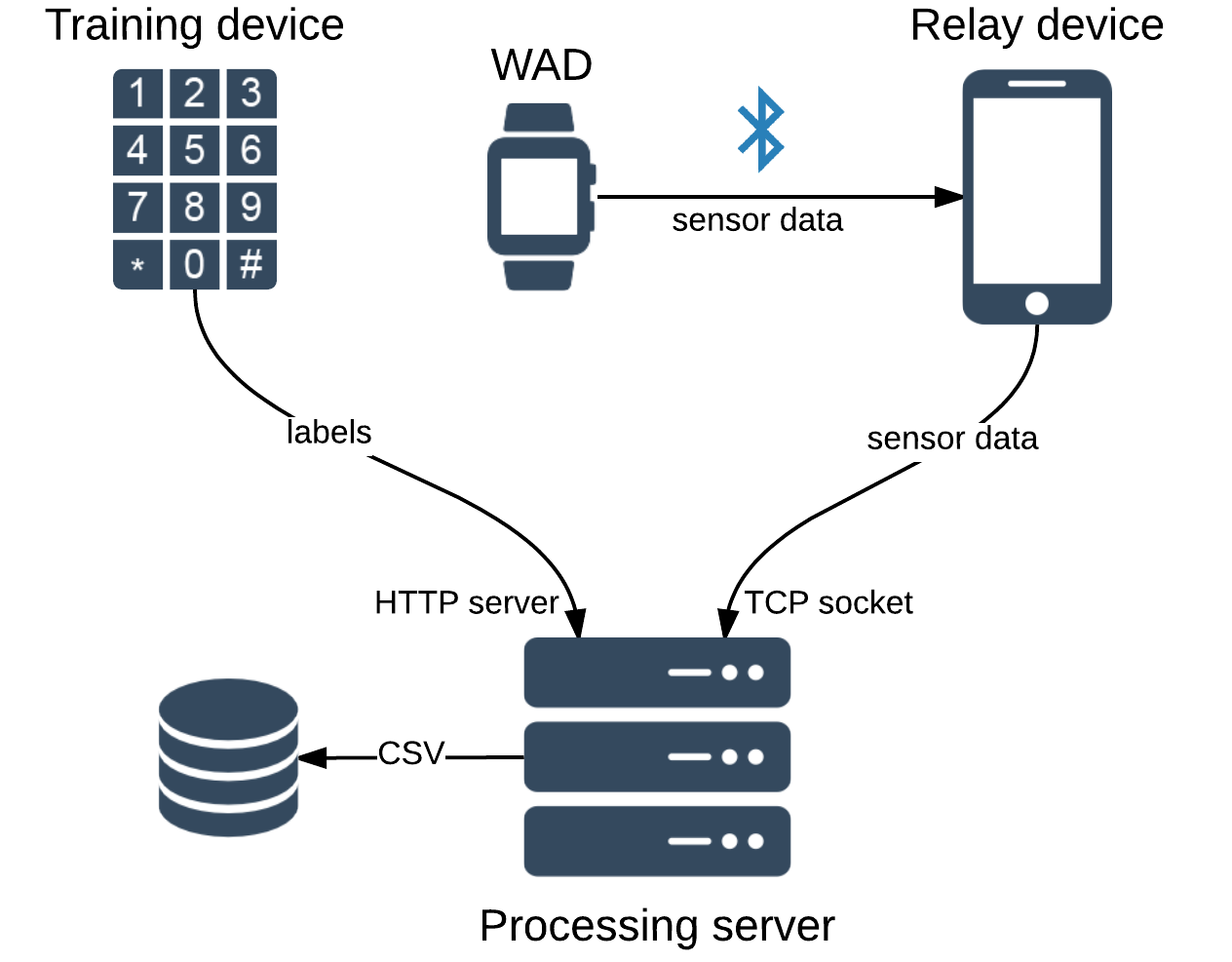}
    \caption{System overview using the Client-Server architectural model.}
    \label{fig:architecture}
\end{figure}

As shown in Figure \ref{fig:architecture}, the system consists of two clients connected to a processing server on the same local network. A smartphone client is acting as a proxy by receiving motion sensor events from the WAD and by relaying the data to the processing server. In the training phase, a second client is in charge of sending the labels to the server. One can see here an important advantage of the Client-Server model. This architecture indeed allows flexibility to experiment with different types of training devices without having to change the rest of the system. That is, the very same services provided by the networked server are directly available to any client connected to the same local network.

\section{Client}

    \subsection{Wearable Application}

A wearable application is implemented to read the sensor values from the WAD and was tested on a Sony SmartWatch 3. The developed software is written in Java and designed to be deployed on devices running Android Wear API level 21. The application primary goal is to listen to accelerometer and gyroscope sensor events and to send them for further processing. Because such WAD has limited networking capabilities, the smartwatch is not able to send motion sensors data directly to the server. Instead, the smartwatch is paired using Bluetooth with a compatible Android device to establish a communication channel allowing the use of the Android Wearable Data Layer API to encapsulate the sensor data. The paired mobile phone is then relaying the data to the server.
        
    \subsection{Mobile Application}

\begin{figure}[ht]
    \centering
    \includegraphics[width=1.\linewidth]{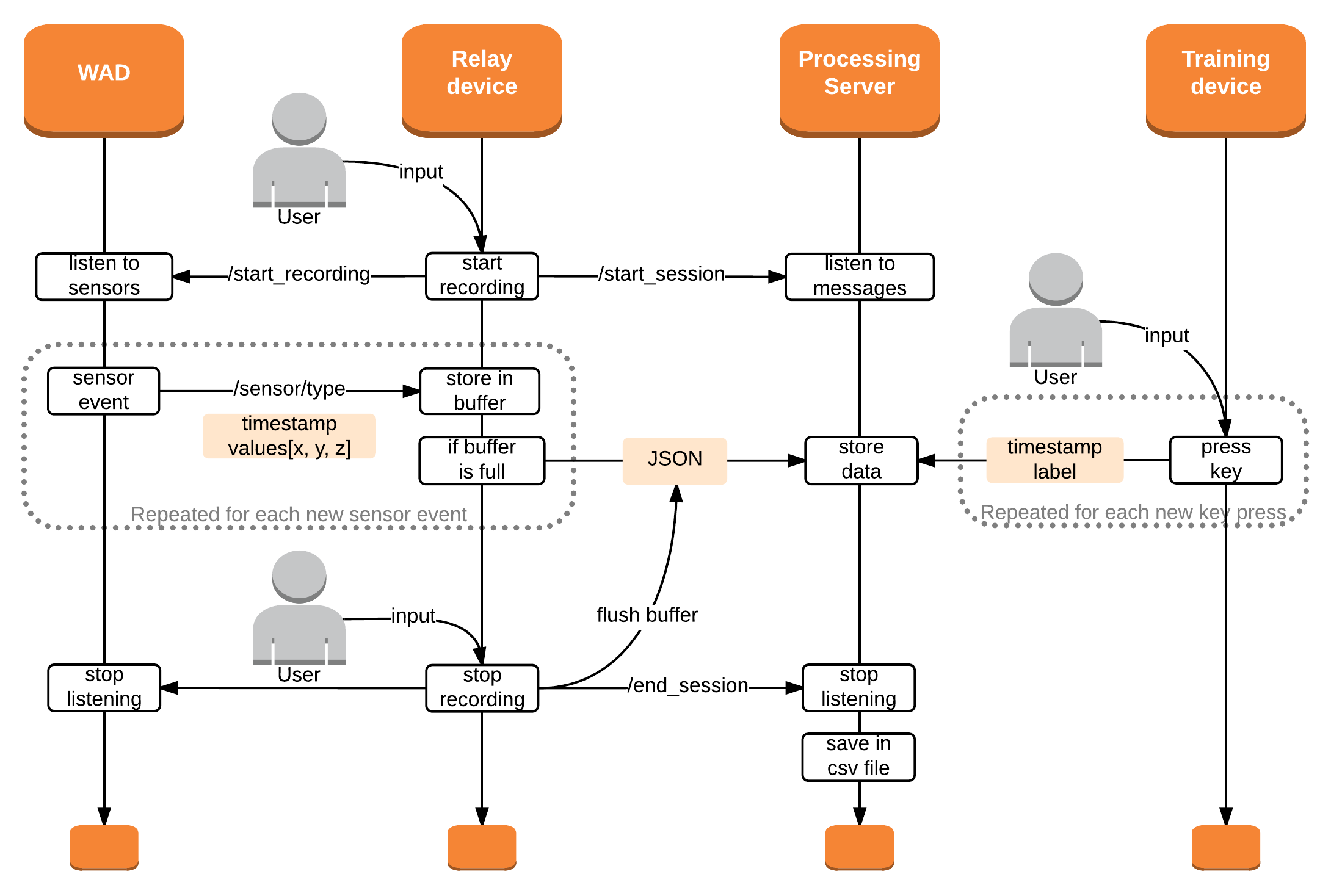}
    \caption{Components communication during data acquisition.}
    \label{fig:acquisition}
\end{figure}

This mobile application is needed for the relay device and was tested on an LG Nexus 4 smartphone and implemented in Java to target devices running Android API level 19. As shown in Figure \ref{fig:acquisition}, a recording session begins when a user requests the mobile client to start recording. The smartphone then sends one message to the server to initiate a new session and one message to the WAD to start listening to motion sensor events. When the user is typing on the training client, labels with timestamps\footnote{The timestamps are measured in $ms$ and require the devices (i.e. the WAD and the training device) to have synchronized clocks.} are sent to the server. In the meantime, the user's hand motions are recorded by the WAD and reported to the relay device as an Android Wearable Data Layer message containing a timestamp value, a three-dimensional array (i.e. one dimension per axis), and the type of sensor that spread the event (i.e. gyroscope or accelerometer). The relay device stores the data locally in a buffer for each sensor type. Once a buffer is full (i.e. reach a defined size limit), the data are serialized to the JSON data-interchange format \cite{jsonformat} and sent to the server through a TCP socket connection. JSON makes it easy for humans to read the data and for machines to parse, which enable fast to implement and less error-prone cross-device communication protocols.

A recording session stops when the user requests the mobile client to stop recording. The relay device then sends a message to the WAD to stop listening to motion sensors and wait until it receives the last message that was waiting to be transmitted.\footnote{As a result of Bluetooth limited throughput and the important amount of motion events to be sent, the relay device sometimes need to wait few minutes for all the packets to be received successfully from the WAD.} Once the WAD receives the last message, the smartphone flushes the buffer by serializing and transmitting all of its remaining data. Finally, the relay smartphone sends a message to the server to close the session.
        
    \subsection{Training Application}
    
Two different training softwares are implemented to experiment with both touchlogging and keylogging. Despite the fact that the technologies used are different for the two training platforms, both applications are communicating with the server using the same communication protocol. During training, labels corresponding to entered keys and their respective timestamps are serialized to JSON and sent to the server using the HTTP protocol \cite{httpprotocol}.
        
\begin{figure}[H]
    \begin{subfigure}{.45\textwidth}
        \centering
        \includegraphics[width=.6\linewidth]{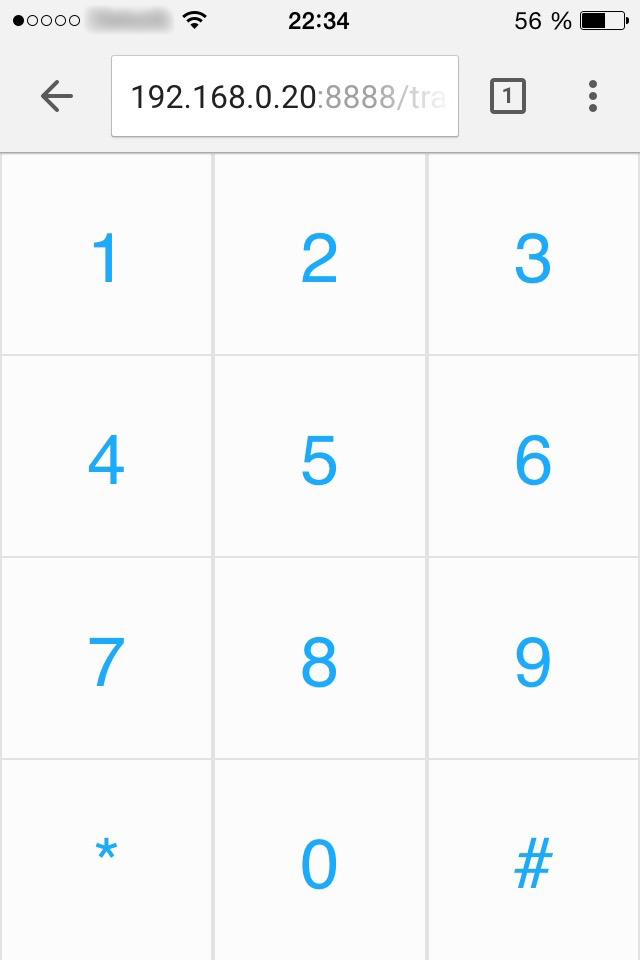}
        \caption{Web application for touchlogging attack scenario.}
    \end{subfigure}
    \begin{subfigure}{.55\textwidth}
        \centering
        \includegraphics[width=.8\linewidth]{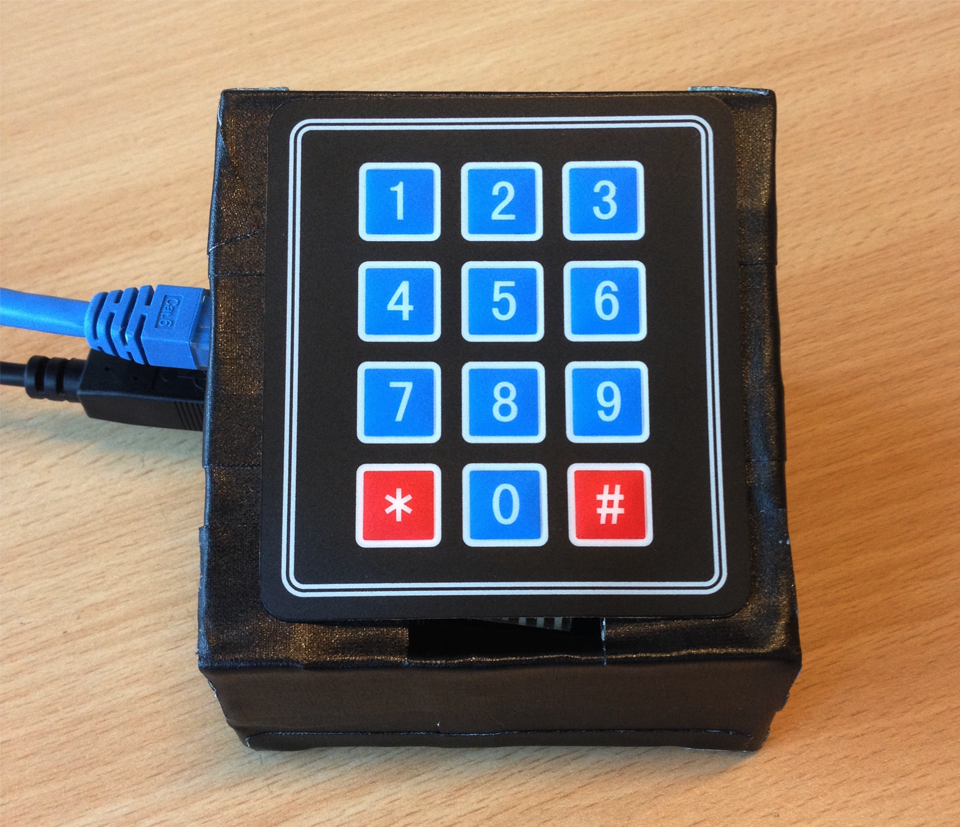}
        \caption{Physical device prototype for keylogging attack scenario.}
    \end{subfigure}
    \caption{Training devices.}
    \label{fig:trainingdevices}
\end{figure}

\noindent
\textbf{Touchlogging:}
To enable any touchscreen device (e.g. smartphone, tablet) to be used as a training application, the touchlogging training application is implemented using web technologies HTML and JavaScript. The UI can thus be scaled and adapted to any screen size as shown in Figure \ref{fig:trainingdevices} (a) where the interface is displayed on an iPhone 4S screen. The virtual keyboard is displayed on a surface of size $65mm \times 50mm$.

\noindent
\textbf{Keylogging:}
The keylogging training application is implemented in C++ for the Open-Source Arduino microcontroller \cite{arduino}. This training device consists of an Arduino UNO board with an Ethernet shield for networking capabilities and a physical keypad for label input. Because of hardware limitations, the application need to request the current timestamp in $ms$ to the server at the beginning of the training session.\footnote{The microcontroller stores $long$ variables on $32 bits$ and lack a Unix epoch clock. The rest of the system is using Unix timestamps in $ms$ consisting of $13$ digits hence require at least $41 bits$ to be stored assuming that the first digit will not change until $2033$ from now. Therefore making the hardware unable to handle such numbers. To solve this problem, it is first assumed that recording sessions are under one hour in duration. Then the reference timestamp received from the server is split: the first $5$ digits are stored in a $char$ array, and the remaining $8$ digits are stored in a $long$ which require now only $27$ bits. When a key is entered by the user, the $long$ value is incremented by the number of $ms$ that have passed since the time at which the training session started. This value is concatenated with the first digits stored in the $char$ array and ready to be sent to the server.} The built physical prototype is depicted in Figure \ref{fig:trainingdevices} (b). The keypad effective surface is $55mm \times 45mm$ in size.

\section{Server}

The server is needed to fulfil two main tasks. First to receive data from the different clients, organize them and save them persistently. Secondly and most importantly to perform data analytics by using the previously saved data; namely keystroke inference from motion sensor measurements.

\begin{figure}[ht]
    \begin{subfigure}{.5\textwidth}
        \centering
        \includegraphics[width=1.\linewidth]{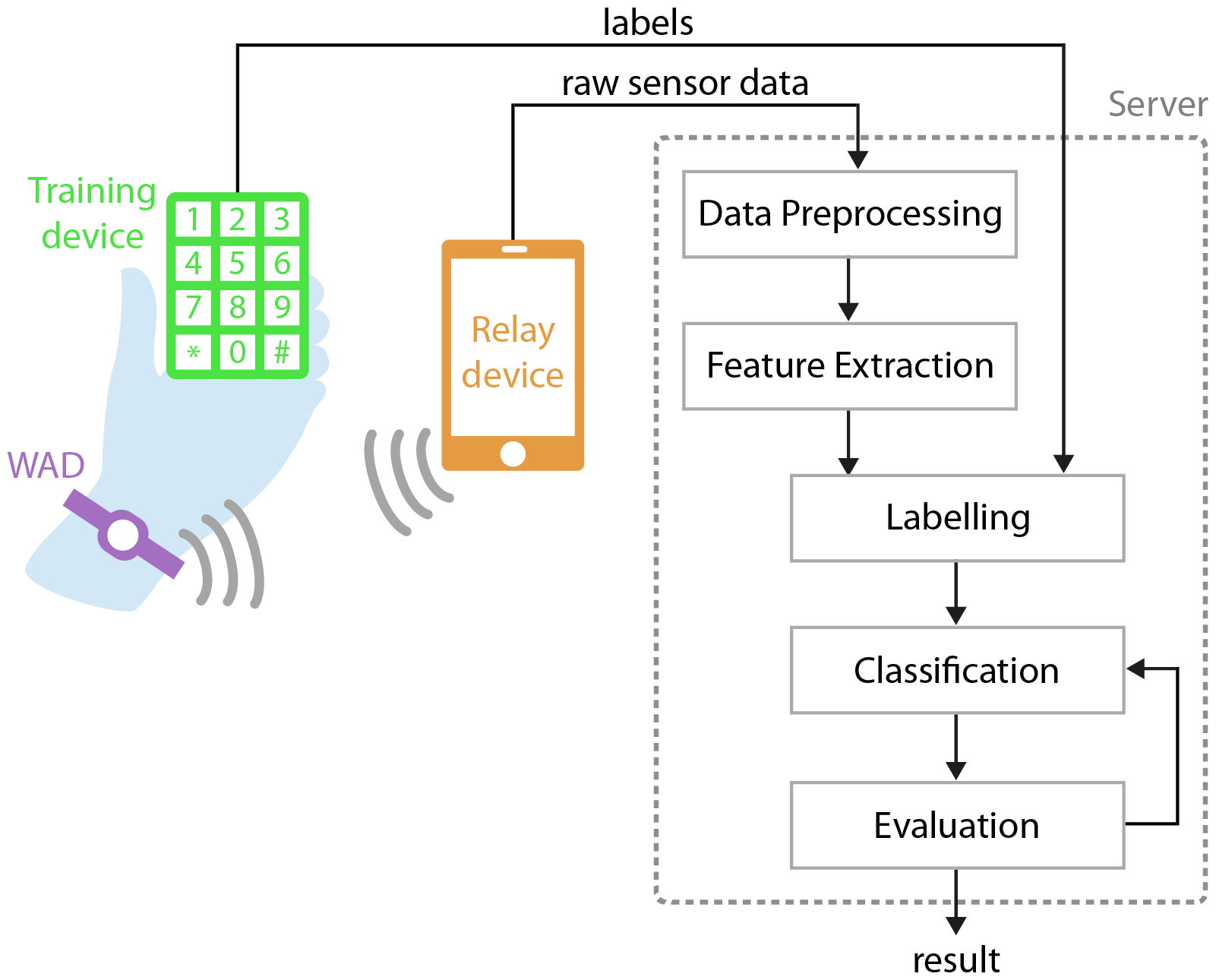}
        \caption{Training phase.}
    \end{subfigure}
    \begin{subfigure}{.5\textwidth}
        \centering
        \includegraphics[width=.9\linewidth]{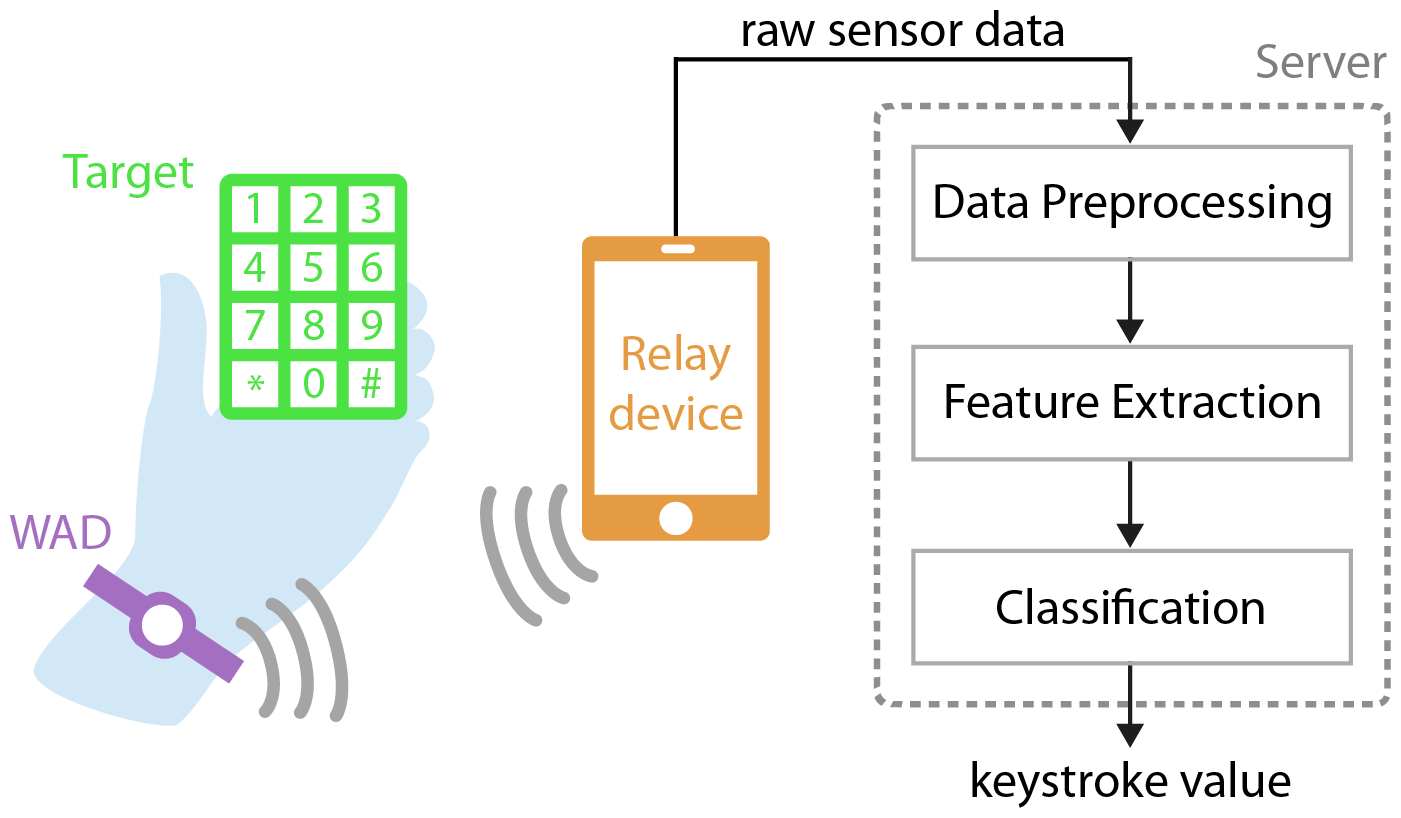}
        \caption{Logging phase.}
    \end{subfigure}
    \caption{Data analytics server main components.}
    \label{fig:archanalytics}
\end{figure}

The TCP socket and HTTP modules used to manage the data acquisition process are implemented in Java, and the data analytics process is implemented in Python to benefit from flexible data structures and scientific computation tools.\footnote{The ANNs are implemented using modules available in the PyBrain Open-Source library \cite{pybrain} with a C++ wrapper additionally used to speed up the computations. Some experiments were also performed using Torch \cite{torch} and the programming language Lua.} When the server receives the end-of-session message from the relay device, the data are first sorted by timestamps because they are not guaranteed to be ordered when received. The server then saved them in one CSV file per sensor. Figure \ref{fig:archanalytics} presents the main components of the data analytics pipeline and their connections. The raw data can initially be pre-processed to mitigate the effect of noise and measurement inaccuracies. Features are then extracted from the data in time windows corresponding to the keystrokes duration. During the training phase, the classification model is trained with the extracted features by iteratively evaluating the prediction outputs until a satisfying accuracy is reached or a maximum number of iterations have occurred. At the end of this phase, the trained model is serialized in XML and saved persistently. During the logging phase, the classification model is deserialized from the file system, and its inputs are activated with newly recorded data to attempt to predict the labels.
\chapter{Data Analytics}

\begin{figure}[H]
    \begin{subfigure}{1.\textwidth}
        \centering
        \includegraphics[width=.9\linewidth]{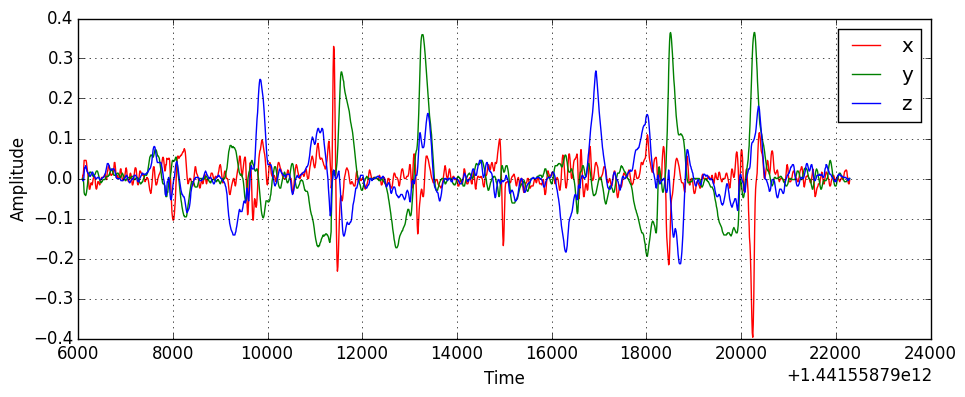}
        \caption{Gyroscope.}
    \end{subfigure}
    \begin{subfigure}{1.\textwidth}
        \centering
        \includegraphics[width=.9\linewidth]{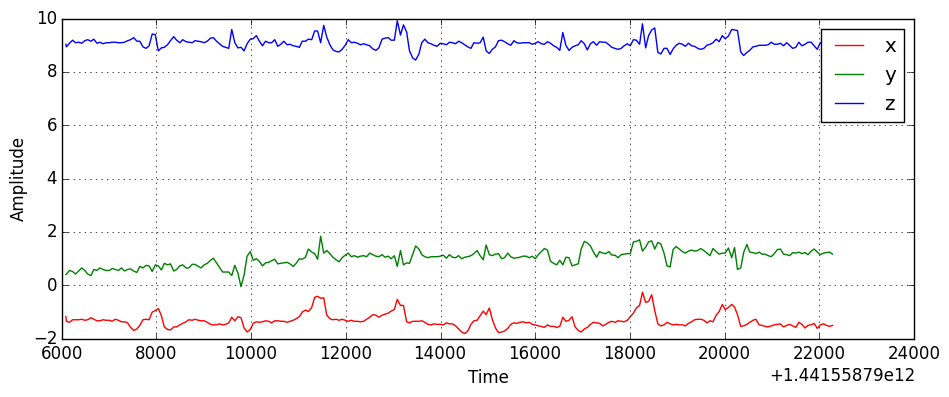}
        \caption{Accelerometer.}
    \end{subfigure}
    \caption{Noisy raw signals recorded from motion sensors.}
    \label{fig:rawsig}
\end{figure}

The first objective of this chapter is to detail how the data are recorded and what methodologies have been used to clean the signal. As shown in Figure \ref{fig:rawsig}, the raw signal is subject to noise and, therefore, can be pre-processed before to be used for data analysis purpose. On one hand, the important amount of noise can potentially obfuscate patterns and largely alter the classification accuracy. On the other hand, a deep neural network architecture would, in theory, be able to handle such noisy data. Pre-processing can thus be applied optionally depending on the experiments.

\begin{figure}[H]
    \centering
    \includegraphics[width=1.\linewidth]{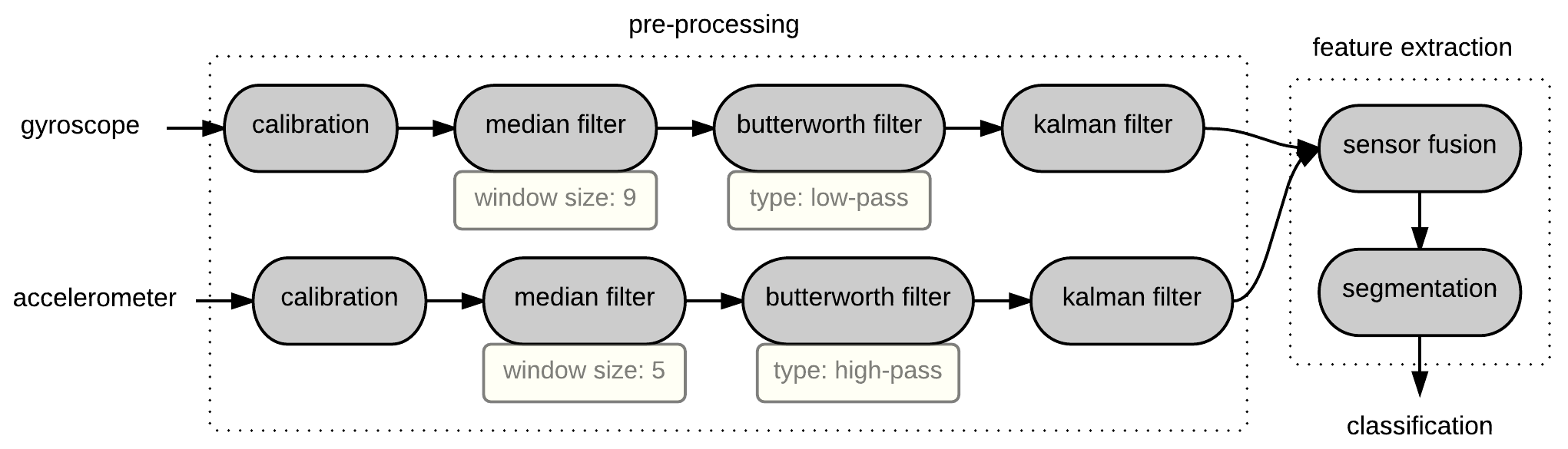}
    \caption{Data processing pipeline.}
    \label{fig:pipeline}
\end{figure}

To remove noise and smooth the signal, four pre-processing operations are applied to both sensors data as illustrated in Figure \ref{fig:pipeline}. The reader is invited to refer to Appendix \ref{ch:appendixprocessing} for visual examples depicting the improvement of the signal after each pre-processing operation on a $12$-keystrokes signal. Secondly, this chapter details the signal segmentation approaches and the feature extraction strategy employed to reduce the data to a meaningful selection. Finally, the statistical models chosen to perform classification are described.
   
\section{Data Acquisition}
\label{sec:dataacquisition}

\noindent
\textbf{Sensor Recording:}
The data are acquired from the gyroscope and accelerometer sensors built-in the smartwatch. Sensor events data are stored in tuples $(t_{i}, x_{i}, y_{i}, z_{i}),i=1...n$, where $t_{i}$ is the time in $ms$ at which the event occurred, $x_{i}$, $y_{i}$, $z_{i}$ are the values along the three axes $x$, $y$, and $z$, respectively, and $n$ is the total number of motion sensor events in an entire recording session. We observed that while the sampling rate was not constant, the delay between sensor events varies slightly enough for us to initially ignore the sampling rate problem during pre-processing.

\noindent
\textbf{Label Recording:}
The training device reports labels in the form of tuples $(t_{j}, l_{j}),j=1...m$, where $t_{j}$ is the time in $ms$ at which the keystroke appended, $l_{j}$ is the label (i.e. the value of the entered key), and $m$ is the total number of keystrokes in the entire recording session.

\section{Pre-processing}

\subsection{Calibration}
\label{ssec:calibration}
Both motion sensors need to be calibrated to align all three axes. In fact, the accelerometer axes contain values in different absolute ranges because of the effect of gravity (as illustrated in Figure \ref{fig:rawsig} (b)). Although the gyroscope axes should average to zero, a small non-zero difference was observed. Calibration is performed by subtracting each sensor value with the mean of its axes, such that:

\begin{equation} \label{eq:calib}
    \underset{v \in \{X, Y, Z\}}{f(v_{i})} = v_{i} - \overline{v}
\end{equation}

\begin{figure}[H]
    \begin{subfigure}{1.\textwidth}
        \centering
        \includegraphics[width=.9\linewidth]{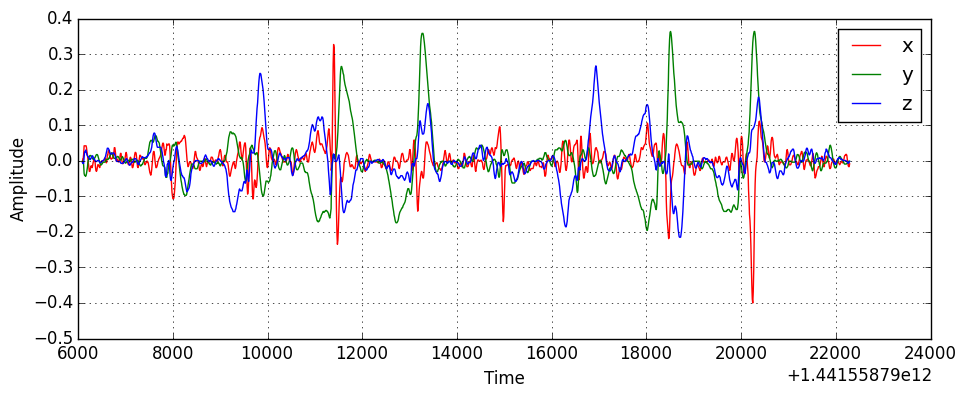}
        \caption{Gyroscope.}
    \end{subfigure}
    \begin{subfigure}{1.\textwidth}
        \centering
        \includegraphics[width=.9\linewidth]{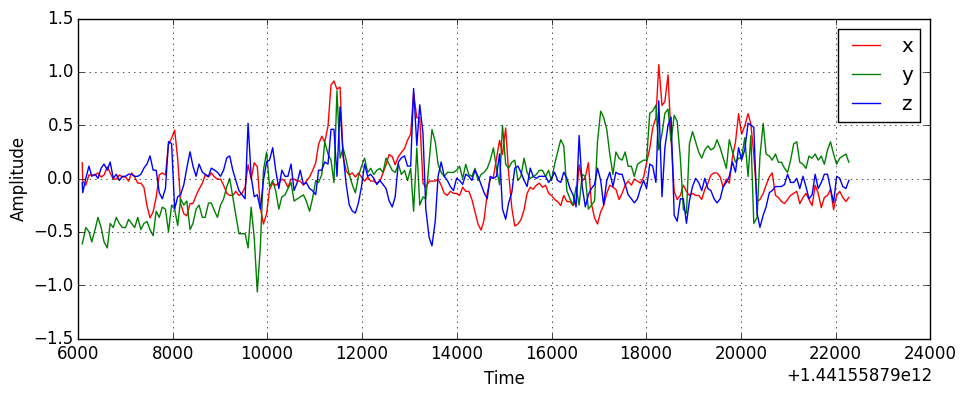}
        \caption{Accelerometer.}
    \end{subfigure}
    \caption{Signals after calibration.}
    \label{fig:calibsig}
\end{figure}

Where $v$ is the amplitude value on the given axis. The result of this operation can be seen in Figure \ref{fig:calibsig}.

\subsection{Median Filter}
The moving median filter is a first pre-processing step used to mitigate the effect of noise in the data. The moving mean filter is a possible alternative but has the disadvantage to attenuate the trends in the data. The moving median removes the noise while preserving the signal pattern and is applied with a sliding window to compute the median value in a fixed range \cite{justusson1981median}, that is:

\begin{equation}
    \underset{v \in \{X, Y, Z\}}{f(v_{i})} = Median(v_{i-\frac{1}{2}(w-1) },...,v_{i},...,v_{i+\frac{1}{2}(w-1)})
\end{equation}

\begin{figure}[H]
    \begin{subfigure}{1.\textwidth}
        \centering
        \includegraphics[width=.9\linewidth]{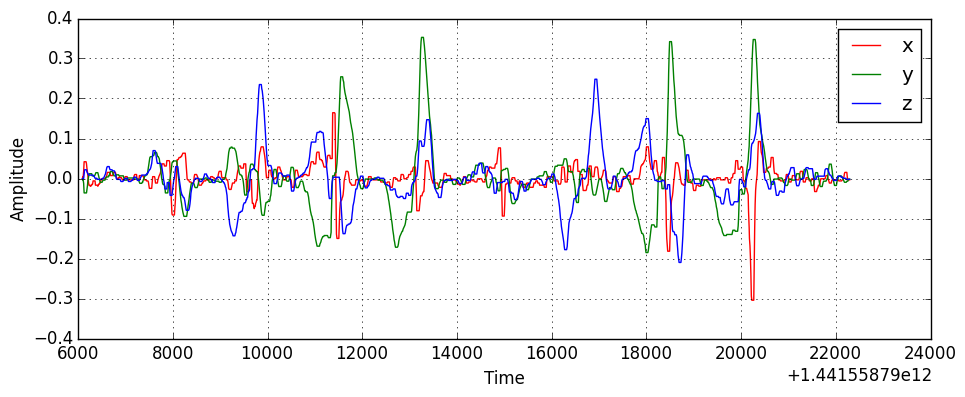}
        \caption{Gyroscope.}
    \end{subfigure}
    \begin{subfigure}{1.\textwidth}
        \centering
        \includegraphics[width=.9\linewidth]{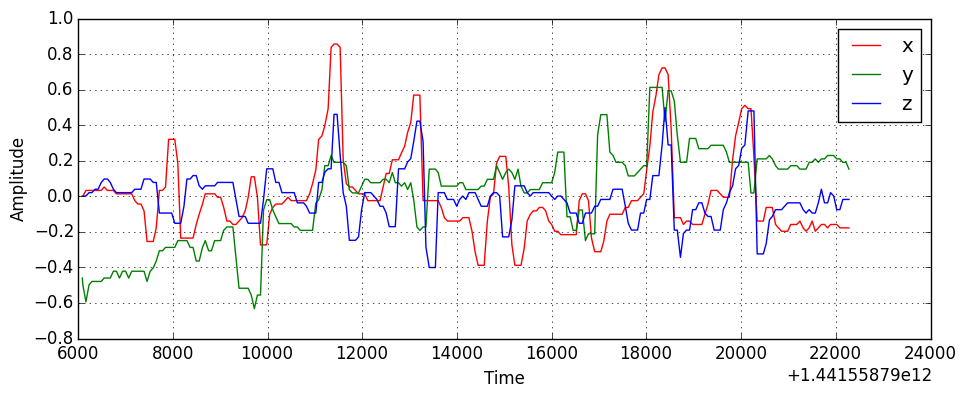}
        \caption{Accelerometer.}
    \end{subfigure}
    \caption{Signals after median filtering.}
    \label{fig:mediansig}
\end{figure}

Where $v$ is the amplitude value on the given axis and $w$ is an odd number representing the size of the sliding window. Since the sensors sampling frequencies are different, the number of data-points at the end of a recording session is different. Hence, the sliding window size has to be different for each sensor to remove noise optimally while preserving the signal as much as possible. Experiments show that $w = 9$ and $w = 5$ provide satisfying filtering results for the gyroscope and the accelerometer, respectively. As shown in Figure \ref{fig:mediansig}, the operation helps to remove noise but the signals need to be further processed to be smoother.

\subsection{Butterworth Filter}
The Android API makes it possible to define a delay at which the events are received. The documentation however clearly stipulates that the specified delay is just a hint to the system and that events can be received faster or slower than the specified rate. In our context, knowing the maximum delay between sensor events in the worth case scenario allow us to optimize the pre-processing algorithm to clean the signal appropriately. The maximum delay for the gyroscope and the accelerometer was found to be $10\;000 \mu s$ and $62\;500 \mu s$, respectively.
Elementary physics tells us that the sampling rate frequency can be computed from the sampling delay such that:
        
\begin{equation} \label{eq:freq}
    f = \frac{1}{d \cdot 10^{-6}}
\end{equation}
        
Where $d$ is the sampling delay in $\mu s$ and $f$ is the frequency in $Hz$. Using the maximum delay measured in the data acquisition step, we can estimate that the sensors will report new data with a frequency of $100 Hz$ and $16 Hz$ for the gyroscope and the accelerometer, respectively.

\begin{figure}[H]
    \begin{subfigure}{1.\textwidth}
        \centering
        \includegraphics[width=.9\linewidth]{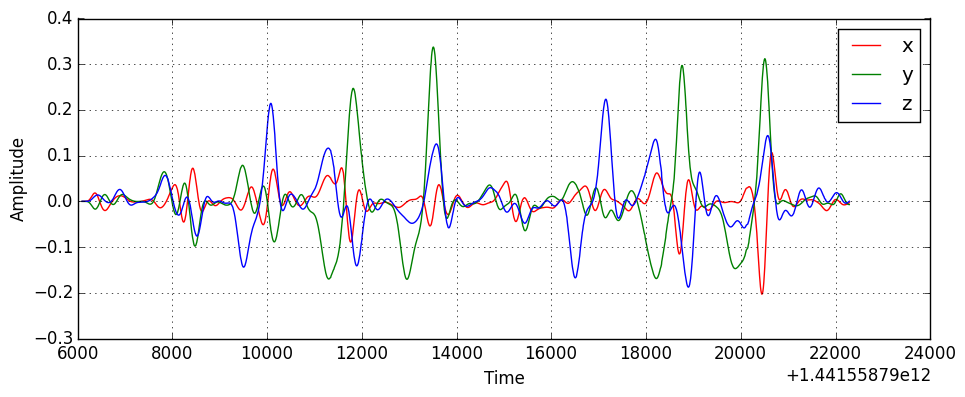}
        \caption{Gyroscope processed with low-pass filter.}
    \end{subfigure}
    \begin{subfigure}{1.\textwidth}
        \centering
        \includegraphics[width=.9\linewidth]{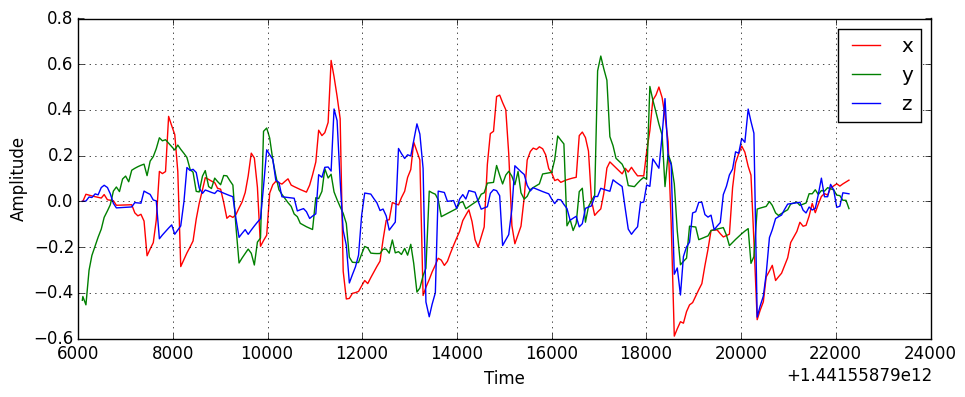}
        \caption{Accelerometer processed with high-pass filter.}
    \end{subfigure}
    \caption{Signals after Butterworth filtering.}
    \label{fig:butterworthsig}
\end{figure}

For the gyroscope, we only want to keep signals with a frequency lower than a certain cutoff frequency and attenuates signals with frequencies higher, hence the use of a low-pass filter. At contrary for the accelerometer, we want to attenuate signals with frequencies lower than the cutoff frequency, thus the need to use a high-pass filter \cite{butterworth1930theory}. Using the frequencies previously calculated, the filters can be applied with \emph{Nyquist frequencies} set to $8 Hz$ and $50 Hz$, for the gyroscope and the accelerometer, respectively. The resulting signals can be seen in Figure \ref{fig:butterworthsig}.

\subsection{Kalman Filter}
\begin{figure}[H]
    \begin{subfigure}{1.\textwidth}
        \centering
        \includegraphics[width=.9\linewidth]{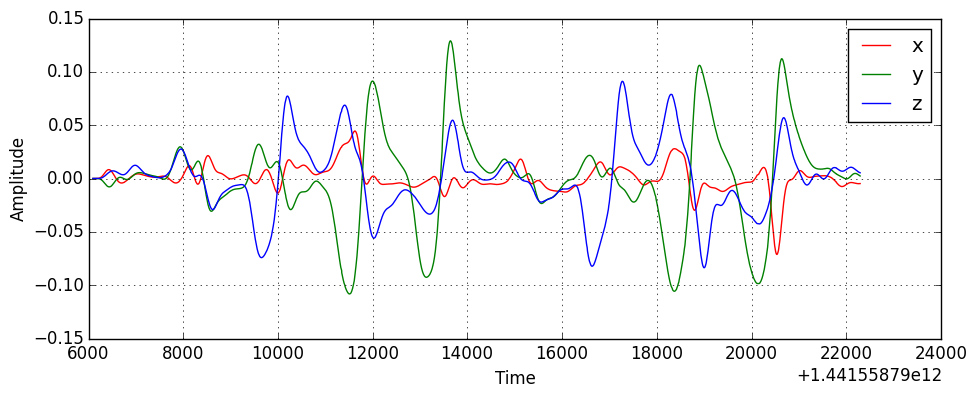}
        \caption{Gyroscope signal after Kalman filtering.}
    \end{subfigure}
    \begin{subfigure}{1.\textwidth}
        \centering
        \includegraphics[width=.9\linewidth]{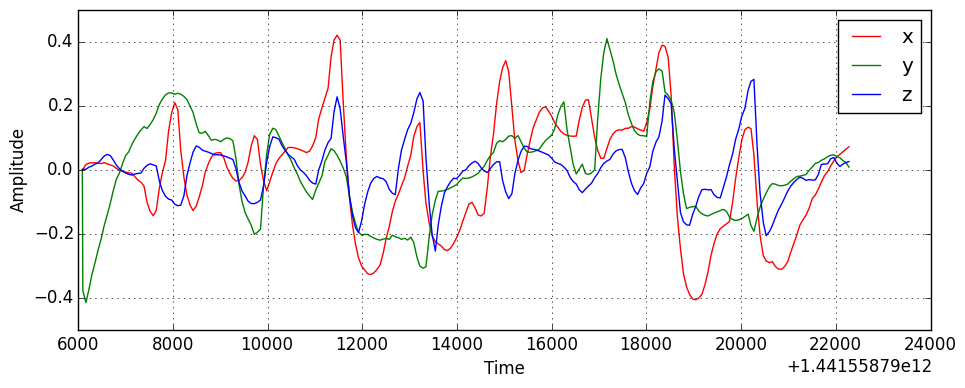}
        \caption{Accelerometer signal after Kalman filtering.}
    \end{subfigure}
    \caption{Signals ready for feature extraction after the last step of the pre-processing pipeline.}
    \label{fig:kalmansig}
\end{figure}

The \emph{Kalman Filter} algorithm produces estimates that minimize Mean Squared Error by observing a series of measurements. Even with data containing statistical noise, the filter can produce estimates allowing patterns to emerge more significantly from the signal \cite{haykin2004kalman}. This advanced filtering technique proved to be useful to our application context by smoothing the signal evenly and attenuating irregular peaks and pits (as shown in Figure \ref{fig:kalmansig}).

\begin{figure}[H]
    \centering
    \includegraphics[width=1.\linewidth]{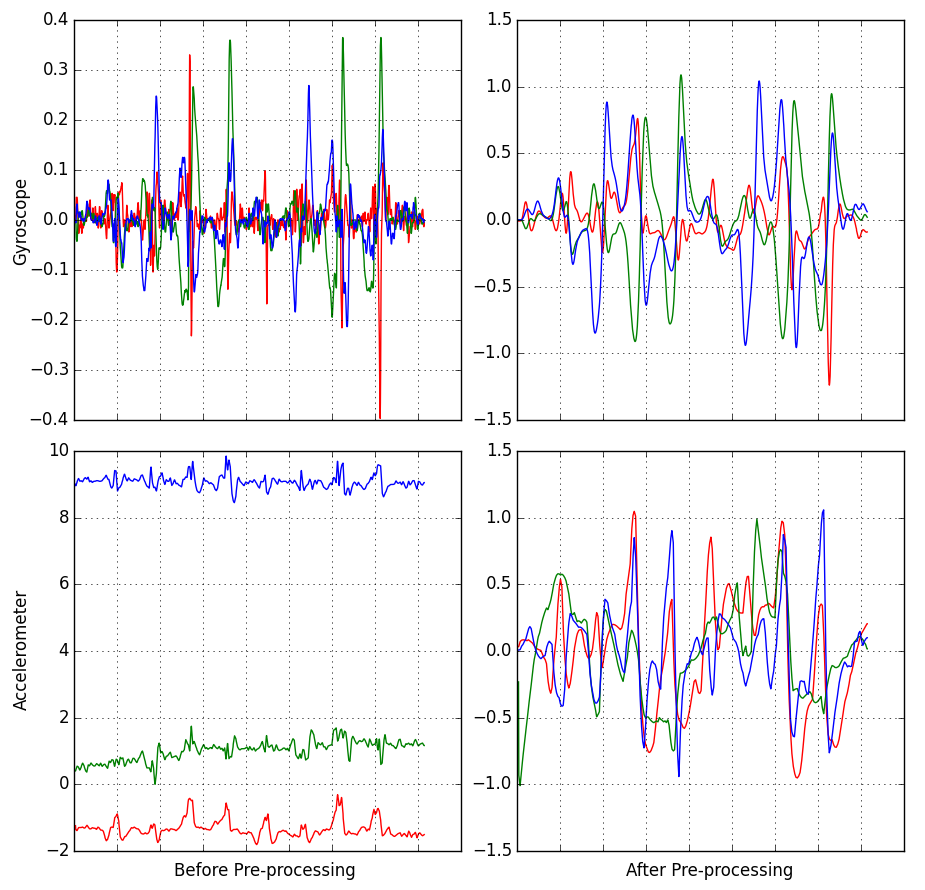}
    \caption{Effect of the pre-processing pipeline on the recorded signal.}
    \label{fig:overview}
\end{figure}

The signals are finally normalized before to be returned. The output of the pre-processing pipeline is a smoother signal where the effect of noise is reduced while the signal patterns are preserved as much as possible. Figure \ref{fig:overview} compares both motion sensor signals before and after being pre-processed.

\section{Feature Extraction}
\label{sec:featureextraction}

\subsection{Sensor Fusion}
\label{ssec:sensorfusion}

As mentioned in Section \ref{sssec:relatedworkclassifymotion}, studies have shown that sensors fusion can increase the robustness of motion sensor outputs classification. In fact, sensors can be subject to inaccurate measurements while recording, merging their outputs with other sensors can minimize uncertainty and provide more accurate measurements. However, the accelerometer is recorded with a sampling frequency significantly lower than the gyroscope\footnote{By a factor of $6.25$ according to the sensors maximum delay measured in the data acquisition step. This factor was confirmed experimentally by dividing the total number of data-point recorded for the gyroscope with the total number of data-point recorded for the accelerometer during a recording session.}, thus making sensor fusion hard using the recorded data as such. Moreover, data-points need to be evenly distributed to allow trends to emerge from the data. The sampling rate can be made constant by distributing the data-points evenly over the complete recording session duration. The implemented algorithm is described as follows:

\begin{enumerate}
    \item First, create a new set $T'$ of $k$ elements with:
    
    \begin{equation}
        k = \frac{1}{\alpha}(t_{\,n} - t_{\,0}) + 1
    \end{equation}
    
    Where $t$ is the timestamp at which a motion sensor event have been recorded, $n$ is the total number of events in an entire recording session, and $\alpha$ is a constant integer referring to the target sampling rate that was defined to be $2 ms$.
    Then populate $T'$ such that:
    
    \begin{equation} 
        \underset{t' \in \;T'}{t'_{\;i}} = t_{\,0} + \alpha\,i
    \end{equation}
    
    Where $i$ is the index position of the time value in $T'$.
 
    \item Secondly, perform a union operation $T'' = T' \cup T$ where $T'$ is the previously generated target set with constant time intervals, and $T$ is a set generated from the measurements with variable time intervals. The operation ensures that no timestamp duplicates exist and return a new set $T''$ of size $k'$.
    
    \item Third, create a new list of tuples $(t'_{\;i}, x'_{\;i}, y'_{\;i}, z'_{\;i}),i=1...k'$ where $t'$ is set to the values in $T''$, while $x'_{\;i}$, $y'_{\;i}$, and $z'_{\;i}$ are respectively set to $x_{i}$, $y_{i}$, and $z_{i}$ when the values are known from the recorded measurements (i.e. when $t'_{\;i} \in T$).
  
    \item Fourth, linear interpolation is used to compute the unknown values in the tuples list, that is:
    
    \begin{equation} \label{eq:linearinterpolation}
        \underset{v \in \{X', Y', Z'\}}{f(v_{i})} = v_{i - 1} + \frac{1}{j - (i - 1)}(v_j - v_{i - 1})
    \end{equation}
    
    Where $v_{i}$ is an unknown value on the given axis at the position index $i$, and $j$ is the position index of the next known value in the time series. Note that the algorithm is computing missing values from left to right so $v_{i - 1}$ is always known when $v_{i}$ is computed.
    
    \item Finally, since the known values have now been used to compute the missing data-points, the last step consists of keeping only the tuples where $t'_{\;i} \in T'$, the previously generated target set with constant time intervals.
\end{enumerate}

\begin{figure}[ht]
    \centering
    \includegraphics[width=.9\linewidth]{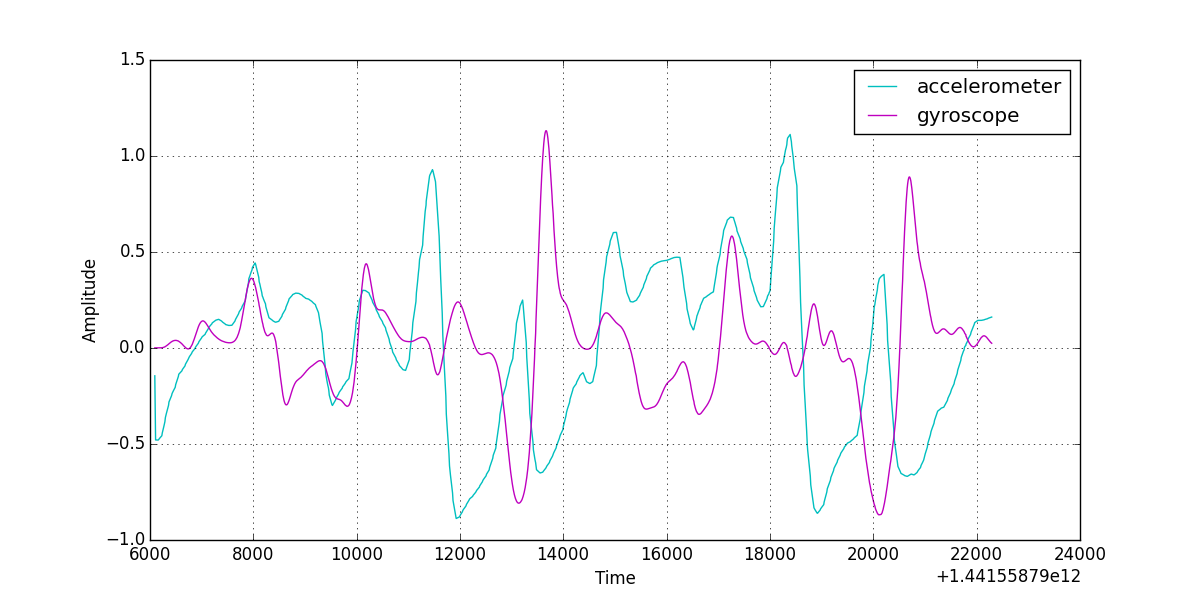}
    \caption{Gyroscope and accelerometer mean signal aligned with time fitting.}
    \label{fig:fitting}
\end{figure}

Linear interpolation allowed both sensor tuple sets to contain data-points evenly distributed over the recording session duration. However, it is not yet possible to fuse the sensors because the accelerometer timestamps values are different from the gyroscope's since the sensors have been reporting new events asynchronously independent of each others. The sensors have thus recorded data in slightly shifted time windows. The accelerometer measurements are therefore processed using the same previously described approach to normalize the sampling rate. That is, linear interpolation (Equation \ref{eq:linearinterpolation}) is used to compute unknown accelerometer values at the same timestamps as the gyroscope's. Figure \ref{fig:fitting} shows the two sensors mean signals after linear interpolation of the accelerometer to fit the gyroscope time values.

The sensor fusion algorithm returns a vector for each time frame $t_{i}$ and allow multiple combinations of axes to perform different classification experiments. For example, $\langle gx_{i}, gy_{i}, gz_{i}, \overline{a_{i}} \rangle$ is a possible four-dimensions vector returned by the fusion algorithm. With $x$, $y$, and $z$ values along the different axes of the accelerometer $a$ and the gyroscope $g$. Mean values $\overline{g_{i}}$ and $\overline{a_{i}}$ are simply the average of the three axes values (i.e. $\overline{g_{i}} = \frac{1}{3}(gx_{i} + gy_{i} + gz_{i})$).

\subsection{Segmentation}

The server is receiving a continuous stream of motion sensor events. Therefore, it is important to segment the data stream into sections corresponding to each keystroke before to perform classification.

\subsubsection{Segmentation from Labels}
\label{ssec:segmentationlabels}

\begin{figure}[H]
    \begin{subfigure}{1.\textwidth}
        \centering
        \includegraphics[width=.9\linewidth]{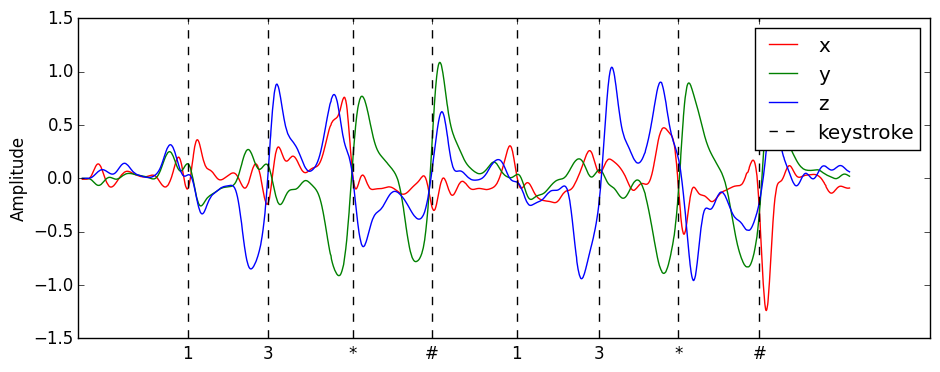}
        \caption{Labels position over sensor signals aligned using timestamps.}
    \end{subfigure}
    \begin{subfigure}{1.\textwidth}
        \centering
        \includegraphics[width=.8\linewidth]{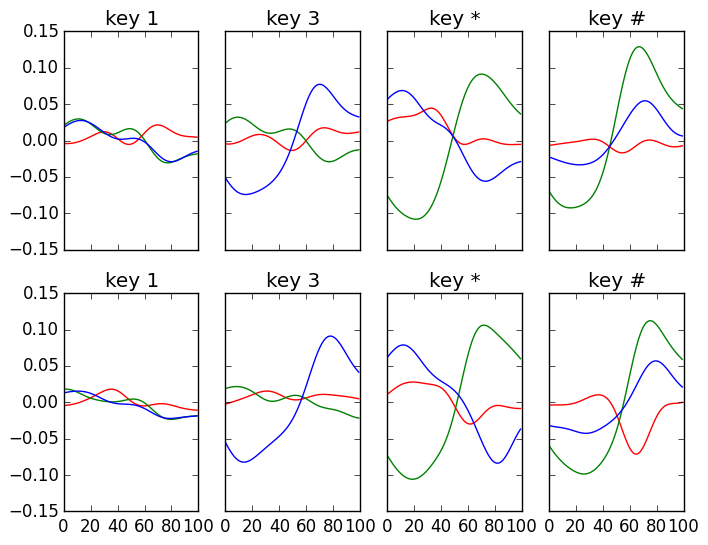}
        \caption{Final segmentation from label positions.}
    \end{subfigure}
    \caption{Segmentation of a gyroscope signal consisting of two sequences of the same four keys.}
    \label{fig:label}
\end{figure}

As mentioned in Section \ref{sec:dataacquisition}, tuples containing labels and timestamps in $ms$ are sent to the server during training. These time values are used as reference points to segment the signal into $m$ pieces, where $m$ is the total number of keystrokes in the entire recording session (Figure \ref{fig:label} shows an example of a signal after segmentation). We defined a fixed-size sampling window to select subsets of data-points occurring during keystrokes as follows: 

\begin{equation} \label{eq:segmentation}
     [t_{i} - \alpha,t_{i} + \alpha[ = \{v_{i}\in \{X,Y,X\} \;\vline\; v_{i-\alpha} \leq v_{i} < v_{i+\alpha}\}
\end{equation}

Where $\alpha$ is half the size of the sampling window. Considering that Asonov et al. \cite{asonov2004keyboard} determined the duration of a key-press to be approximately $100 ms$ and knowing that our target sampling rate was defined to be $2 ms$ (see Section \ref{ssec:sensorfusion}), we defined a sampling window of $50$ data-points. Thus $\alpha = 25$ in our implementation.

\subsubsection{Heuristic Segmentation}
\label{sec:heuristicsegment}

As explained in Chapter \ref{ch:attack}, it is assumed that the attacker has not compromised the target device on which the user is entering keys. Therefore, both the training phase and the logging phase cannot realistically rely on keystroke timestamps to segment the signal. Observations have showed that keystrokes tend to cause high peaks in the signal. Therefore, the \emph{Peak-to-Average Power Ratio} can be used to detect such peaks as follows:

\begin{enumerate}
    \item First of all, we observed that the gyroscope's signal peaks were better aligned with the actual keystroke timestamps than the accelerometer's. Considering that the sensors data are three-dimensional, the signals on the three axes are merged by simply calculating the gyroscope's mean signal $\overline{g}$ such that:

    \begin{equation}
        \overline{g_i} = \frac{1}{3}(gx_{i} + gy_{i} + gz_{i})
    \end{equation}
    
    \item Secondly, the mean signal can now be used to compute the \emph{Peak-to-Average Power Ratio} defined as the square \emph{Crest Factor} as follows:

    \begin{equation} \label{eq:crestfactor}
       f(v_{i}) = \left( \dfrac {v_i}{r(\overline{g})} \right)^{2}
    \end{equation}
    
    Where $v_{i}$ is the amplitude of the mean signal $\overline{g}$ at the index position $i$ and $r(\overline{g})$ return the \emph{Root Mean Square} of the signal such that:

    \begin{equation}
        r(\overline{g}) = \sqrt {\dfrac {1}{n} \sum ^{n}_{i=1} \overline{g_i}^{2}}
    \end{equation}

    Where $n$ is the total number of data-points in the gyroscope's average signal $\overline{g}$.
    
    \item Finally, the peaks can be detected by applying the following \emph{First-Order Logic} rule:
    
    \begin{equation}
        peak(r_{i}) \rightarrow (r_{i} > r_{i-1}) \land (r_{i} > r_{i+1}) \land (r_{i} > \alpha)
    \end{equation}
    
    Where $r_{i}$ is the \emph{Peak-to-Average Power Ratio} at the index position $i$, and $\alpha$ is a constant value $\alpha = 0.4$ used to discard peaks too small to be the result of a keystroke. If the rule is evaluated to $true$, then there is a peak in the signal at position $i$.
\end{enumerate}

\begin{figure}[H]
    \centering
    \includegraphics[width=.9\linewidth]{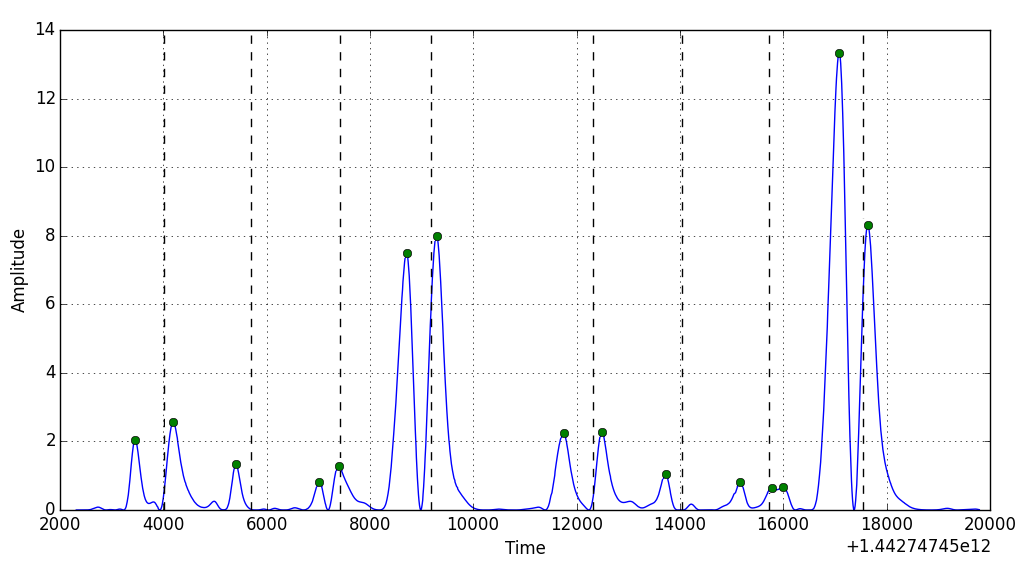}
    \caption{Example of peaks detected (shown as green circles) from the gyroscope Peak-to-Average Power Ratios with actual keystroke positions (shown as vertical dashed lines).}
    \label{fig:peakdetection}
\end{figure}

As shown in Figure \ref{fig:peakdetection}, this heuristic allows the detection of potential keystroke positions. Once the peak positions are known, it is possible to segment the signal by following the same approach described in Section \ref{ssec:segmentationlabels}.

\subsection{Statistical Features}

Statistical models such as RNNs are designed to process sequential data by performing Unsupervised Feature Learning. However, traditional FNNs usually need engineered features to model the data efficiently. The following statistical vector is therefore calculated:

\begin{equation}
    \langle \min, \max, RMS, \rho, \lambda, \sigma, \kappa, \gamma \rangle
\end{equation}

With $RMS$ the Root Mean Square, $\rho$ the number of peaks in the signal detected using the approach described in Section \ref{sec:heuristicsegment}, $\lambda$ the Crest Factor computed from Equation \ref{eq:crestfactor}, $\sigma$ the skewness, $\kappa$ the kurtosis, and $\gamma$ the variance.
This statistical vector is computed for all three axes for both sensors. Thus returning a statistical feature vector of length $48$.

\section{Classifier Model}
\label{sec:classifier}

The classifier takes accelerometer and gyroscope data as input and output classes corresponding to keystrokes. A typical input vector consists of values normalized in the range $[-1, 1]$ and the output is a binary vector of the same length as the number of labels. That is, each label is assigned a binary representation stored in a look-up table. Different multi-class classification models were implemented to compare their respective efficiency to process the data. Each of the models is trained with supervised learning thanks to a dataset containing labeled data-points. An online approach is used to update the weights each time a training example is shown to the network. The weights are updated using an improved variant of the Backpropagation iterative gradient descent termed \emph{Rprop-} \cite{riedmiller1993direct, igel2003empirical}. For result reproducibility, the weights are initialized with a pseudo-random number generator seeded with a hard-coded constant integer. All models rely on the same loss function to compute the network error. Namely, the \emph{Mean Squared Error} defined as follows:

\begin{equation} \label{eq:lossfunction}
    E = \frac{1}{n} \sum\limits^{n}_{i=1}(T_i - y_i)^{2}
\end{equation}

With $n$ the number of neurons in the output layer, $T$ the target expected output, and $y$ the predicted output.

\subsection{Performance Evaluation}

Measuring how well a classifier performs allows the selection of an optimal solution for the problem at hand. Some metrics thus need to be defined to assess the performance and the effectiveness of a classification model.

\noindent
\textbf{Precision, Recall, and F1 Score:}
The precision $P$ is important to measure the overall quality of the classification results while the recall $R$ determines the classifier capacity to analyse efficiently the majority of the data it is exposed to, such that:

\begin{equation}
    P = \frac{TP}{TP + FP},
    R = \frac{TP}{TP + FN}
\end{equation}

Where $TP$ is the number of true positives (i.e. the number of correctly classified items), $FP$ is the number of false positives (i.e. the number of incorrectly classified items, with $TP + FP$ equal to the collection size), and $FN$ the number of false negatives (i.e. the number of items incorrectly misclassified). A more accurate performance metrics termed \emph{F-score} combines both precision and recall to provide an overall performance score such that:

\begin{equation}
    F = (1 + \beta)^2 \frac{P \cdot R}{\beta^2 \cdot P + R} 
\end{equation}

With $\beta^2 = 1$ to compute the harmonic mean of precision and recall (i.e. F1 Score) \cite{manning2008introduction}. 

\noindent
\textbf{Reliability:}
The F1 Score is an efficient method to measure the quality of the classification results but does not provide any information about how reliable the results are. H{\"u}sken and Stagge \cite{husken2003recurrent} proposed a method to assess the reliability of a classification algorithm based on the value distribution of the output neurons. That is, if the predicted values of all the output neurons are numerically close, the classifier is not totally convinced by its classification decision. On the contrary, if the value of one output neuron is high while the rest of the neurons produce lower values, the classifier is very confident with its classification decision. The reliability $R$ of a classification result can be computed from the \emph{Entropy} $S$ such that:

\begin{equation}
    R = 1 - \frac{1}{\log n}S
\end{equation}
\begin{equation}
    S = - \sum ^{n}_{i=1} y_i \log y_i
\end{equation}

Where $n$ is the total number of output neurons (i.e. classes), and $y_i$ is the output value of the output neuron $i$. If all the output neurons generate similar values, $R$ will tend to $0$ and by opposition tends to $1$ if one of the output neuron generates a value numerically far from the others.

\noindent
\textbf{K-Fold Cross-Validation:}
Testing a classifier traditionally involve splitting the dataset into two parts: one used for training and one used for evaluation. However, the main downside of this method, termed holdout, is that the data used for evaluation are never used to train the classifier and vice versa. Thus leading to a less general performance assessment. Different techniques have been developed to address this issue. A popular solution termed \emph{K-Fold Cross-Validation} is used in this project to assess the quality of the different classification models. This method first consists of shuffling the dataset and splitting it into $k$ partitions (i.e. folds) approximately equal in size. The classifier is then trained with $k - 1$ datasets and evaluated on the remaining partition. This process is repeated $k$ times by selecting a different training set and evaluation set such that every partition is used at most one time for evaluation and $k - 1$ times for training. The evaluation results are finally averaged to represent the global performance of the classifier. All data are thus used for both training and evaluation to provide a more general and accurate performance assessment \cite{kohavi1995study}.

\subsection{Sensor Fusion Benchmark}
\label{ssec:fusionbenchmark}

\begin{figure}[H]
    \centering
    \includegraphics[width=.45\linewidth]{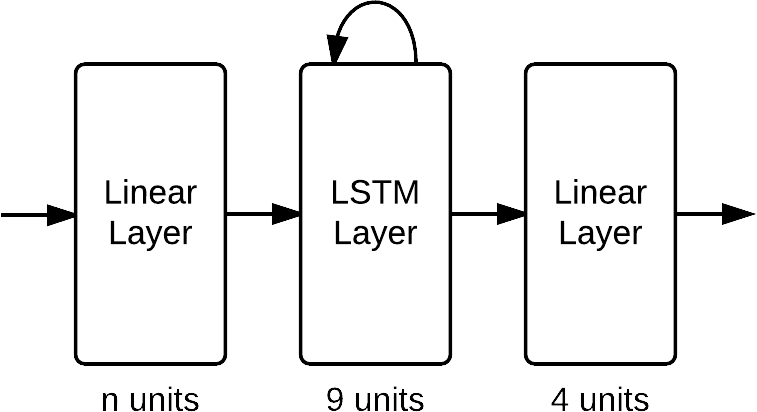}
    \caption{Neural network used for sensor fusion benchmark.}
    \label{fig:netfusion}
\end{figure}

Experiments were performed to select the sensor fusion strategy holding the best classification results. Observations showed that signals resulting from motions occurring while typing on keys in opposite corners (i.e. $1$, $3$, $*$, $\#$) were differentiable enough to be recognized with the naked eye by simply looking at the resulting signal patterns (example in Figure \ref{fig:label}). Thus, motions from keystrokes on these specific keys were recorded to construct a toy dataset based on the assumption that a good classifier model should be able to do as well as the human eye on these simple patterns. The toy dataset contains a total of $120$ keystrokes targeting $4$ labels, resulting in $30$ instances per key.

\begingroup
\renewcommand{\arraystretch}{1.5}
\begin{table}[ht]
    \centering
    \begin{tabular}{| l | l | l |} \hline
        \textbf{Vector} & \textbf{F1 Score} & \textbf{Reliability} \\ \hline
$\langle gx_{i}, gy_{i}, gz_{i} \rangle$ & $0.94$ & $0.57$ \\ \hline
$\langle ax_{i}, ay_{i}, az_{i} \rangle$ & $0.75$ & $0.44$ \\ \hline
$\langle \overline{g_{i}} \rangle$ & $0.68$ & $0.43$ \\ \hline
$\langle \overline{a_{i}} \rangle$ & $0.62$ & $0.35$ \\ \hline
$\langle \overline{g_{i}}, \overline{a_{i}} \rangle$ & $0.76$ & $0.41$ \\ \hline
$\langle \overline{g_{i}}, ax_{i}, ay_{i}, az_{i} \rangle$ & $0.81$ & $0.45$ \\ \hline
$\langle gx_{i}, gy_{i}, gz_{i}, \overline{a_{i}} \rangle$ & $0.95$ & $0.54$ \\ \hline
$\langle gx_{i}, gy_{i}, gz_{i}, ax_{i}, ay_{i}, az_{i} \rangle$ & $0.99$ & $0.64$ \\ \hline
    \end{tabular}
    \caption{Fusion strategy benchmark results (average values for $100$ training Epochs).}
    \label{tab:fusion}
\end{table}
\endgroup

To assess the quality of the different fusion strategies, the vectors were used to train an RNN with one hidden LSTM layer of $9$ units (as illustrated in Figure \ref{fig:netfusion}) for $100$ \emph{Epochs} using the toy dataset. That is, $100$ passes through the entire training dataset. The number of units in the linear input layer depends on the length of the feature vector returned by the fusion algorithm. The output layer is a standard linear layer and the evaluation is performed with the same dataset used for training. Although this would be a bad approach for evaluating the performance of the classifier itself, the goal here is only to assess the quality of the features returned from the fusion algorithm. In fact, a good fusion strategy allows the generation of feature vectors that the classifier should be able to memorize and remember. Thus, a fusion strategy is satisfying if the classifier can learn to generate the appropriate output if the same input vector is seen again. As shown in Table \ref{tab:fusion}, the best sensor fusion strategy is a six-dimensions vector consisting of the three axes of both the gyroscope and the accelerometer.

\subsection{Model Benchmark}
\label{ssec:modelbenchmark}

\begin{figure}[H]
    \centering
    \includegraphics[width=.55\linewidth]{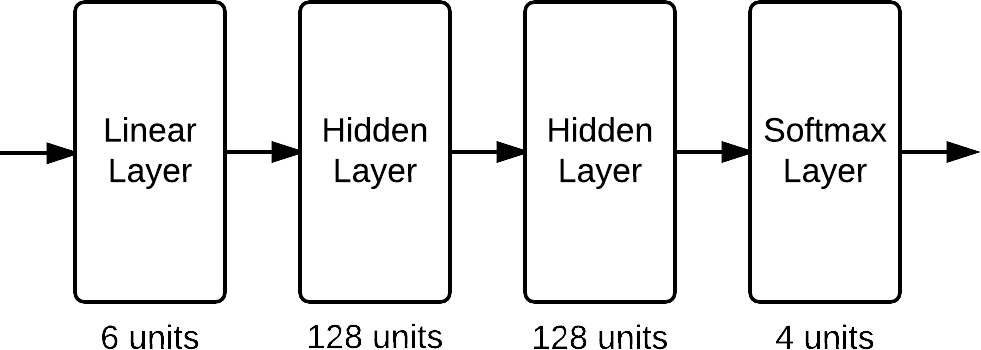}
    \caption{Neural network template.}
    \label{fig:nettemplate}
\end{figure}

Now that a fusion strategy has been chosen to leverage the quality of the predictions, it is possible to compare different types of models. Each ANN is built following the architecture template illustrated in Figure \ref{fig:nettemplate} and trained for $100$ Epochs on the same toy dataset used in Section \ref{ssec:fusionbenchmark}. A linear layer is first used to forward the input vector to the internal structure of the network. Each layer is then fully connected to the next layer in the network. Since classification is performed, a Softmax layer is finally used as output (see Table \ref{tab:activationfunction} for mathematical definition of Softmax). The compared models only differ from the type of hidden layer employed and the type of features they are trained on. The hidden layers each contains $128$ hidden units. The following results are measured using \emph{K-Fold Cross-Validation} with $k = 5$. The results are therefore averaged mean and averaged standard deviation. Confusion matrices generated during this benchmark can be seen in Appendix \ref{ch:appendixbenchmark}.

\begin{figure}[!ht]
    \centering
    \includegraphics[width=1.\linewidth]{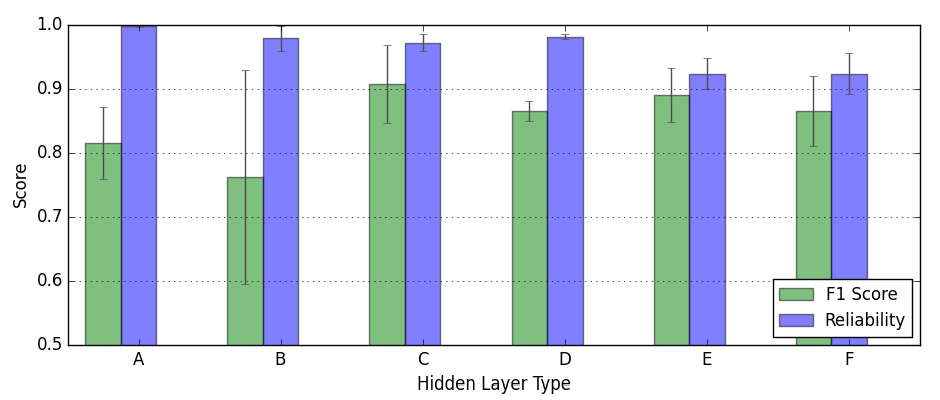}
    \caption{Hidden layer benchmark (see Table \ref{tab:benchmarkhiddenlayer} for references).}
    \label{fig:benchmarkhiddenlayer}
\end{figure}

\noindent
\textbf{Multilayer FNN:}
Standard FNN is one of the most simple ANN architecture and is interesting for comparing their performance with more advanced ANNs such as networks with recurrent architecture. Layers with different activation functions are compared to select an appropriate FNN architecture for the problem at hand. Standard \emph{Rprop-} is used for training. As presented in Table \ref{tab:benchmarkhiddenlayer} and Figure \ref{fig:benchmarkhiddenlayer}, FNN with Sigmoid hidden layers can process statistical features significantly more efficiently than Tanh  layers. However, when the network is trained with data segments directly, Tanh layers are able to make more reliable predictions with a smaller standard deviation.

\begingroup
\renewcommand{\arraystretch}{1.5}
\begin{table}[ht]
    \centering
    \begin{tabular}{| c | l | l | l | l | l | l |} \hline
        \multirow{2}{*}{\textbf{Ref.}} &
        \multirow{2}{*}{\textbf{Hidden Layer}} &
        \multirow{2}{*}{\textbf{Features}} &
        \multicolumn{2}{c|}{\textbf{F1 Score}} & \multicolumn{2}{c|}{\textbf{Reliability}} \\ \cline{4-7}
        & & & Mean & Std. Dev & Mean & Std. Dev \\ \hline
        A & Sigmoid & Statistical & $0.816$ & $0.056$ & $0.999$ & $0.0014$ \\ \hline
        B & Tanh & Statistical & $0.762$ & $0.167$ & $0.979$ & $0.0192$ \\ \hline
        C & Sigmoid & Segment & $0.908$ & $0.061$ & $0.972$ & $0.0134$ \\ \hline
        D & Tanh & Segment & $0.866$ & $0.016$ & $0.982$ & $0.0045$ \\ \hline
        E & LSTM & Segment & $0.891$ & $0.042$ & $0.924$ & $0.0247$ \\ \hline
        F & LSTM peephole & Segment & $0.866$ & $0.055$ & $0.924$ & $0.0315$ \\ \hline
    \end{tabular}
    \caption{Hidden layer benchmark (see Figure \ref{fig:benchmarkhiddenlayer} for graphical representation).}
    \label{tab:benchmarkhiddenlayer}
\end{table}
\endgroup

\noindent
\textbf{Multilayer LSTM:}
Since LSTMs are designed to process sequential data (as explained in Section \ref{sec:backgroundann}), they are especially suitable to process motion sensors from WADs. The training algorithm is however slightly different from the one used to train multilayer FNNs. To allow time-related patterns to emerge, \emph{Backpropagation Through Time} is used to pass along the LSTM internal state to the next recurrent LSTM instance (i.e. the previous cell state $c_{t - 1}$, the previous output vector $y_{t - 1}$, and the new  datapoint $x_{t}$). The LSTM initial cell state is a vector consisting of null values that is passed together with the first data-point of the sequence. During evaluation, the predictions from an FNN are returned as generated by the neural network. However, because LSTM units first need to be activated to initialize their memory cell internal structure, the outputs generated for the whole sequence are contributing to the final prediction result. The output vectors are simply added together and normalized before to be returned. Two different LSTM implementations were studied: standard recurrent LSTM consisting of a forget gate, and LSTM with peephole connections. Table \ref{tab:benchmarkhiddenlayer} and Figure \ref{fig:benchmarkhiddenlayer} show that while LSTM with peepholes is able to make remarkably good predictions in some cases, the standard deviation remains higher than when a standard recurrent LSTM layer is used. Thus making the peephole implementation less robust in this specific application context.

\chapter{Evaluation}
\label{ch:evaluation}

This chapter first goal is to describe the different experiments setup to collect empirical data from various people. Secondly, to describe the different results returned by the system. Finally, to interpret the results and their relations with the research questions enunciated in Section \ref{sec:problemstatement}.

\section{Empirical Data Collection}

Seven persons external to this research and aged between $23$ and $30$ have participated in the following experiments. Each person was asked to enter multiple series of keystrokes on a touchscreen and a keypad while wearing a WAD on their wrist. To prevent the influence of external motions, the participants were required to sit in a comfortable position allowing them to stay still for the entire duration of the recording session. Each dataset contains $240$ keystrokes with $20$ instances of each of the $12$ labels (i.e. $1$, $2$, $3$, $4$, $5$, $6$, $7$, $8$, $9$, $0$, $*$, $\#$). If a classifier makes its predictions at random, the probability of correctly classifying a keystroke $K$ is $P(K) = \frac{1}{12}$.

\section{Experiments}

\begin{figure}[H]
    \begin{subfigure}{1.\textwidth}
        \centering
        \includegraphics[width=.55\linewidth]{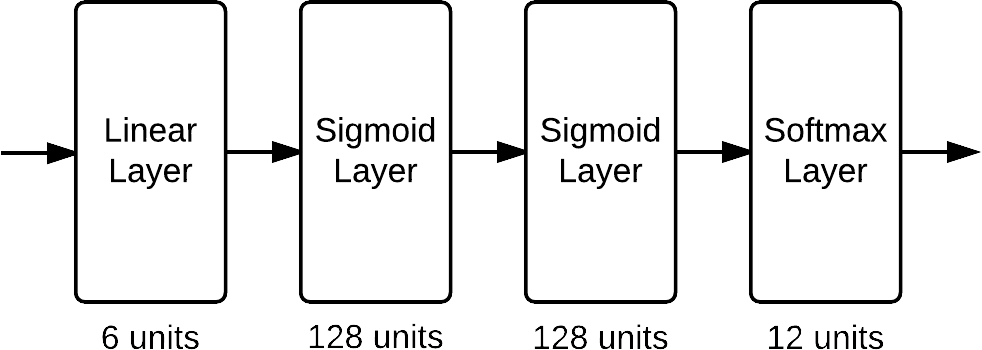}
        \caption{FNN-Sigmoid.}
    \end{subfigure}
    \begin{subfigure}{1.\textwidth}
        \centering
        \includegraphics[width=.55\linewidth]{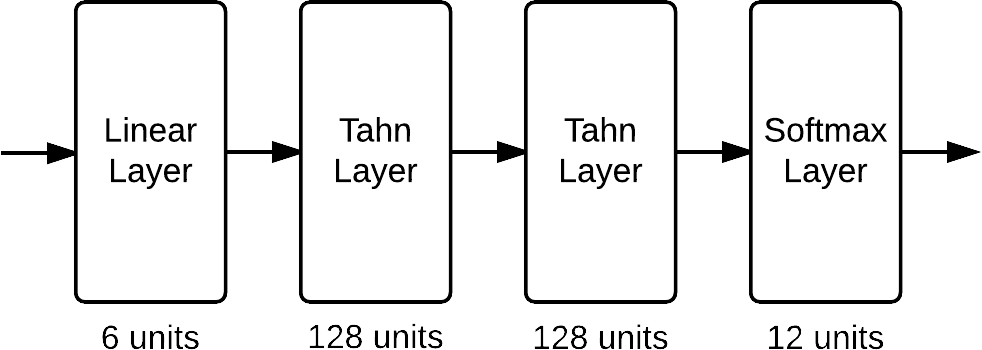}
        \caption{FNN-Tanh.}
    \end{subfigure}
    \begin{subfigure}{1.\textwidth}
        \centering
        \includegraphics[width=.55\linewidth]{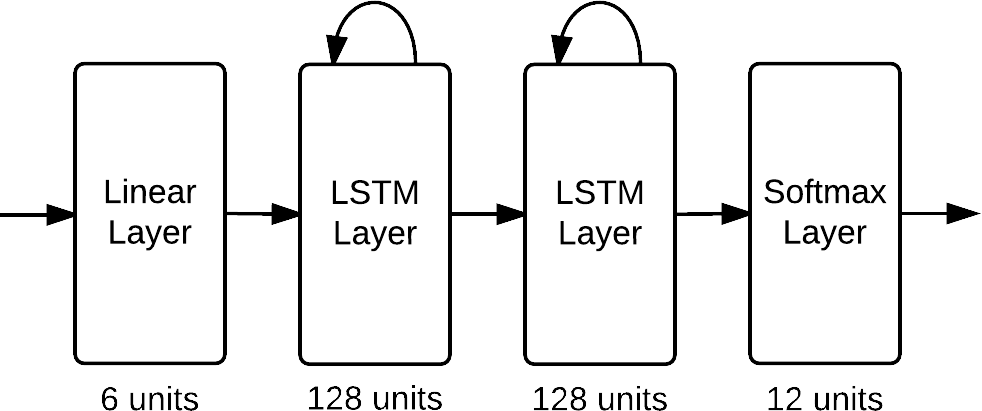}
        \caption{RNN-LSTM.}
    \end{subfigure}
    \caption{Neural network architectures selected for experiments.}
    \label{fig:netexperiments}
\end{figure}

\noindent
\textbf{Classification Model:}
As detailed in Section \ref{sec:classifier}, different ANN architectures were compared to select appropriate models for experimenting with real data. The number of output units is increased to reflect the number of labels in the collected measurements. The terms \emph{FNN-Sigmoid} and \emph{FNN-Tanh} are used to refer to the feedforward implementations depicted in Figure \ref{fig:netexperiments} (a) and (b), respectively. FNN-Sigmoid is preferred to process statistical features while FNN-Tanh is the feedforward architecture chosen to process data segments as features. The chosen recurrent implementation to process data sequences is shown in Figure \ref{fig:netexperiments} (c) and referred to as \emph{RNN-LSTM}. 

\noindent
\textbf{Data Preparation Schemes:}
The experiments require many different data preparation schemes to compare the performance of the classifiers in various contexts. Notably, we want to observe how well deep models can process both pre-processed data and noisy raw data. We consider the data raw when non-trivial pre-processing operations are completely ignored (only calibration as described in Section \ref{ssec:calibration} is allowed, which is used mostly to ignore the effect of gravity on the accelerometer measurements). The feature extraction steps detailed in Section \ref{sec:featureextraction} are subsequently applied indifferently to the data being pre-processed or raw (including normalizing the sampling rate during sensor fusion as described in Section \ref{ssec:sensorfusion}). The following data preparation schemes are used to train and evaluate the three classifiers:

\begin{itemize}
    \item \emph{P-T}: Pre-processed data with timestamp-based segmentation.
    \item \emph{P-H}: Pre-processed data with heuristic segmentation.
    \item \emph{R-T}: Raw data with timestamp-based segmentation.
    \item \emph{R-H}: Raw data with heuristic segmentation.
\end{itemize}

The results are measured using \emph{K-Fold Cross-Validation} with $k = 5$ by training every classifier for $100$ Epochs. The results are averaged over all folds and the seven participants. Each one of the three classifiers is thus trained for five folds on four different data schemes. Leading to a total of $60$ training sessions for each participant in every respective experiment. These computationally intensive tasks were performed on a dedicated machine and took several days to complete. The reader is invited to refer to Chapter \ref{ch:attack} to read about the motivations behind the following experiments. Appendix \ref{ch:appendixresults} provides confusion matrices and loss graphs for further details.

\subsection{Experiment 1: Touchlogging Attack}

\begin{figure}[H]
    \centering
    \includegraphics[width=1.\linewidth]{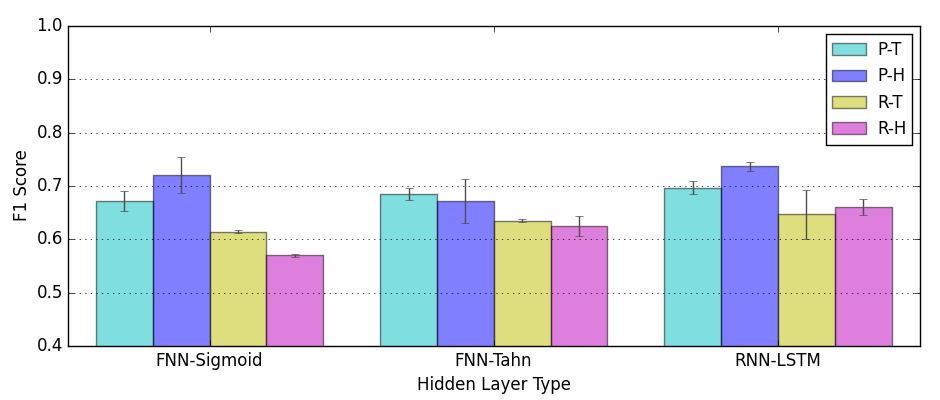}
    \caption{F1 Score for the different ANN architectures during the touchlogging experiment.}
    \label{fig:touchloggingresult}
\end{figure}

The goal of this experiment is to assess the likelihood of a touchlogging attack on a smartphone touchscreen using motion sensor outputs from a WAD. The participants were invited to enter keystrokes on our training application running on an iPhone 4S (interface shown in Figure \ref{fig:trainingdevices} (a)). Figure \ref{fig:touchloggingresult} provides a graphical representation comparing the performances of the three classifiers. The results are detailed in the subsequent Tables.

The results presented in Table \ref{tab:keyloggingfnnsigmoid} shows that FNN-Sigmoid is better at analysing pre-processed data than raw data. Despite performing worse on unprocessed data, it is a surprise that the model can still classify such features with reliable predictions and low standard deviation. The lower F1 Score on raw data can be explained by the fact that the model relies on statistical features. As a consequence, raw data displays unstable statistical properties, thus demonstrating the benefit of pre-processing the measurements beforehand to train such classifier.

\begingroup
\renewcommand{\arraystretch}{1.5}
\begin{table}[H]
    \centering
    \begin{tabular}{| c | l | l | l | l |} \hline
        \multirow{2}{*}{\textbf{Data Prep.}} &
        \multicolumn{2}{c|}{\textbf{F1 Score}} & \multicolumn{2}{c|}{\textbf{Reliability}} \\ \cline{2-5}
        & Mean & Std. Dev & Mean & Std. Dev \\ \hline
        P-T & $0.672$ & $0.0193$ & $0.982$ & $0.0029$ \\ \hline
        P-H & $0.720$ & $0.0339$ & $0.969$ & $0.0094$ \\ \hline
        R-T & $0.614$ & $0.0029$ & $0.980$ & $0.0022$ \\ \hline
        R-H & $0.570$ & $0.0029$ & $0.982$ & $0.0024$ \\ \hline
    \end{tabular}
    \caption{Touchlogging using FNN-Sigmoid with statistical features.}
    \label{tab:touchloggingfnnsigmoid}
\end{table}
\endgroup

Unexpectedly, the FNN-Tanh implementation is able to learn from data segment in a relatively acceptable way. Thus, allowing the model to make predictions similar in quality to the FNN-Sigmoid model. Even though being very basic, the two-layered model seems to be able to perform  Unsupervised Feature Learning to some extent. Table \ref{tab:touchloggingfnntahn} additionally shows that FNN-Tanh can make slightly better predictions than FNN-Sigmoid on raw data, probably because the former can learn features from the data it is exposed to without relying on engineered statistical features sensible to noise. However, the evolution of the loss during training (as illustrated in Figures \ref{fig:touchloggingtahnpt} and \ref{fig:touchloggingtahnph} in Appendix \ref{ch:appendixresults}) suggests that FNN-Tanh is harder to train on raw data when patterns are difficult to learn by the network naive internal structure.

\begingroup
\renewcommand{\arraystretch}{1.5}
\begin{table}[H]
    \centering
    \begin{tabular}{| c | l | l | l | l |} \hline
        \multirow{2}{*}{\textbf{Data Prep.}} &
        \multicolumn{2}{c|}{\textbf{F1 Score}} & \multicolumn{2}{c|}{\textbf{Reliability}} \\ \cline{2-5}
        & Mean & Std. Dev & Mean & Std. Dev \\ \hline
        P-T & $0.685$ & $0.0106$ & $0.948$ & $0.0019$ \\ \hline
        P-H & $0.672$ & $0.0415$ & $0.912$ & $0.0080$ \\ \hline
        R-T & $0.635$ & $0.0029$ & $0.840$ & $0.0021$ \\ \hline
        R-H & $0.625$ & $0.0183$ & $0.864$ & $0.0170$ \\ \hline
    \end{tabular}
    \caption{Touchlogging using FNN-Tanh with data segment as features.}
    \label{tab:touchloggingfnntahn}
\end{table}
\endgroup

As depicted in Figure \ref{fig:touchlogginglstmpt} and as expected, RNN-LSTM can be trained efficiently on time-series data. In fact, this architecture outperforms the other ANNs in most cases as illustrated in Figure \ref{fig:touchloggingresult}. Although the model displays no difficulty processing raw data, it nevertheless seems to have trouble analysing heuristically-segmented raw data. Thanks to its capacity to process sequence data, RNN-LSTM is the champion of the touchlogging task.

\begingroup
\renewcommand{\arraystretch}{1.5}
\begin{table}[H]
    \centering
    \begin{tabular}{| c | l | l | l | l |} \hline
        \multirow{2}{*}{\textbf{Data Prep.}} &
        \multicolumn{2}{c|}{\textbf{F1 Score}} & \multicolumn{2}{c|}{\textbf{Reliability}} \\ \cline{2-5}
        & Mean & Std. Dev & Mean & Std. Dev \\ \hline
        P-T & $0.697$ & $0.0128$ & $0.902$ & $0.0097$ \\ \hline
        P-H & $0.737$ & $0.0088$ & $0.935$ & $0.0057$ \\ \hline
        R-T & $0.647$ & $0.0460$ & $0.808$ & $0.0118$ \\ \hline
        R-H & $0.660$ & $0.0147$ & $0.822$ & $0.0074$ \\ \hline
    \end{tabular}
    \caption{Touchlogging using RNN-LSTM with data segment as features.}
    \label{tab:touchloggingrnnlstm}
\end{table}
\endgroup
                
\subsection{Experiment 2: Keylogging Attack}

This experiment's purpose is to determine the feasibility of a keylogging attack by analysing WAD motions while typing on an ATM-like physical keypad. The keystrokes were entered by the participants on our training device (depicted in Figure \ref{fig:trainingdevices} (b)) by using the same method to type on both the touchscreen and the keypad. That is, if the index finger were preferred to enter keystrokes on the smartphone, the index finger should also be used to enter keys on the physical keypad to use the same data for \emph{Experiment 3} (see Chapter \ref{ch:attack} for motivation details). 

\begin{figure}[H]
    \centering
    \includegraphics[width=1.\linewidth]{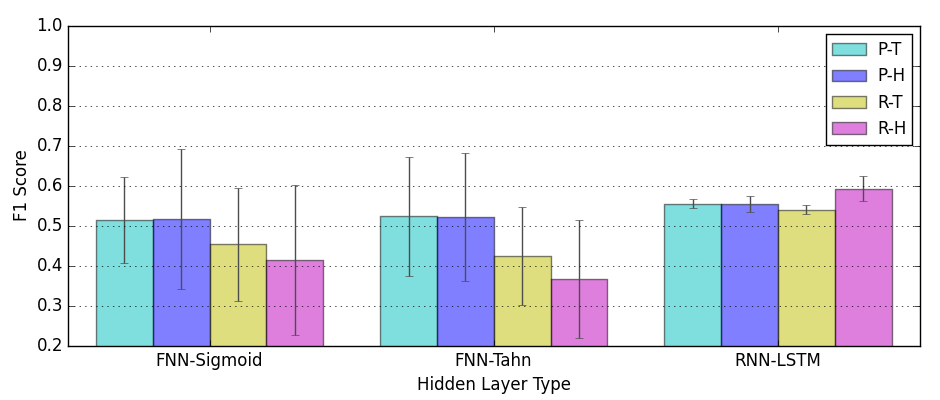}
    \caption{F1 Score for the different ANN architectures on the keylogging experiment.}
    \label{fig:keyloggingresult}
\end{figure}

Despite the fact that the FNN-Sigmoid is convinced by its predictions, the quality of its decisions is far from satisfying. In fact, Table \ref{tab:keyloggingfnnsigmoid} shows that the average reliability score is high with a small standard deviation even thought the standard deviation of the F1 Score is very important. The predictions average close to and below average.

\begingroup
\renewcommand{\arraystretch}{1.5}
\begin{table}[H]
    \centering
    \begin{tabular}{| c | l | l | l | l |} \hline
        \multirow{2}{*}{\textbf{Data Prep.}} &
        \multicolumn{2}{c|}{\textbf{F1 Score}} & \multicolumn{2}{c|}{\textbf{Reliability}} \\ \cline{2-5}
        & Mean & Std. Dev & Mean & Std. Dev \\ \hline
        P-T & $0.515$ & $0.1078$ & $0.973$ & $0.0153$ \\ \hline
        P-H & $0.518$ & $0.1745$ & $0.973$ & $0.0088$ \\ \hline
        R-T & $0.454$ & $0.1417$ & $0.964$ & $0.0149$ \\ \hline
        R-H & $0.415$ & $0.1868$ & $0.969$ & $0.0101$ \\ \hline
    \end{tabular}
    \caption{Keylogging using FNN-Sigmoid with statistical features.}
    \label{tab:keyloggingfnnsigmoid}
\end{table}
\endgroup
        
Similarly to the FNN-Sigmoid model, FNN-Tanh is not performing optimally during the keylogging task as conveyed by Table \ref{tab:keyloggingfnntahn}. Its predictions for raw data classification are indeed below average. FNN-Tanh reliability is fluctuating with a great standard deviation, showing that the model has difficulties to generate strong predictions. This ANN architecture clearly struggles to achieve Unsupervised Feature Learning on the keypad dataset.

\begingroup
\renewcommand{\arraystretch}{1.5}
\begin{table}[H]
    \centering
    \begin{tabular}{| c | l | l | l | l |} \hline
        \multirow{2}{*}{\textbf{Data Prep.}} &
        \multicolumn{2}{c|}{\textbf{F1 Score}} & \multicolumn{2}{c|}{\textbf{Reliability}} \\ \cline{2-5}
        & Mean & Std. Dev & Mean & Std. Dev \\ \hline
        P-T & $0.524$ & $0.1493$ & $0.913$ & $0.0276$ \\ \hline
        P-H & $0.523$ & $0.1594$ & $0.908$ & $0.0244$ \\ \hline
        R-T & $0.425$ & $0.1234$ & $0.775$ & $0.0179$ \\ \hline
        R-H & $0.368$ & $0.1468$ & $0.759$ & $0.0275$ \\ \hline
    \end{tabular}
    \caption{Keylogging using FNN-Tanh with data segment as features.}
    \label{tab:keyloggingfnntahn}
\end{table}
\endgroup

The graphical comparison represented in Figure \ref{fig:keyloggingresult} distinctly shows the performance of the RNN-LSTM implementation surpassing the other models in a significant way. In fact, the predictions are above average with a small standard deviation. RNN-LSTM performs equally on the four different data preparation schemes and is even able to make its best predictions when exposed to the most chaotic data scheme, namely, raw data with heuristic segmentation (i.e. R-H). This ANN is undeniably the best performing ANN architecture for the keylogging task.

\begingroup
\renewcommand{\arraystretch}{1.5}
\begin{table}[H]
    \centering
    \begin{tabular}{| c | l | l | l | l |} \hline
        \multirow{2}{*}{\textbf{Data Prep.}} &
        \multicolumn{2}{c|}{\textbf{F1 Score}} & \multicolumn{2}{c|}{\textbf{Reliability}} \\ \cline{2-5}
        & Mean & Std. Dev & Mean & Std. Dev \\ \hline
        P-T & $0.556$ & $0.0110$ & $0.871$ & $0.0201$ \\ \hline
        P-H & $0.554$ & $0.0200$ & $0.858$ & $0.0184$ \\ \hline
        R-T & $0.541$ & $0.0110$ & $0.758$ & $0.0021$ \\ \hline
        R-H & $0.593$ & $0.0310$ & $0.776$ & $0.0225$ \\ \hline
    \end{tabular}
    \caption{Keylogging using RNN-LSTM with data segment as features.}
    \label{tab:keyloggingrnnlstm}
\end{table}
\endgroup
    
\subsection{Experiment 3: from Touchlogging to Keylogging}

For this experiment, the classifier is trained for $200$ Epochs with the complete dataset recorded during \emph{Experiment 1} when users are entering keys on a smartphone touchscreen. The logging phase is later performed using the full measurement collection recorded for \emph{Experiment 2} with users typing on an ATM-like keypad. It is worth noting that the classifier is trained and later evaluated with features generated by the same data scheme. That is, when the model is trained with pre-processed data with timestamp-based segmentation (i.e. P-T), the same data scheme is applied to the evaluation dataset. This experimentation is completed using the \emph{RNN-LSTM} model because it is the model yielding the best results in both previous experiments.

The results presented in Table \ref{tab:exp3rnnlstm} are calculated at once when the evaluation dataset is shown to the classifier. Although the returned predictions are far from excellent, RNN-LSTM is still able to recognize patterns from unknown signals recorded when users are typing on a different keyboard than the one used for training. In this application context, it is worth noting that the classifier can perform better when it is both trained and evaluated with pre-processed data. Similarly to the results presented in Figure \ref{fig:touchloggingresult} and Table \ref{tab:touchloggingrnnlstm}, RNN-LSTM have trouble learning from raw data with timestamp-based segmentation (i.e. R-T).\footnote{The F1 Score is $0.082$ and corresponds to a random guess over $12$ possible labels. Thus showing that RNN-LSTM is completely unable to make educated predictions from R-T.}

\begingroup
\renewcommand{\arraystretch}{1.5}
\begin{table}[H]
    \centering
    \begin{tabular}{| l | l | l |} \hline
        \textbf{Data Prep.} & \textbf{F1 Score} & \textbf{Reliability} \\ \hline
        P-T & $0.184$ & $0.795$ \\ \hline
        P-H & $0.191$ & $0.808$ \\ \hline
        R-T & $0.082$ & $0.646$ \\ \hline
        R-H & $0.122$ & $0.622$ \\ \hline
    \end{tabular}
    \caption{Results from RNN-LSTM trained for touchlogging and evaluated for keylogging with data segments used as features.}
    \label{tab:exp3rnnlstm}
\end{table}
\endgroup

\section{Discussions}

The LSTM implementation is undeniably the best classification model used in the experiments. In fact, this architecture can achieve touchlogging and keylogging with a maximum accuracy of $73\%$ and $59\%$, respectively. The LSTM model can also successfully classify signals with an accuracy of $19\%$ when the dataset used for training and logging are originated from two different keypads.

Across all the different results, the heuristically-based segmentation usually leverages better classification decisions than timestamp-based segmentation. This might be caused by two factors. First, the accuracy of the timestamp sent to the server by the training devices is questionable. As detailed in Section \ref{ssec:segmentationlabels}, keystrokes are thought to happen in a $100 ms$ time window on average and the timestamps used by the system are measured in $ms$. Even though these time values are expected to match actual keystrokes when aligned with the signal (as shown on Figure \ref{fig:label}), it is likely that small time measuring inaccuracies can lead to worse classification results. Second, the heuristic segmentation works by measuring physical properties of the signal (as explained in Section \ref{sec:heuristicsegment}). With this in mind and given the experiments results, it is reasonable to assume that these physical properties are resilient across keystrokes and consequently a robust method for signal segmentation.

As a reminder, both the touchlogging and the keylogging datasets contains the same number of measurements with each keystroke being equally represented. The important difference between the FNNs performances during the two first experiments is thus thought to be caused by the data themselves. In fact, the size of the keypad built for \emph{Experiment 2} is smaller than the size of the touchscreen used for \emph{Experiment 1}. As a consequence, this slight difference in size produces tinier motions with less extreme value variation. Thus leading naive models to struggle finding subtle patterns in the signal, even when manually selected statistical features are used.

Using a recurrent model such as LSTM have proved to yield in general far better results than simple feedforward models. Thanks to its internal structure designed to process sequential data, the RNN-LSTM model behaves robustly regardless of the data collection used and can even learn to distinguish patterns in noisy measurements. An impressive characteristic of the recurrent model is indeed its ability to process raw data as well as pre-processed data as demonstrated in the experiments.

\chapter{Conclusion}

The purpose of this chapter is mainly to summarize and reflect on our findings. Conceivable future works are also considered to extend this project.

\section{Summary}

The system developed in this work can perform touchlogging and keylogging with an accuracy of $73\%$ and $59\%$, respectively. Despite the fact that these results are smaller than the ones claimed in related works, our classifier can perform equally successfully when confronted with raw unprocessed data. Thus demonstrating that deep neural networks are capable of making keystroke inference attacks based on motion sensors easier to achieve by removing the need for non-trivial pre-processing pipelines and carefully engineered feature extraction strategies. All related works rely heavily on such techniques as presented in Chapter \ref{ch:relatedwork}.

Moreover, the system is still able to infer keystrokes with an accuracy of $19\%$ when trained and evaluated with datasets recorded from different keypads. This result suggests that an attacker could log keys from a wide range of devices even if its classifier is trained with measurements from a different compromised device.

Dramatically, these observations imply that a cyber-criminal would be able, in theory, to eavesdropped on any device operated by the user while wearing a WAD. Thus granting access to sensitive and highly valuable information and possibly causing important damages.

To minimize the risk of such attacks, users should always wear their WAD on their less preferred hand for device interaction. For example, a right-handed person should wear the WAD on its left arm. Because of the demonstrated risks, the different operations systems powering wearable technologies should require user permissions before any application is allowed to use the accelerometer and the gyroscope. Furthermore, a permission system should restrict or allow access to the motion sensors in specific contexts or for trusted applications only.

\section{Future Work}

\emph{Convolutional Neural Network (CNN)} is a class of powerful deep models designed to process multi-dimensional arrays such as images and video frames \cite{lecun1995convolutional}. A possible addition to this project could involve experiments with such ANN architecture to classify motion sensors measurement sequences. It would indeed be interesting to compare the performance of CNN with LSTM in our application context.

A significant extension of this work could be the implementation of a more dynamic system allowing automatic signal segmentation and classification in real-time. For example, LSTM layers could be trained to identify keystrokes in the measurement flow, segment the signal automatically, and pass the resulting data forward to other layers responsible for further processing and classification.

An analogous attack targeting WAD users could be implemented to reconstruct hand-written messages by recording and analysing the hand's motion. For example, gesture-based password (e.g. Android lock screen) could potentially be cracked using such an attack.

In order to assess the security threat in specific contexts, it would be interesting to perform further experiments using standard hardware such as actual ATM keypads, electronic building access system keypads, or hotel room safe keypads.

User's moves can create important motion interferences that can potentially obfuscate keystroke signal patterns. Experiments in such conditions could be conducted to compare the results and the likelihood of such an attack in these contexts (e.g. controlled environment when the user is sitting, uncontrolled environment when she is walking).

Benchmarking different smartwatch and fitness tracker models could potentially show that some products allow more accurate keystroke inference than others. Thus possibly proving that users are at greater risk when using specific WAD models.

The motion of WAD could also potentially be used for the identification and tracking of users as studied in similar research \cite{monrose2000keystroke, ilonen2003keystroke}.

Some WAD models come built-in with a wide range of sensors including \emph{Galvanic Skin Response (GSR)}, heart rate sensor, \emph{Electromyography (EMG)}, or ambient light sensor. Fusing motion sensors with one or many of these additional sensors might further improve the accuracy and the robustness of the keystroke predictions.

\clearpage
\makebibliography

\begin{appendices}
\chapter{Backpropagation}
\label{ch:appendixbackprop}

This appendix provides additional details about the Backpropagation algorithm \cite{han2011data, bishop2006pattern, rumelhart1985learning, haykin2004comprehensive} to support Section \ref{sec:backgroundann}. The total network error $E$ is computed from a loss function, such as the Mean Squared Error formalized in Equation \ref{eq:lossfunction}. According to the chain rule, the gradient can be expressed as:

\begin{equation}
    \frac{\partial E}{\partial W_{ij}} = \frac{\partial E}{\partial y_{i}} \frac{\partial y_{i}}{\partial x_{i}} \frac{\partial x_{i}}{\partial W_{ij}}
\end{equation}

First, the partial derivative with respect to $W_{ij}$ can be computed from Equation \ref{eq:ouput} as follows:

\begin{equation}
    \frac{\partial x_{i}}{\partial W_{ij}} = y_{j}
\end{equation}

Secondly, the partial derivative with respect to $x_{i}$ is:

\begin{equation}
    \frac{\partial y_{i}}{\partial x_{i}} = \frac{\partial \phi(x_i)}{\partial x_{i}}
\end{equation}

Which is the derivative of the activation function of neuron $i$.
Thirdly, the partial derivative with respect to $y_{i}$ can be computed as follows:

\begin{equation}
    \frac{\partial E}{\partial y_{i}} =
    \begin{cases}
        \begin{aligned}
            \frac{\partial}{\partial y_{i}} (T_i - y_i)
        \end{aligned} &
        \text{if $i \in$ output layer,}
        \\
        \begin{aligned}
            \frac{\partial}{\partial y_{i}} \left(\sum\limits^{n}_{j=1}W_{ij}\frac{\partial E}{\partial y_{j}}\right)
        \end{aligned} &
        \text{otherwise;}
    \end{cases}
\end{equation}

Fourthly, combining the partial derivatives allow the computation of the error at a given neuron $i$ such that:

\begin{equation}
    \frac{\partial E}{\partial y_{i}} \frac{\partial y_{i}}{\partial x_{i}} = e_{i} =
    \begin{cases}
        \begin{aligned}
            \frac{\partial \phi(x_i)}{\partial x_{i}} (T_i - y_i)
        \end{aligned} &
        \text{if $i \in$ output layer,}
        \\
        \begin{aligned}
            \frac{\partial \phi(x_i)}{\partial x_{i}} \left(\sum\limits^{n}_{j=1}W_{ij} e_{j}\right)
        \end{aligned} &
        \text{otherwise;}
    \end{cases}
\end{equation}

Finally, the weight can be updated from the gradient as follows:

\begin{equation}
    \frac{\partial E}{\partial W_{ij}} = \frac{\partial E}{\partial y_{i}} \frac{\partial y_{i}}{\partial x_{i}} \frac{\partial x_{i}}{\partial W_{ij}} = e_{i}y_{j}
\end{equation}
\begin{equation}
    W_{ij} = W_{ij} - \eta\,e_{i}y_{j}
\end{equation}

With $\eta$ the learning rate.
\chapter{Signal Pre-processing}
\label{ch:appendixprocessing}

This appendix illustrates the different pre-processing operations applied to the sensor signals. On the following figures, both sensors data have been recorded during the same typing session and the values along the three axis (i.e. x, y, and z) are processed.

\section{Gyroscope}

\begin{figure}[H]
    \centering
    \includegraphics[width=.9\linewidth]{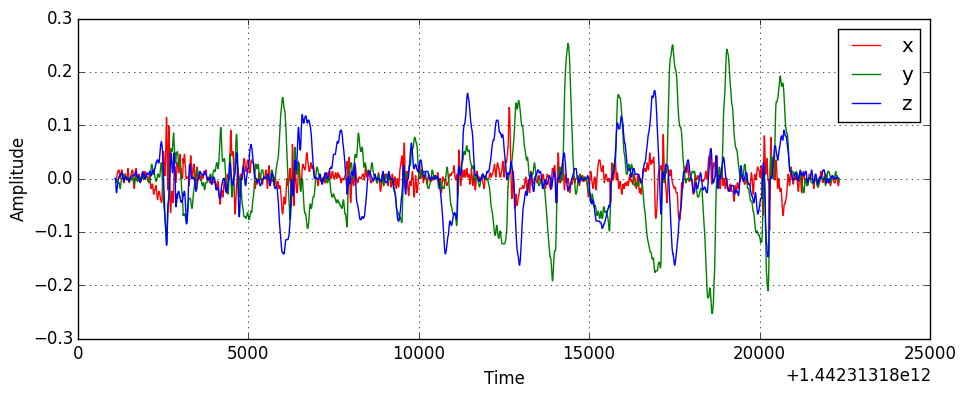}
    \caption{Gyroscope raw signal.}
\end{figure}
    
\begin{figure}[H]
    \centering
    \includegraphics[width=.9\linewidth]{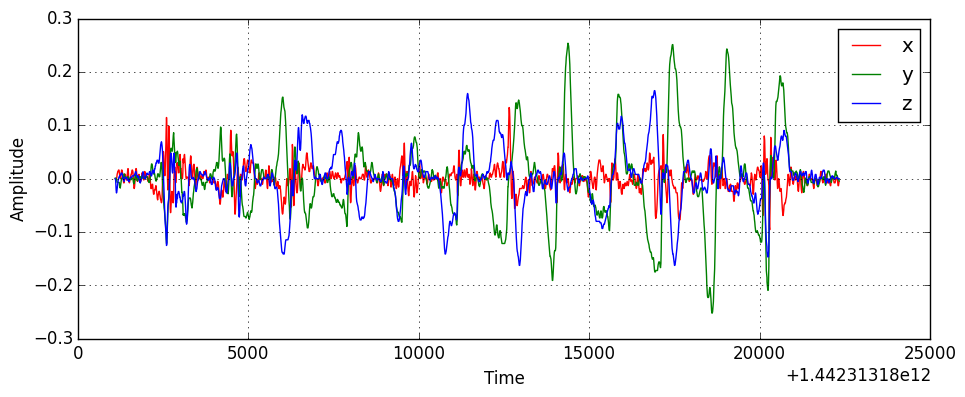}
    \caption{Gyroscope signal after calibration.}
\end{figure}
    
\begin{figure}[H]
    \centering
    \includegraphics[width=.9\linewidth]{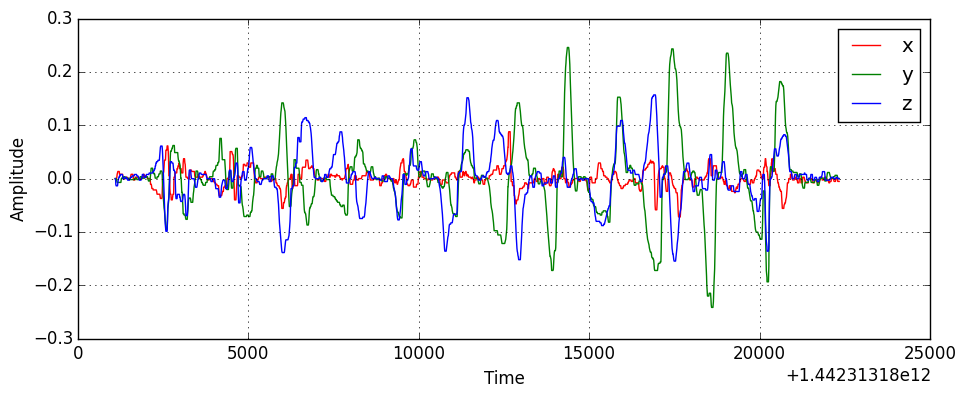}
    \caption{Gyroscope signal after median filtering.}
\end{figure}
    
\begin{figure}[H]
    \centering
    \includegraphics[width=.9\linewidth]{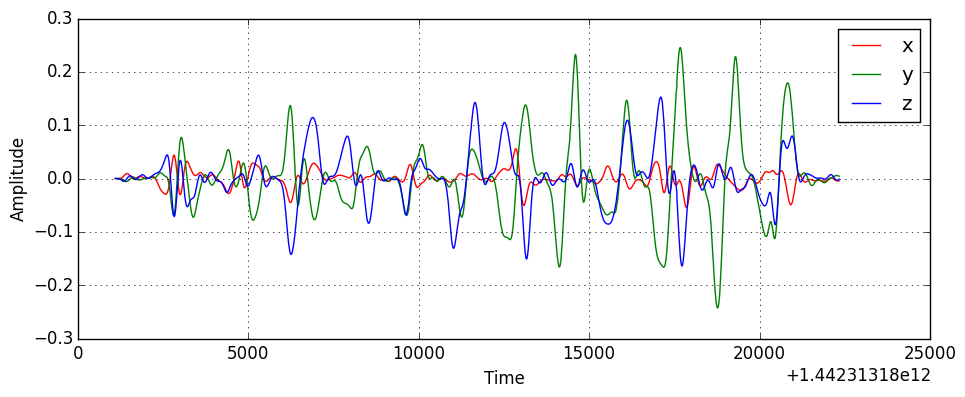}
    \caption{Gyroscope signal after low-pass Butterworth filtering.}
\end{figure}
    
\begin{figure}[H]
    \centering
    \includegraphics[width=.9\linewidth]{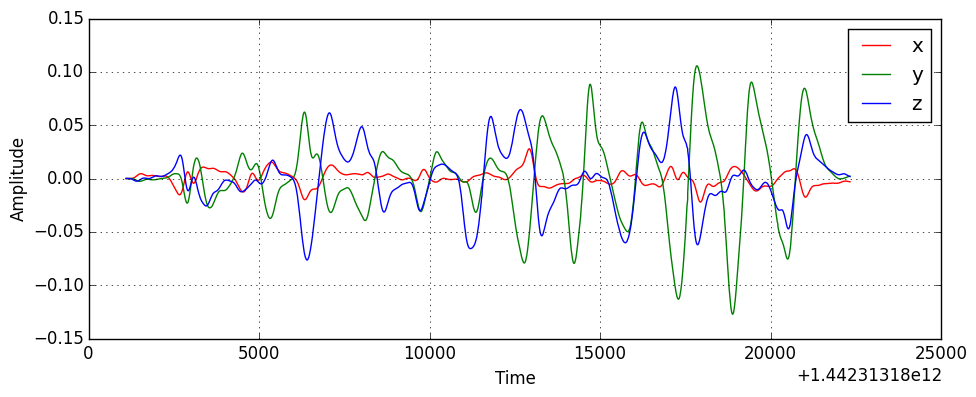}
    \caption{Gyroscope signal after Kalman filtering.}
\end{figure}

\begin{figure}[H]
    \centering
    \includegraphics[width=.9\linewidth]{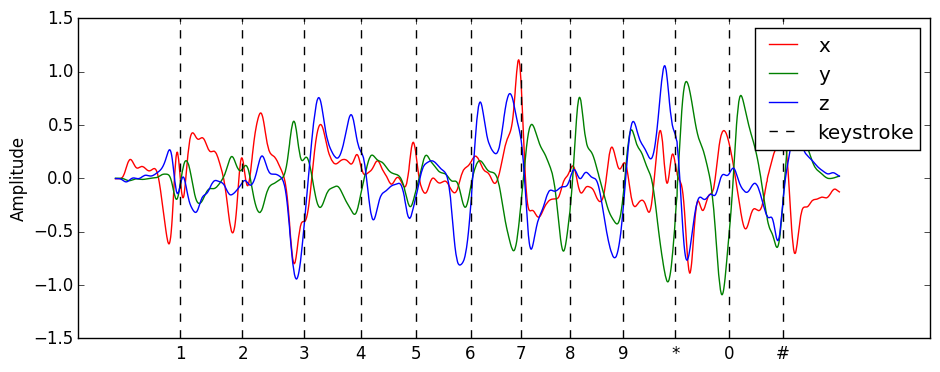}
    \caption{Pre-processed gyroscope signal with labels.}
\end{figure}

\section{Accelerometer}

\begin{figure}[H]
    \centering
    \includegraphics[width=.9\linewidth]{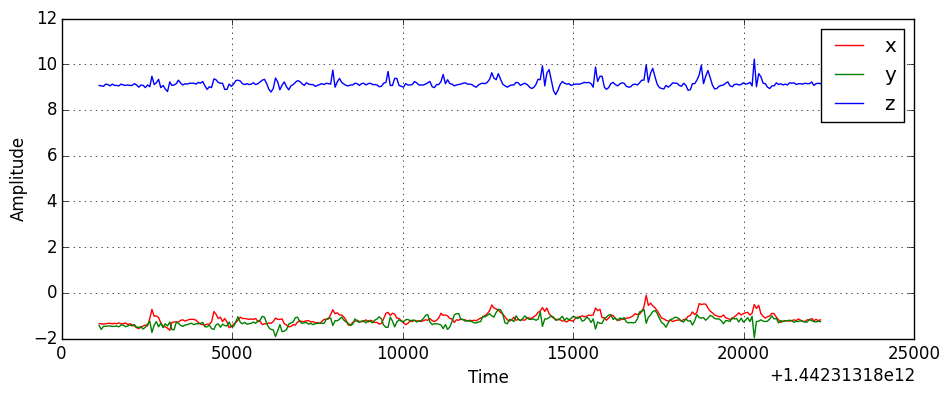}
    \caption{Accelerometer raw signal.}
\end{figure}

\begin{figure}[H]
    \centering
    \includegraphics[width=.9\linewidth]{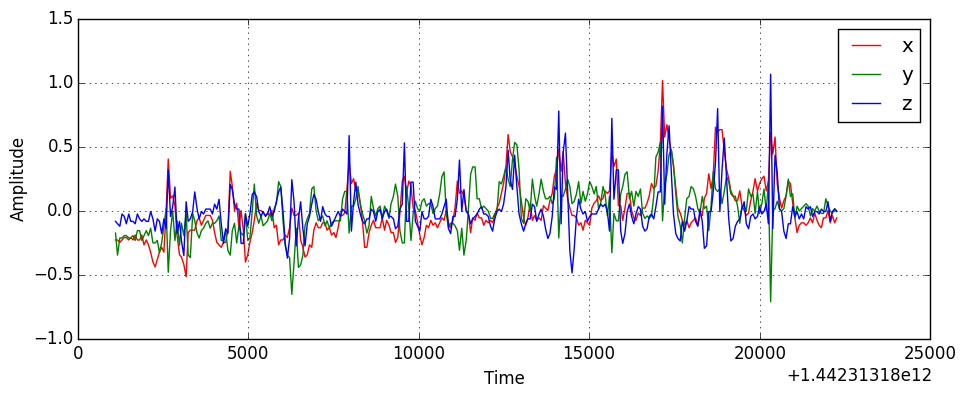}
    \caption{Accelerometer signal after calibration.}
\end{figure}

\begin{figure}[H]
    \centering
    \includegraphics[width=.9\linewidth]{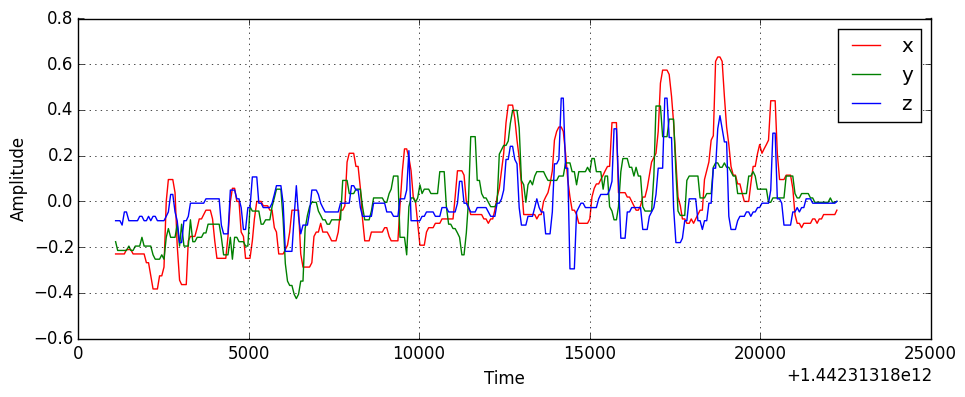}
    \caption{Accelerometer signal after median filtering.}
\end{figure}

\begin{figure}[H]
    \centering
    \includegraphics[width=.9\linewidth]{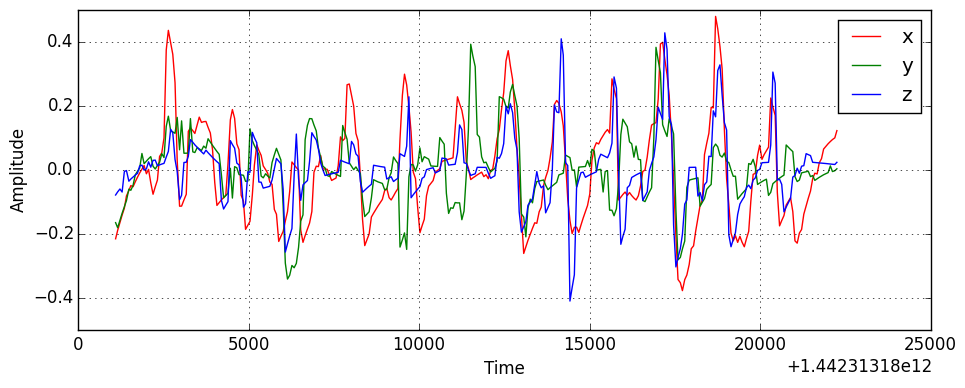}
    \caption{Accelerometer signal after high-pass Butterworth filtering.}
\end{figure}

\begin{figure}[H]
    \centering
    \includegraphics[width=.9\linewidth]{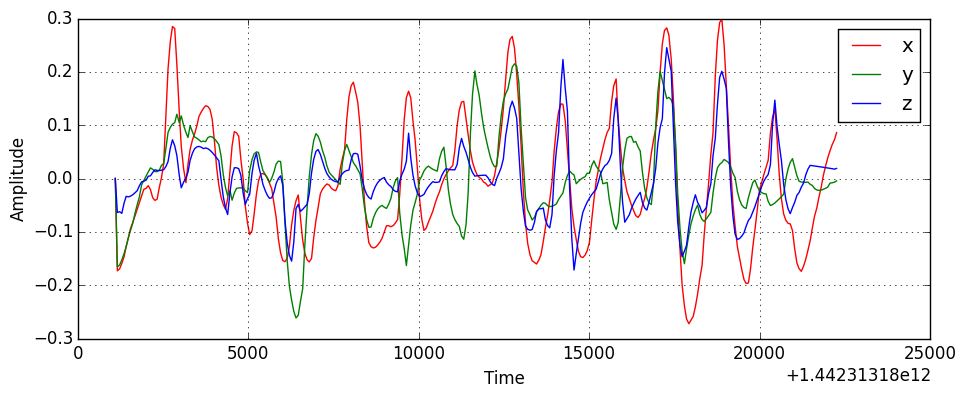}
    \caption{Accelerometer signal after Kalman filtering.}
\end{figure}

\begin{figure}[H]
    \centering
    \includegraphics[width=.9\linewidth]{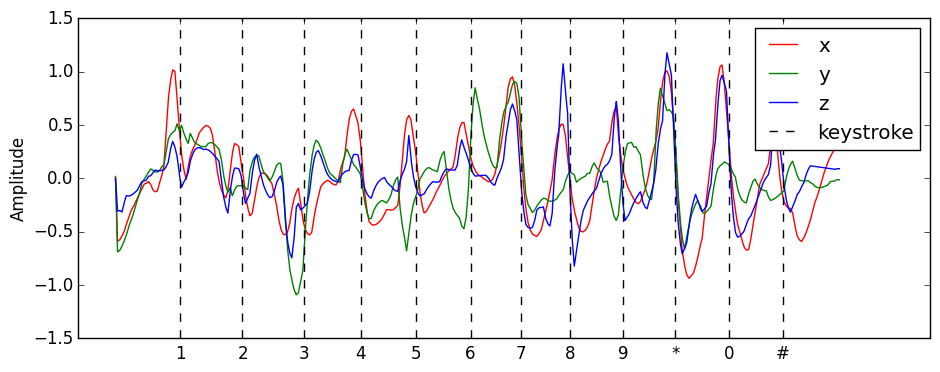}
    \caption{Pre-processed accelerometer signal with labels.}
\end{figure}
\chapter{Confusion Matrices from Model Benchmark}
\label{ch:appendixbenchmark}

This appendix shows the confusion matrices generated during the benchmark detailed in Section \ref{ssec:modelbenchmark} and performed to compare different neural network architectures on different types of features.

\section{Model Training with Statistical Features}

\begin{figure}[H]
    \centering
    \includegraphics[width=.5\linewidth]{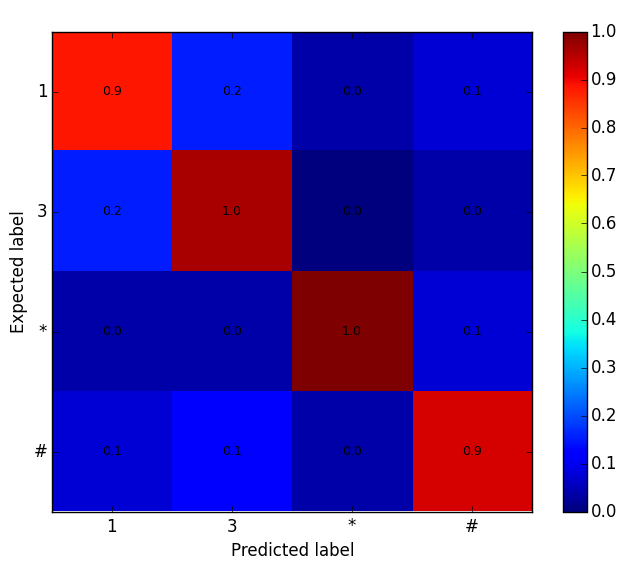}
    \caption{Feedforward Sigmoid.}
\end{figure}

\begin{figure}[H]
    \centering
    \includegraphics[width=.5\linewidth]{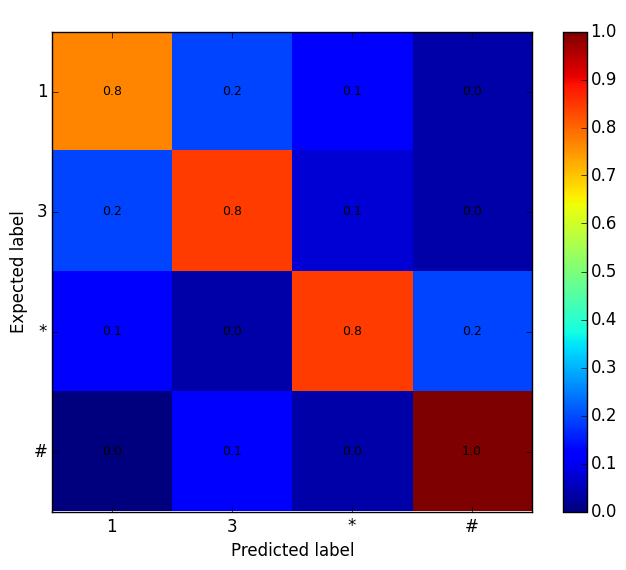}
    \caption{Feedforward Tanh.}
\end{figure}

\section{Model Training with Data Segment as Features}

\begin{figure}[H]
    \centering
    \includegraphics[width=.5\linewidth]{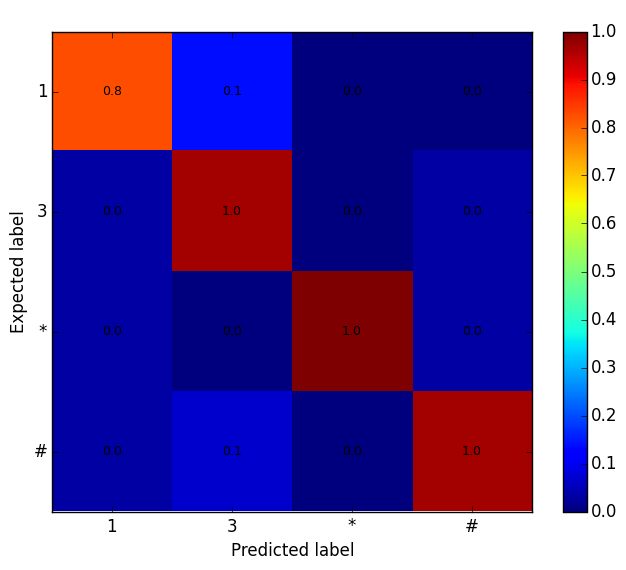}
    \caption{Feedforward Sigmoid.}
\end{figure}

\begin{figure}[H]
    \centering
    \includegraphics[width=.5\linewidth]{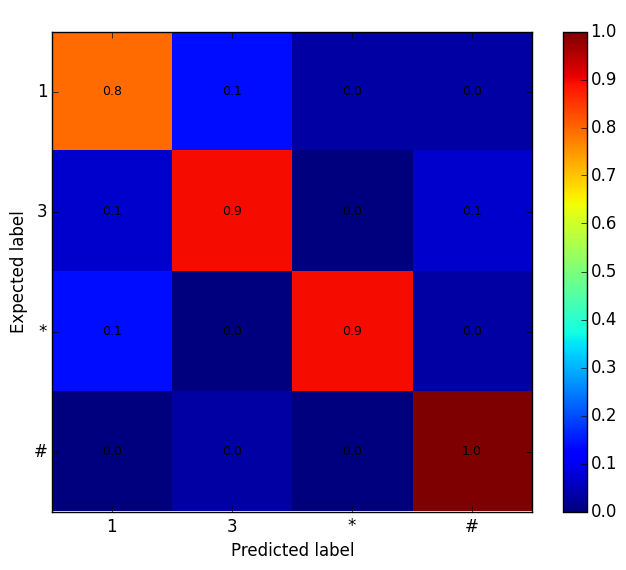}
    \caption{Feedforward Tanh.}
\end{figure}

\begin{figure}[H]
    \centering
    \includegraphics[width=.5\linewidth]{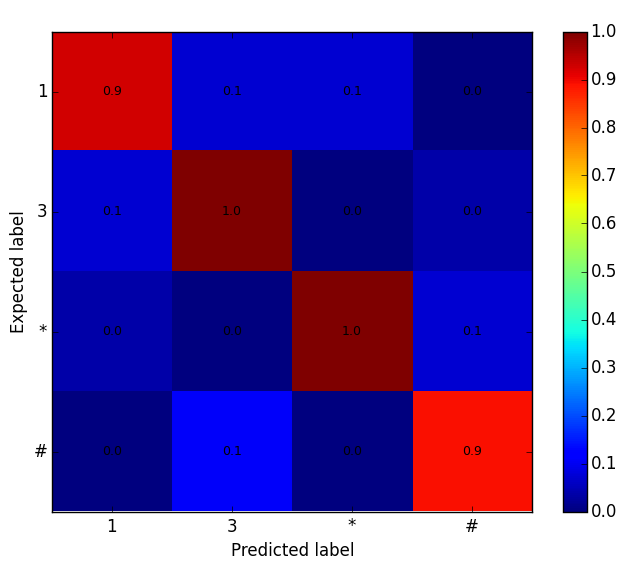}
    \caption{Recurrent LSTM with forget gate.}
\end{figure}

\begin{figure}[H]
    \centering
    \includegraphics[width=.5\linewidth]{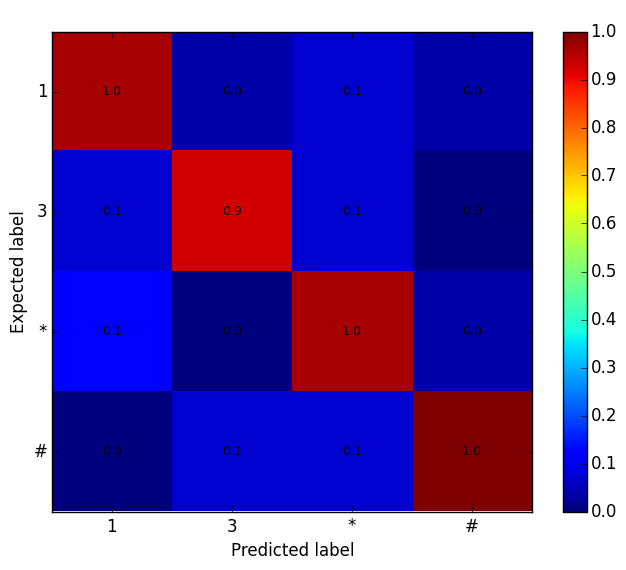}
    \caption{Recurrent LSTM with peephole connections.}
\end{figure}

\chapter{Experiment Results}
\label{ch:appendixresults}

This appendix provides result details (i.e. classifier loss during training, confusion matrices from evaluation) to support Chapter \ref{ch:evaluation}.

\section{Results for Experiment 1: Touchlogging Attack}

\subsection{FNN-Sigmoid}

\begin{figure}[H]
    \begin{subfigure}{.5\textwidth}
        \centering
        \includegraphics[width=1.\linewidth]{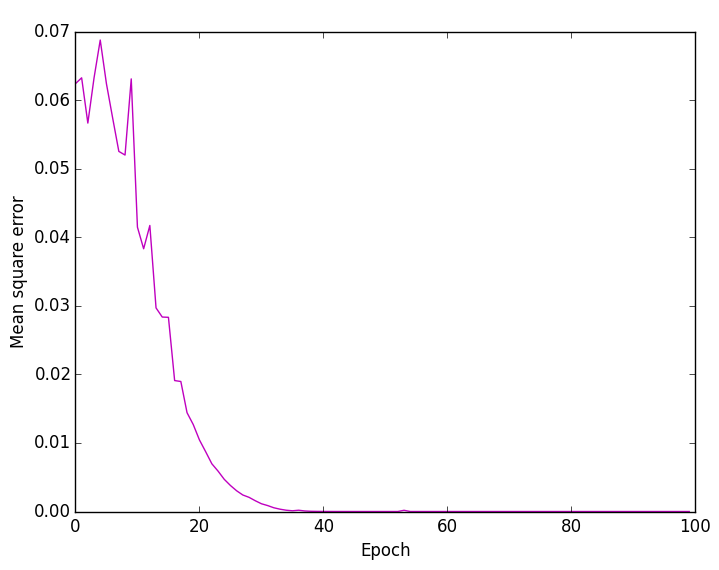}
        \caption{Mean Squared Error optimization.}
    \end{subfigure}
    \begin{subfigure}{.5\textwidth}
        \centering
        \includegraphics[width=1.\linewidth]{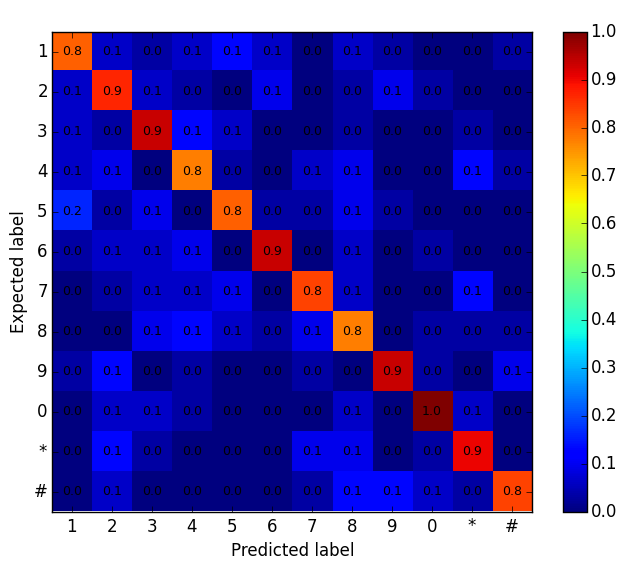}
        \caption{Confusion Matrix.}
    \end{subfigure}
    \caption{Touchlogging with FNN-Sigmoid P-T}
    \label{fig:touchloggingsigmoidpt}
\end{figure}

\begin{figure}[H]
    \begin{subfigure}{.5\textwidth}
        \centering
        \includegraphics[width=1.\linewidth]{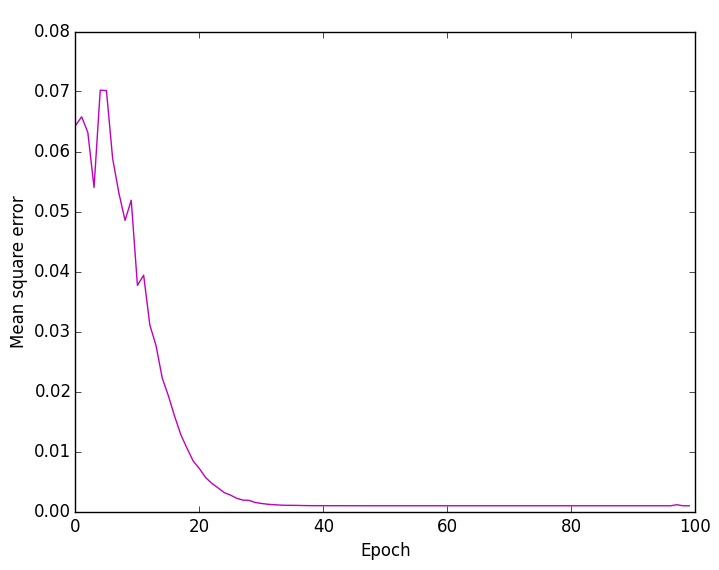}
        \caption{Mean Squared Error optimization.}
    \end{subfigure}
    \begin{subfigure}{.5\textwidth}
        \centering
        \includegraphics[width=1.\linewidth]{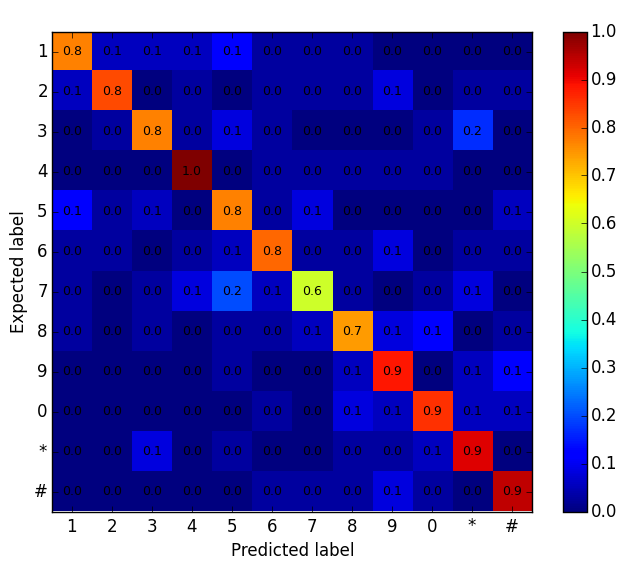}
        \caption{Confusion Matrix.}
    \end{subfigure}
    \caption{Touchlogging with FNN-Sigmoid P-H}
    \label{fig:touchloggingsigmoidph}
\end{figure}

\begin{figure}[H]
    \begin{subfigure}{.5\textwidth}
        \centering
        \includegraphics[width=1.\linewidth]{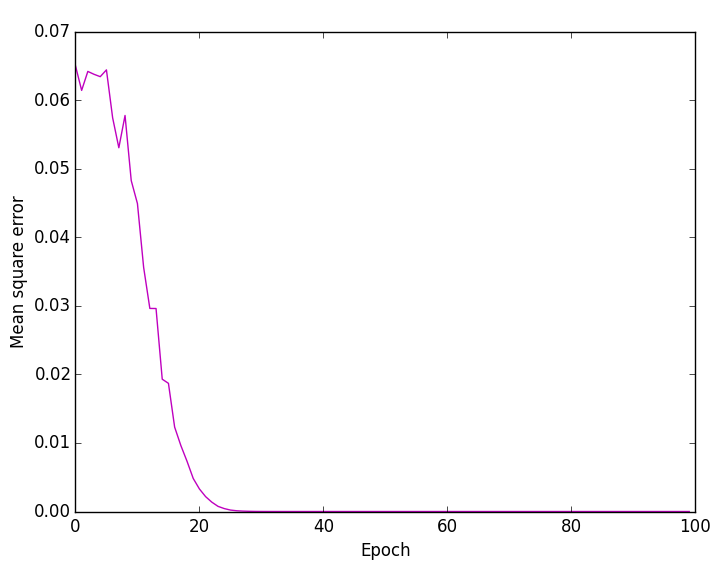}
        \caption{Mean Squared Error optimization.}
    \end{subfigure}
    \begin{subfigure}{.5\textwidth}
        \centering
        \includegraphics[width=1.\linewidth]{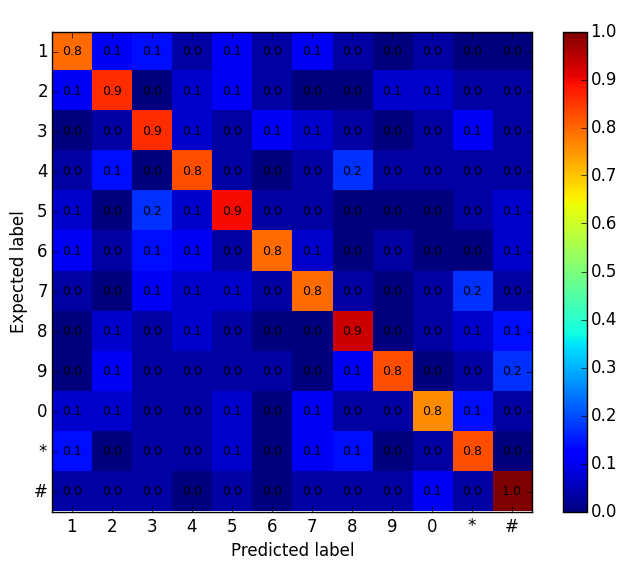}
        \caption{Confusion Matrix.}
    \end{subfigure}
    \caption{Touchlogging with FNN-Sigmoid R-T}
    \label{fig:touchloggingsigmoidrt}
\end{figure}

\begin{figure}[H]
    \begin{subfigure}{.5\textwidth}
        \centering
        \includegraphics[width=1.\linewidth]{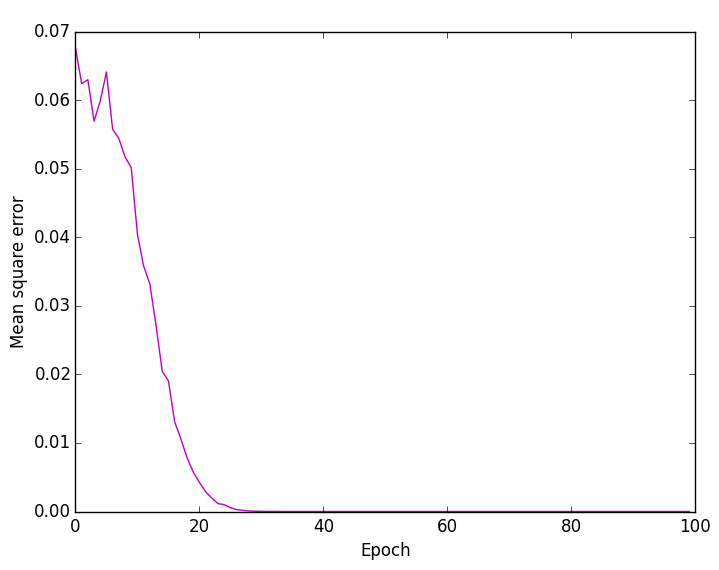}
        \caption{Mean Squared Error optimization.}
    \end{subfigure}
    \begin{subfigure}{.5\textwidth}
        \centering
        \includegraphics[width=1.\linewidth]{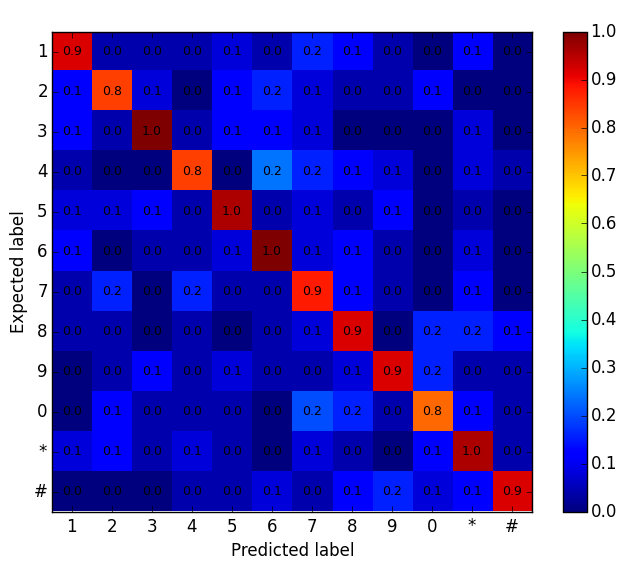}
        \caption{Confusion Matrix.}
    \end{subfigure}
    \caption{Touchlogging with FNN-Sigmoid R-H}
    \label{fig:touchloggingsigmoidrh}
\end{figure}

\subsection{FNN-Tanh}

\begin{figure}[H]
    \begin{subfigure}{.5\textwidth}
        \centering
        \includegraphics[width=1.\linewidth]{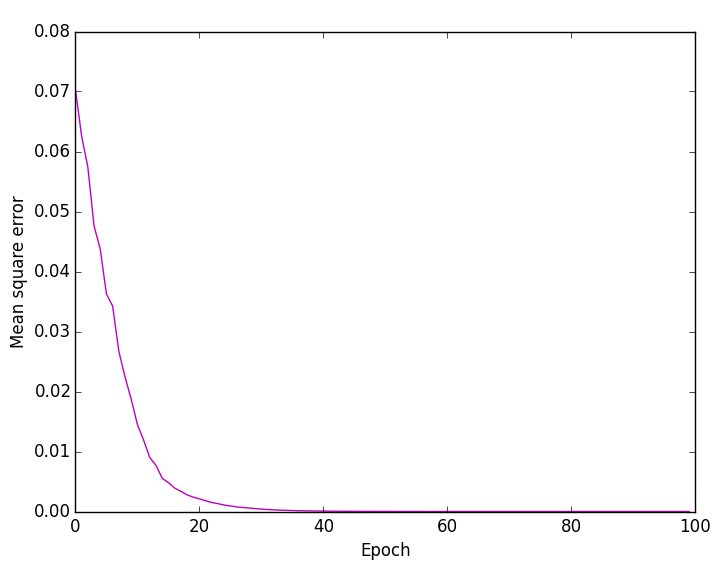}
        \caption{Mean Squared Error optimization.}
    \end{subfigure}
    \begin{subfigure}{.5\textwidth}
        \centering
        \includegraphics[width=1.\linewidth]{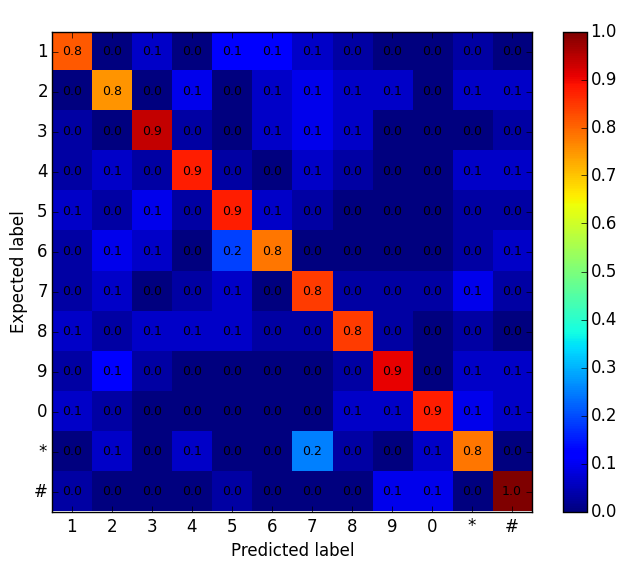}
        \caption{Confusion Matrix.}
    \end{subfigure}
    \caption{Touchlogging with FNN-Tanh P-T}
    \label{fig:touchloggingtahnpt}
\end{figure}

\begin{figure}[H]
    \begin{subfigure}{.5\textwidth}
        \centering
        \includegraphics[width=1.\linewidth]{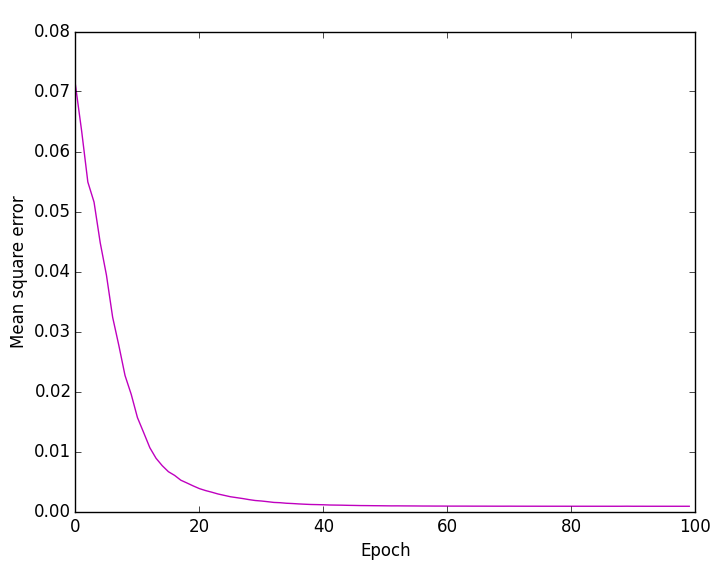}
        \caption{Mean Squared Error optimization.}
    \end{subfigure}
    \begin{subfigure}{.5\textwidth}
        \centering
        \includegraphics[width=1.\linewidth]{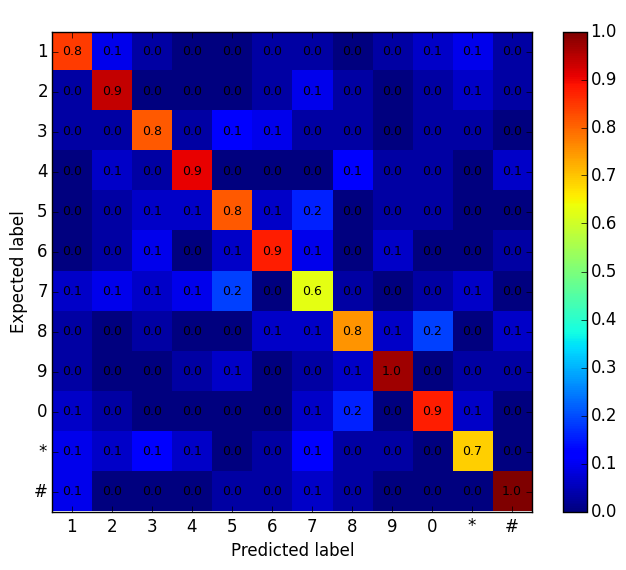}
        \caption{Confusion Matrix.}
    \end{subfigure}
    \caption{Touchlogging with FNN-Tanh P-H}
    \label{fig:touchloggingtahnph}
\end{figure}

\begin{figure}[H]
    \begin{subfigure}{.5\textwidth}
        \centering
        \includegraphics[width=1.\linewidth]{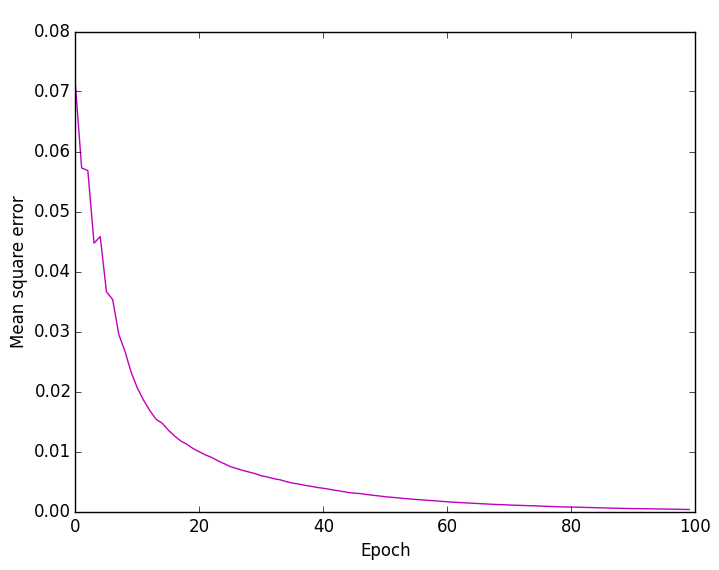}
        \caption{Mean Squared Error optimization.}
    \end{subfigure}
    \begin{subfigure}{.5\textwidth}
        \centering
        \includegraphics[width=1.\linewidth]{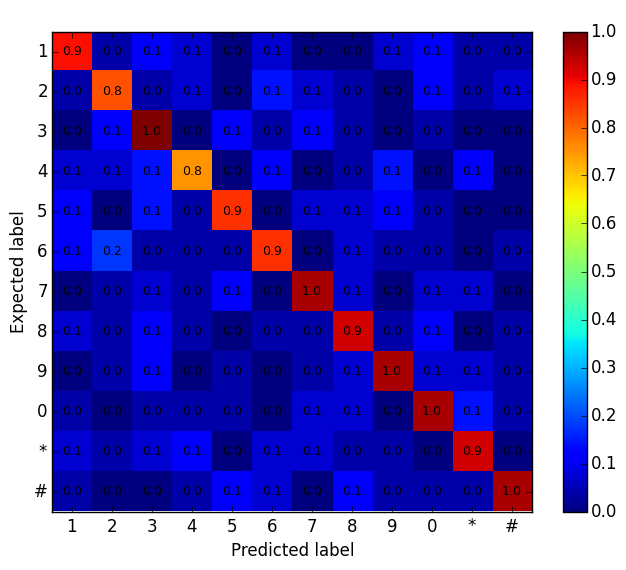}
        \caption{Confusion Matrix.}
    \end{subfigure}
    \caption{Touchlogging with FNN-Tanh R-T}
    \label{fig:touchloggingtahnrt}
\end{figure}

\begin{figure}[H]
    \begin{subfigure}{.5\textwidth}
        \centering
        \includegraphics[width=1.\linewidth]{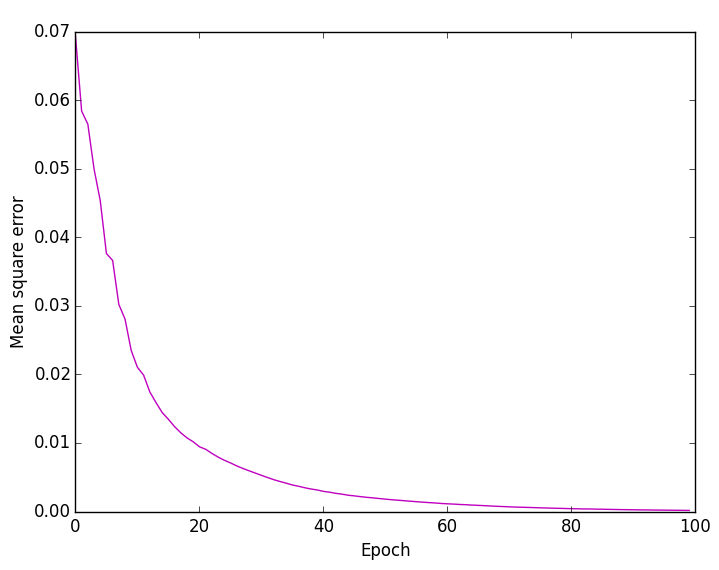}
        \caption{Mean Squared Error optimization.}
    \end{subfigure}
    \begin{subfigure}{.5\textwidth}
        \centering
        \includegraphics[width=1.\linewidth]{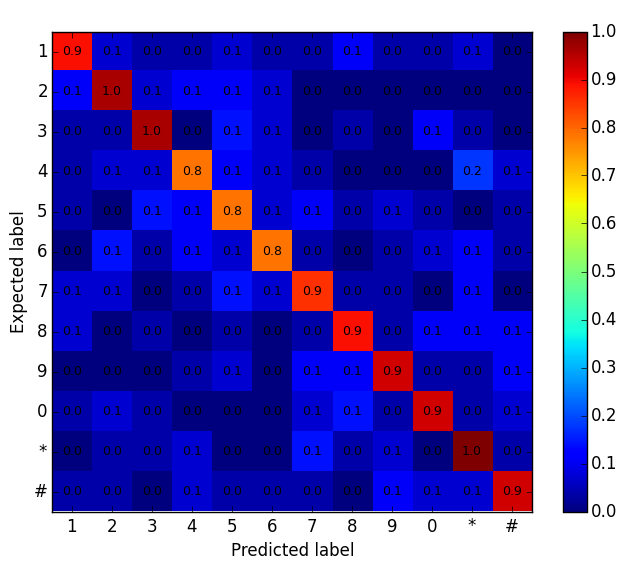}
        \caption{Confusion Matrix.}
    \end{subfigure}
    \caption{Touchlogging with FNN-Tanh R-H}
    \label{fig:touchloggingtahnrh}
\end{figure}

\subsection{RNN-LSTM}

\begin{figure}[H]
    \begin{subfigure}{.5\textwidth}
        \centering
        \includegraphics[width=1.\linewidth]{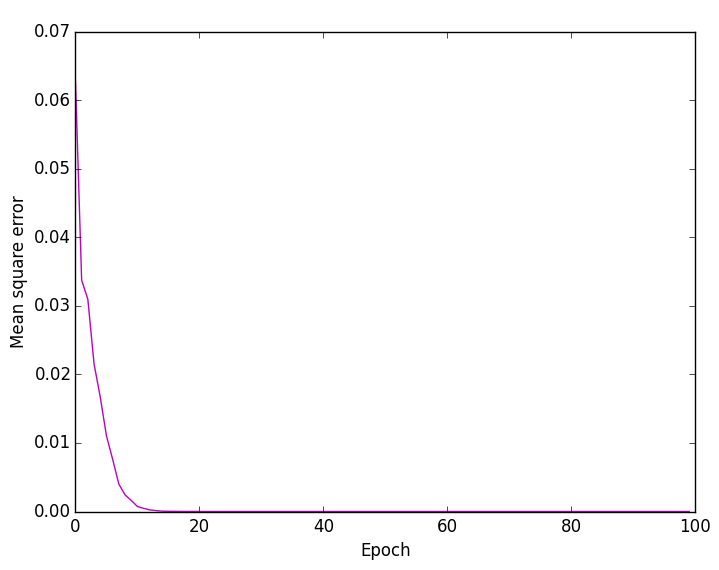}
        \caption{Mean Squared Error optimization.}
    \end{subfigure}
    \begin{subfigure}{.5\textwidth}
        \centering
        \includegraphics[width=1.\linewidth]{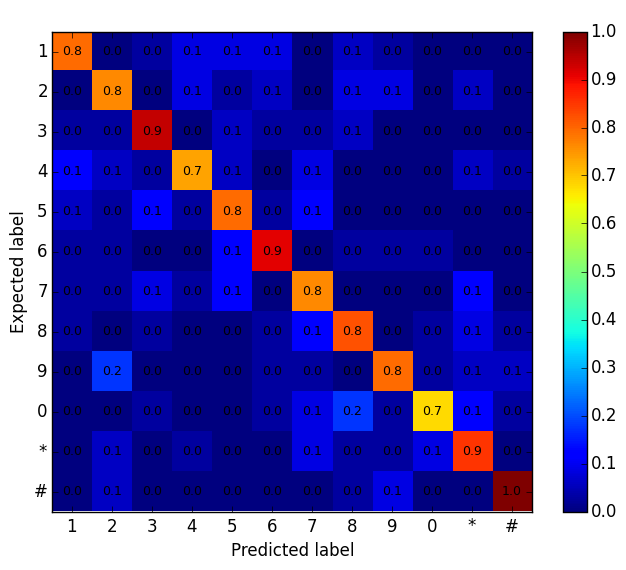}
        \caption{Confusion Matrix.}
    \end{subfigure}
    \caption{Touchlogging with RNN-LSTM P-T}
    \label{fig:touchlogginglstmpt}
\end{figure}

\begin{figure}[H]
    \begin{subfigure}{.5\textwidth}
        \centering
        \includegraphics[width=1.\linewidth]{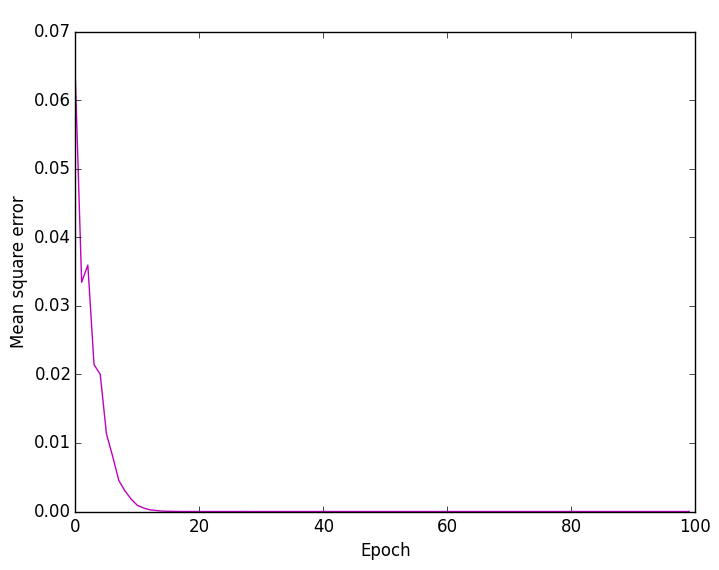}
        \caption{Mean Squared Error optimization.}
    \end{subfigure}
    \begin{subfigure}{.5\textwidth}
        \centering
        \includegraphics[width=1.\linewidth]{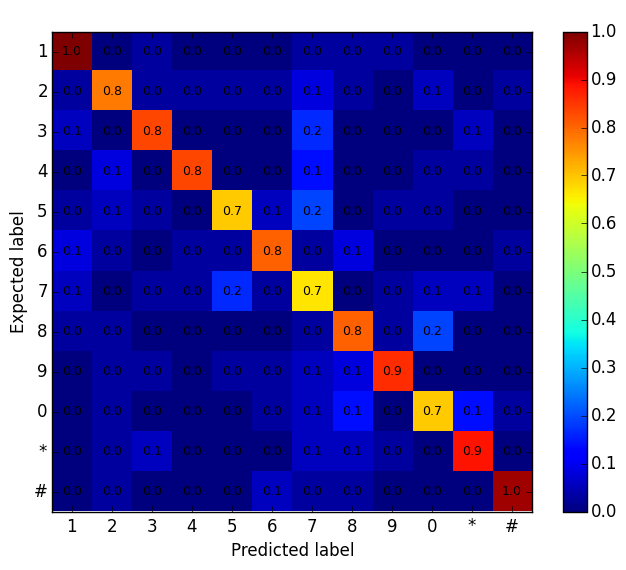}
        \caption{Confusion Matrix.}
    \end{subfigure}
    \caption{Touchlogging with RNN-LSTM P-H}
    \label{fig:touchlogginglstmph}
\end{figure}

\begin{figure}[H]
    \begin{subfigure}{.5\textwidth}
        \centering
        \includegraphics[width=1.\linewidth]{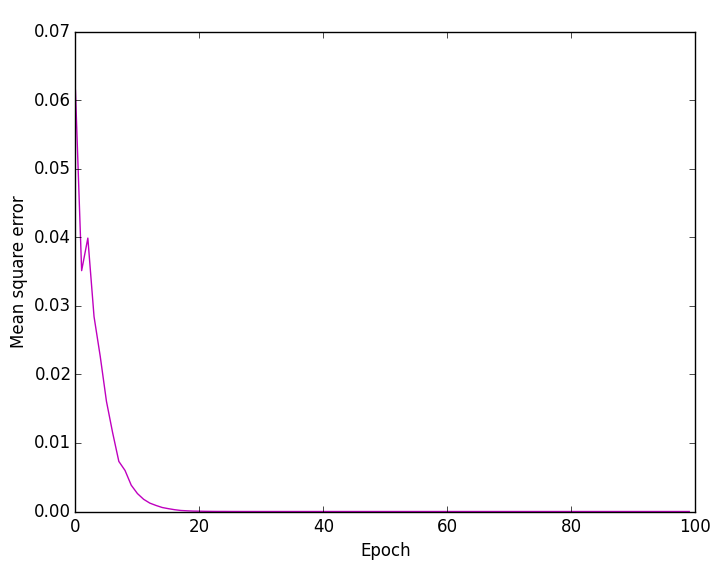}
        \caption{Mean Squared Error optimization.}
    \end{subfigure}
    \begin{subfigure}{.5\textwidth}
        \centering
        \includegraphics[width=1.\linewidth]{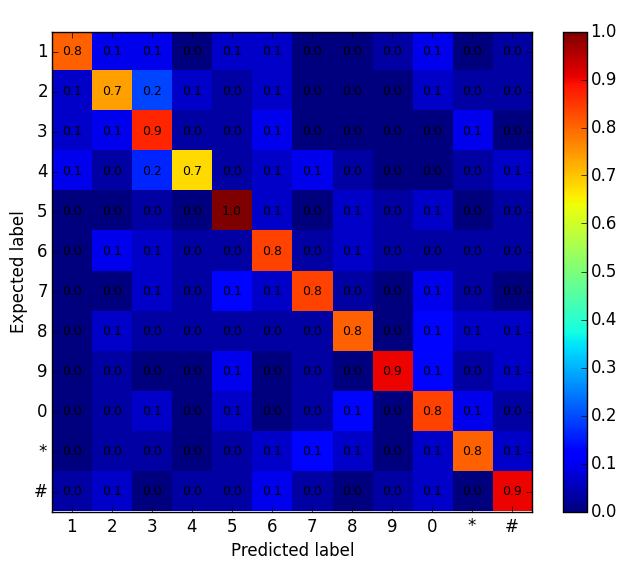}
        \caption{Confusion Matrix.}
    \end{subfigure}
    \caption{Touchlogging with RNN-LSTM R-T}
    \label{fig:touchlogginglstmrt}
\end{figure}

\begin{figure}[H]
    \begin{subfigure}{.5\textwidth}
        \centering
        \includegraphics[width=1.\linewidth]{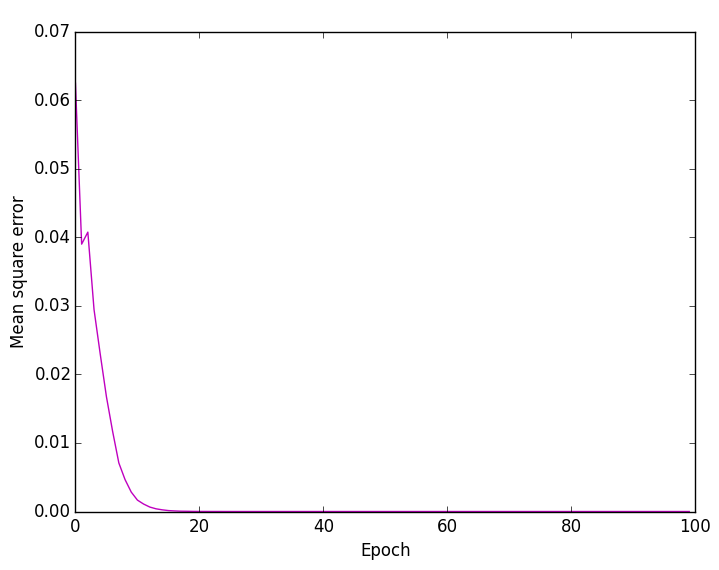}
        \caption{Mean Squared Error optimization.}
    \end{subfigure}
    \begin{subfigure}{.5\textwidth}
        \centering
        \includegraphics[width=1.\linewidth]{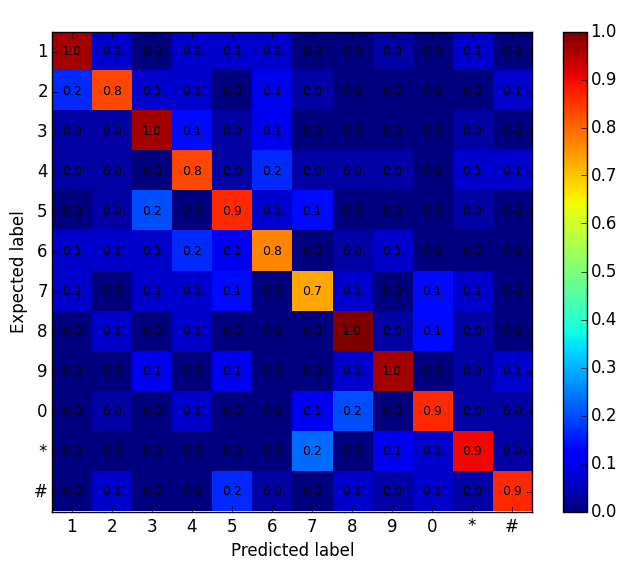}
        \caption{Confusion Matrix.}
    \end{subfigure}
    \caption{Touchlogging with RNN-LSTM R-H}
    \label{fig:touchlogginglstmrh}
\end{figure}

\section{Results for Experiment 2: Keylogging Attack}

\subsection{FNN-Sigmoid}

\begin{figure}[H]
    \begin{subfigure}{.5\textwidth}
        \centering
        \includegraphics[width=1.\linewidth]{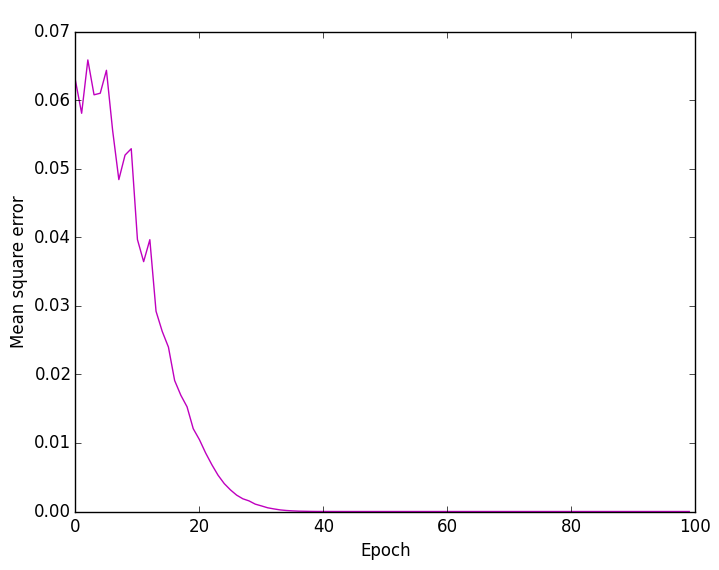}
        \caption{Mean Squared Error optimization.}
    \end{subfigure}
    \begin{subfigure}{.5\textwidth}
        \centering
        \includegraphics[width=1.\linewidth]{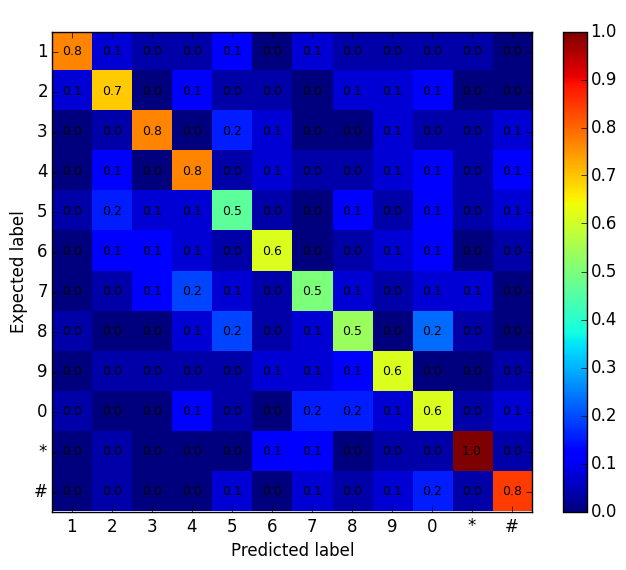}
        \caption{Confusion Matrix.}
    \end{subfigure}
    \caption{Keylogging with FNN-Sigmoid P-T}
    \label{fig:keyloggingsigmoidpt}
\end{figure}

\begin{figure}[H]
    \begin{subfigure}{.5\textwidth}
        \centering
        \includegraphics[width=1.\linewidth]{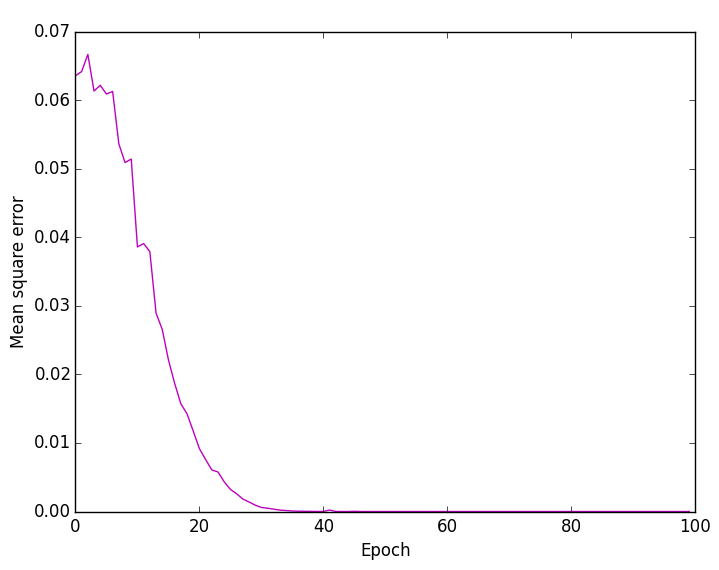}
        \caption{Mean Squared Error optimization.}
    \end{subfigure}
    \begin{subfigure}{.5\textwidth}
        \centering
        \includegraphics[width=1.\linewidth]{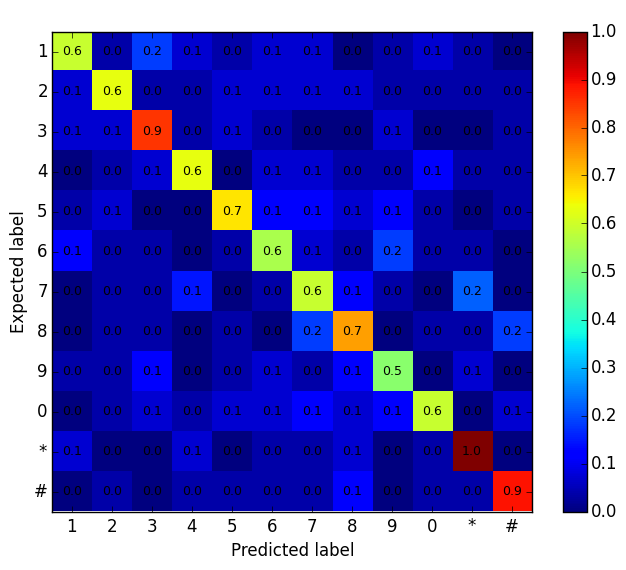}
        \caption{Confusion Matrix.}
    \end{subfigure}
    \caption{Keylogging with FNN-Sigmoid P-H}
    \label{fig:keyloggingsigmoidph}
\end{figure}

\begin{figure}[H]
    \begin{subfigure}{.5\textwidth}
        \centering
        \includegraphics[width=1.\linewidth]{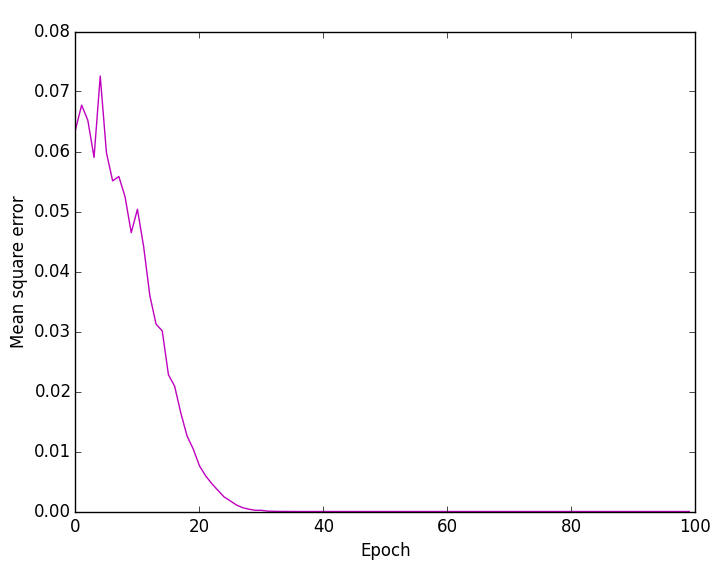}
        \caption{Mean Squared Error optimization.}
    \end{subfigure}
    \begin{subfigure}{.5\textwidth}
        \centering
        \includegraphics[width=1.\linewidth]{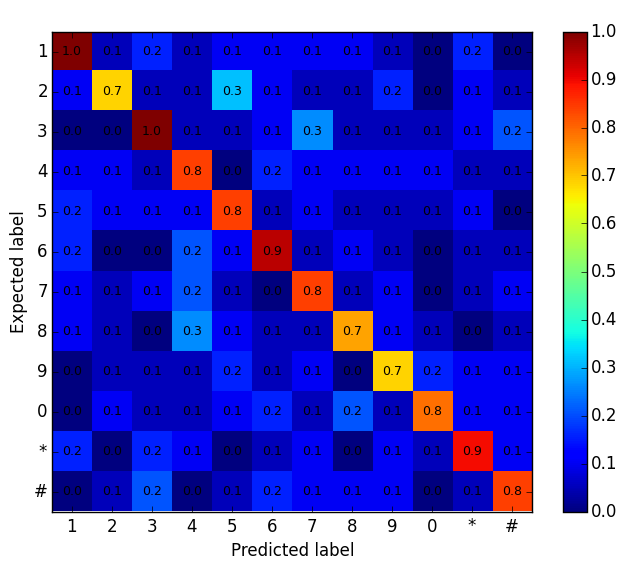}
        \caption{Confusion Matrix.}
    \end{subfigure}
    \caption{Keylogging with FNN-Sigmoid R-T}
    \label{fig:keyloggingsigmoidrt}
\end{figure}

\begin{figure}[H]
    \begin{subfigure}{.5\textwidth}
        \centering
        \includegraphics[width=1.\linewidth]{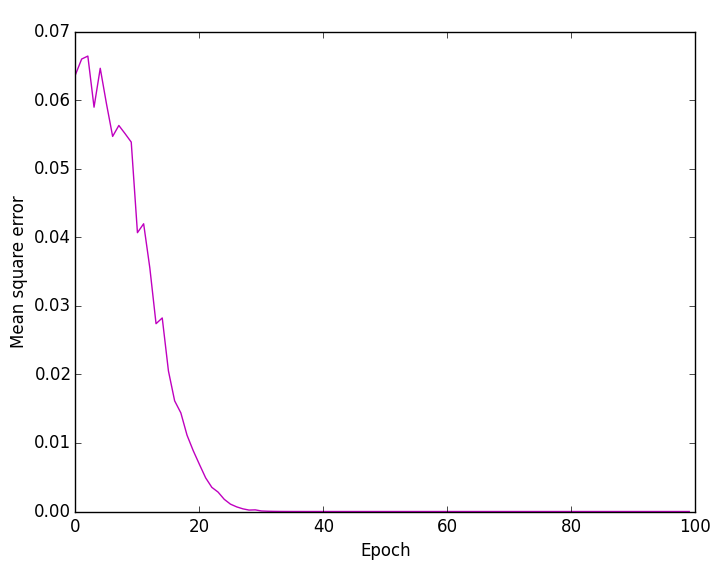}
        \caption{Mean Squared Error optimization.}
    \end{subfigure}
    \begin{subfigure}{.5\textwidth}
        \centering
        \includegraphics[width=1.\linewidth]{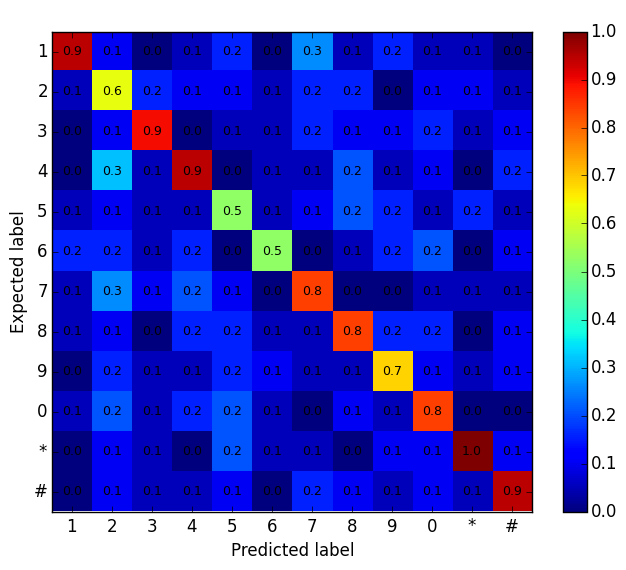}
        \caption{Confusion Matrix.}
    \end{subfigure}
    \caption{Keylogging with FNN-Sigmoid R-H}
    \label{fig:keyloggingsigmoidrh}
\end{figure}

\subsection{FNN-Tanh}

\begin{figure}[H]
    \begin{subfigure}{.5\textwidth}
        \centering
        \includegraphics[width=1.\linewidth]{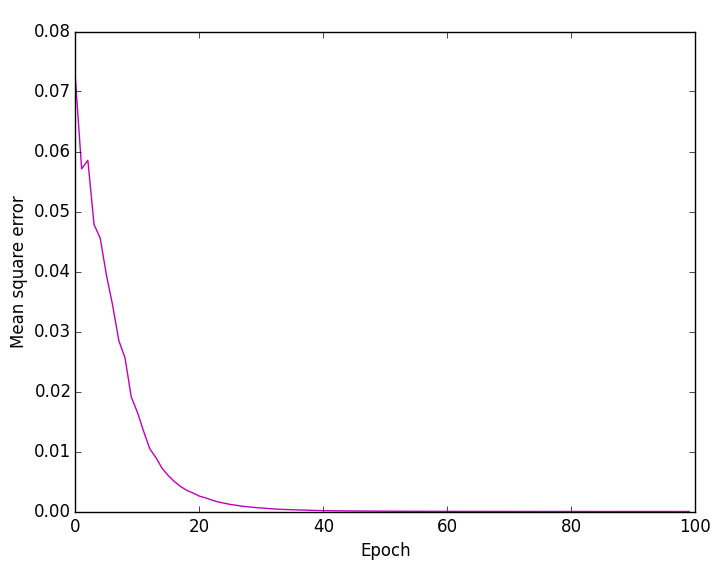}
        \caption{Mean Squared Error optimization.}
    \end{subfigure}
    \begin{subfigure}{.5\textwidth}
        \centering
        \includegraphics[width=1.\linewidth]{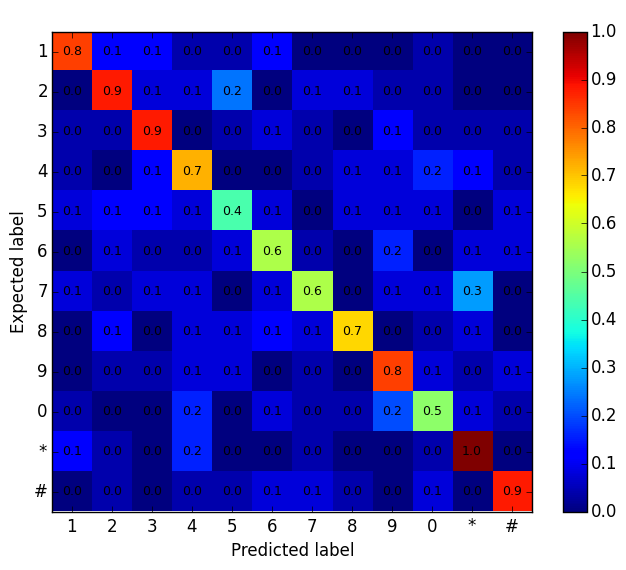}
        \caption{Confusion Matrix.}
    \end{subfigure}
    \caption{Keylogging with FNN-Tanh P-T}
    \label{fig:keyloggingthanpt}
\end{figure}

\begin{figure}[H]
    \begin{subfigure}{.5\textwidth}
        \centering
        \includegraphics[width=1.\linewidth]{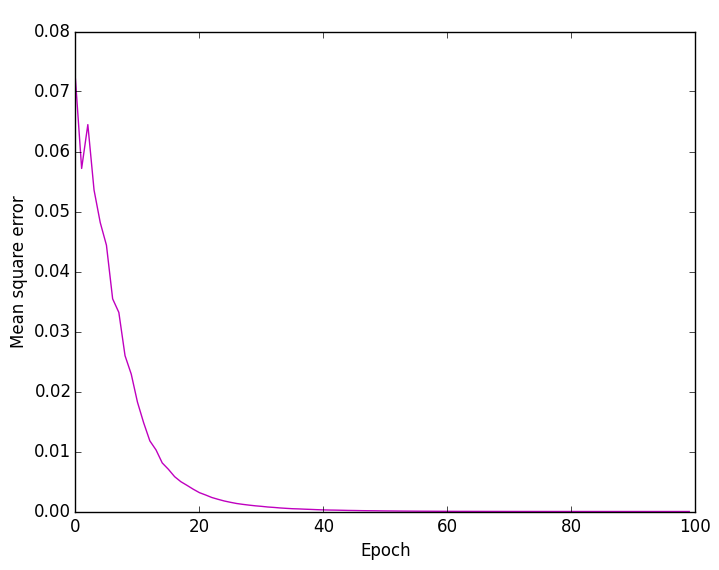}
        \caption{Mean Squared Error optimization.}
    \end{subfigure}
    \begin{subfigure}{.5\textwidth}
        \centering
        \includegraphics[width=1.\linewidth]{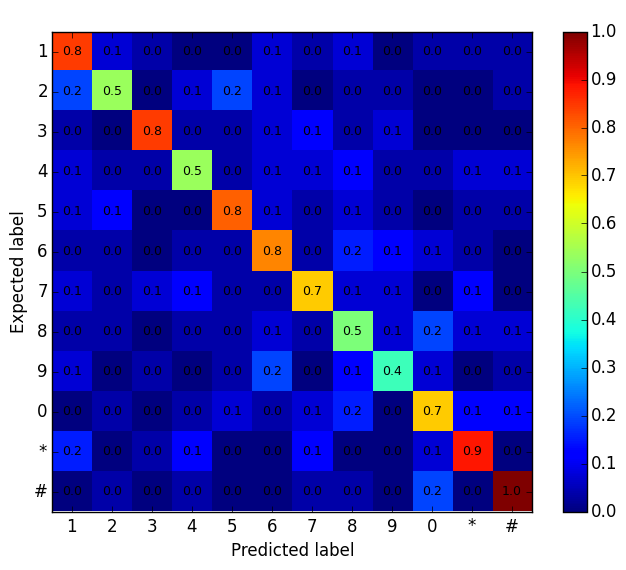}
        \caption{Confusion Matrix.}
    \end{subfigure}
    \caption{Keylogging with FNN-Tanh P-H}
    \label{fig:keyloggingthanph}
\end{figure}

\begin{figure}[H]
    \begin{subfigure}{.5\textwidth}
        \centering
        \includegraphics[width=1.\linewidth]{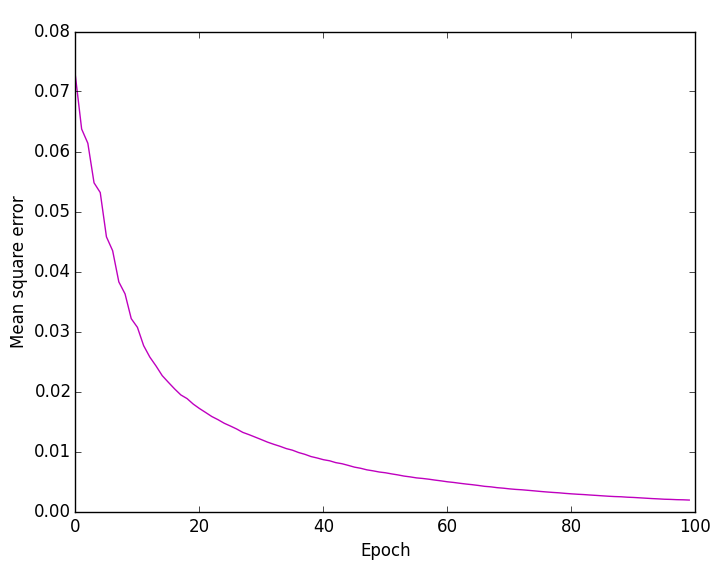}
        \caption{Mean Squared Error optimization.}
    \end{subfigure}
    \begin{subfigure}{.5\textwidth}
        \centering
        \includegraphics[width=1.\linewidth]{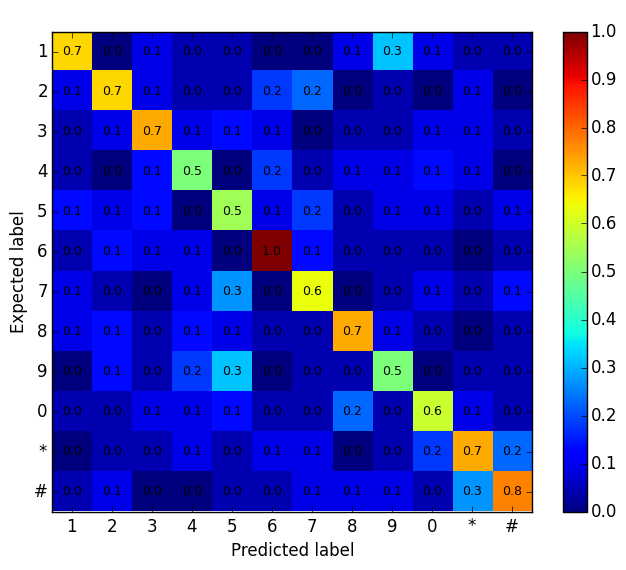}
        \caption{Confusion Matrix.}
    \end{subfigure}
    \caption{Keylogging with FNN-Tanh R-T}
    \label{fig:keyloggingthanrt}
\end{figure}

\begin{figure}[H]
    \begin{subfigure}{.5\textwidth}
        \centering
        \includegraphics[width=1.\linewidth]{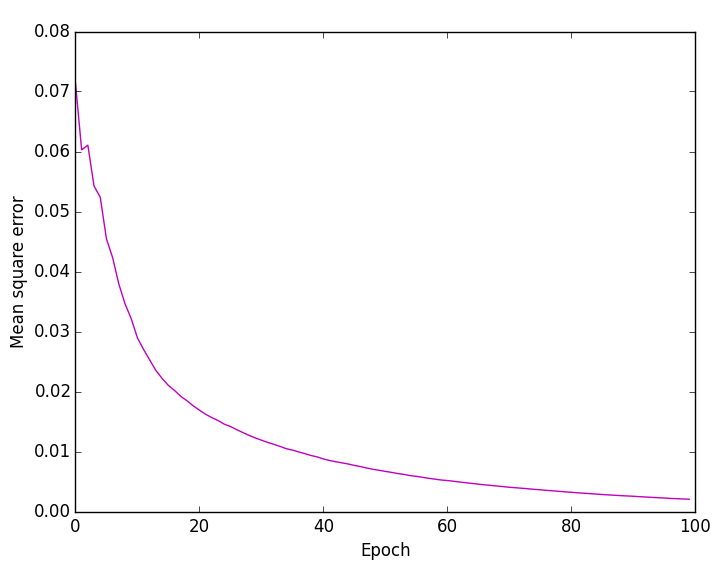}
        \caption{Mean Squared Error optimization.}
    \end{subfigure}
    \begin{subfigure}{.5\textwidth}
        \centering
        \includegraphics[width=1.\linewidth]{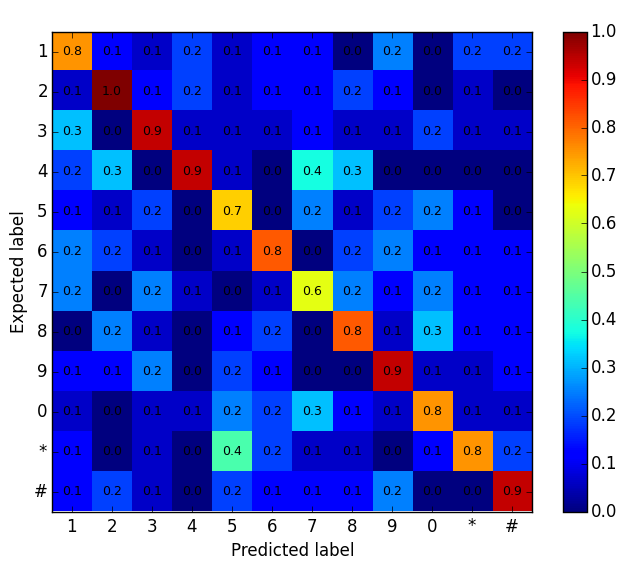}
        \caption{Confusion Matrix.}
    \end{subfigure}
    \caption{Keylogging with FNN-Tanh R-H}
    \label{fig:keyloggingthanrh}
\end{figure}

\subsection{RNN-LSTM}

\begin{figure}[H]
    \begin{subfigure}{.5\textwidth}
        \centering
        \includegraphics[width=1.\linewidth]{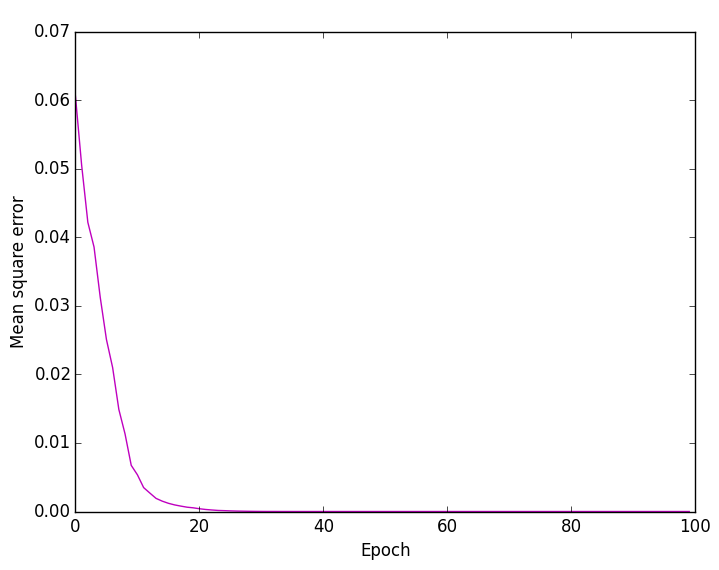}
        \caption{Mean Squared Error optimization.}
    \end{subfigure}
    \begin{subfigure}{.5\textwidth}
        \centering
        \includegraphics[width=1.\linewidth]{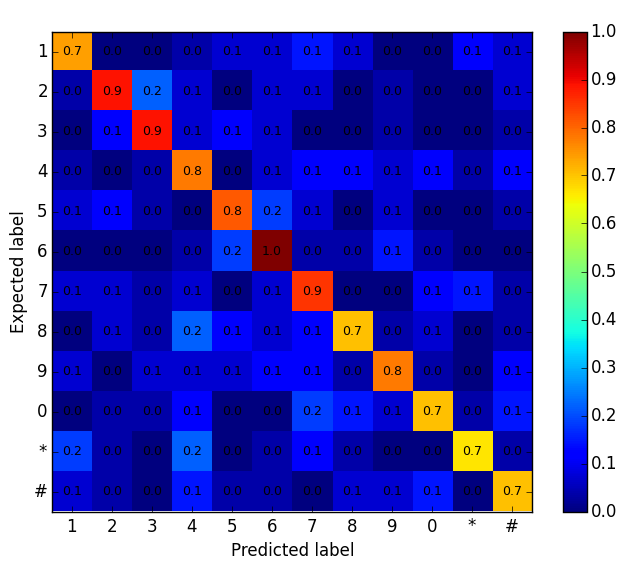}
        \caption{Confusion Matrix.}
    \end{subfigure}
    \caption{Keylogging with RNN-LSTM P-T}
    \label{fig:keylogginglstmpt}
\end{figure}

\begin{figure}[H]
    \begin{subfigure}{.5\textwidth}
        \centering
        \includegraphics[width=1.\linewidth]{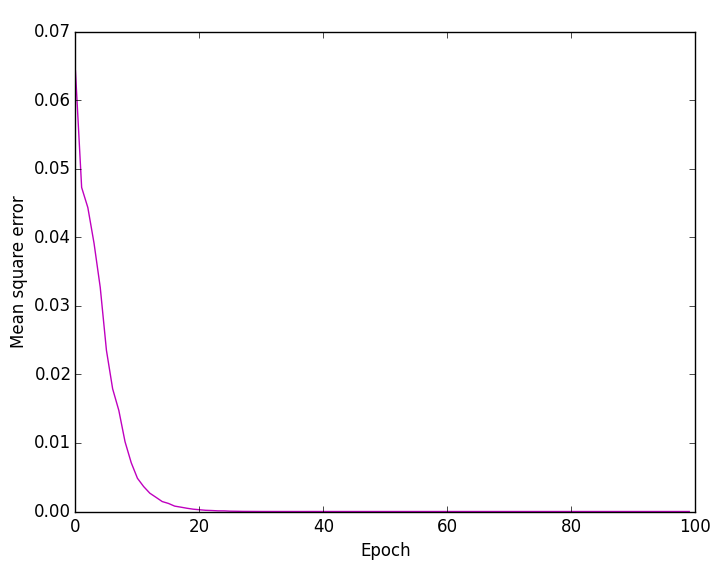}
        \caption{Mean Squared Error optimization.}
    \end{subfigure}
    \begin{subfigure}{.5\textwidth}
        \centering
        \includegraphics[width=1.\linewidth]{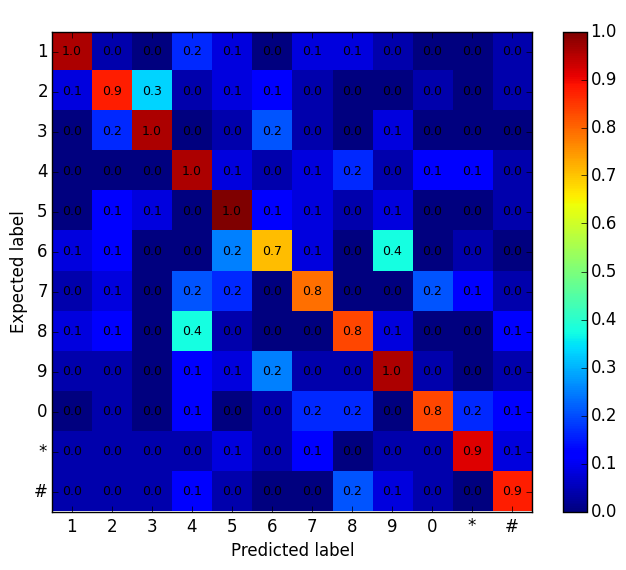}
        \caption{Confusion Matrix.}
    \end{subfigure}
    \caption{Keylogging with RNN-LSTM P-H}
    \label{fig:keylogginglstmph}
\end{figure}

\begin{figure}[H]
    \begin{subfigure}{.5\textwidth}
        \centering
        \includegraphics[width=1.\linewidth]{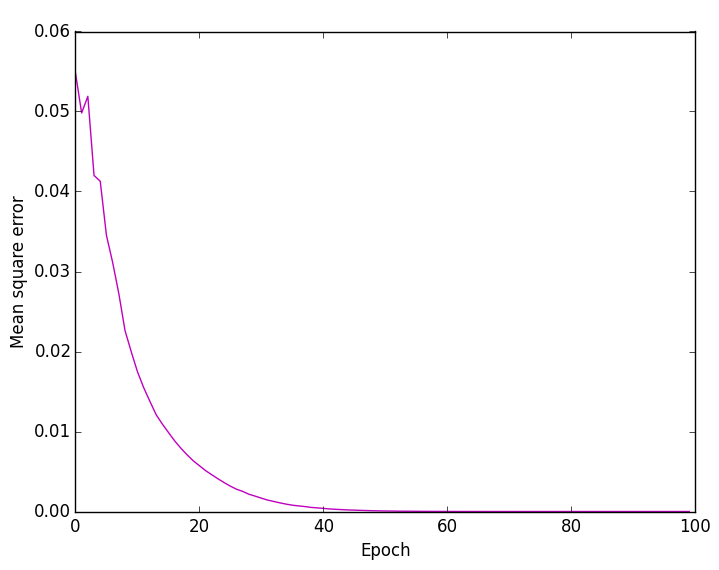}
        \caption{Mean Squared Error optimization.}
    \end{subfigure}
    \begin{subfigure}{.5\textwidth}
        \centering
        \includegraphics[width=1.\linewidth]{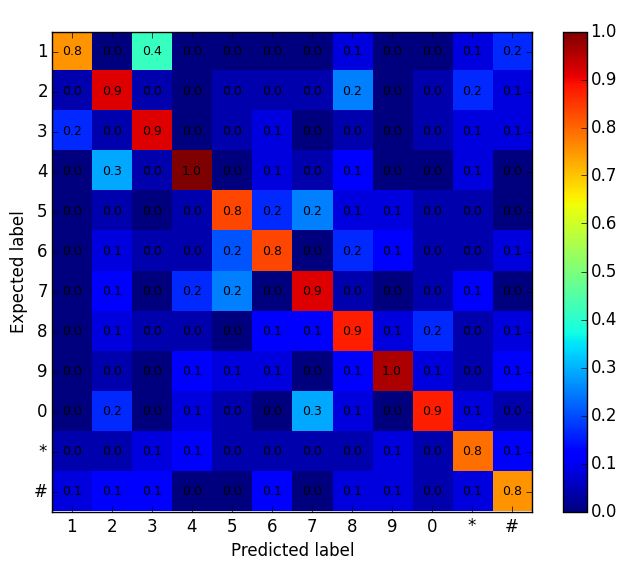}
        \caption{Confusion Matrix.}
    \end{subfigure}
    \caption{Keylogging with RNN-LSTM R-T}
    \label{fig:keylogginglstmrt}
\end{figure}

\begin{figure}[H]
    \begin{subfigure}{.5\textwidth}
        \centering
        \includegraphics[width=1.\linewidth]{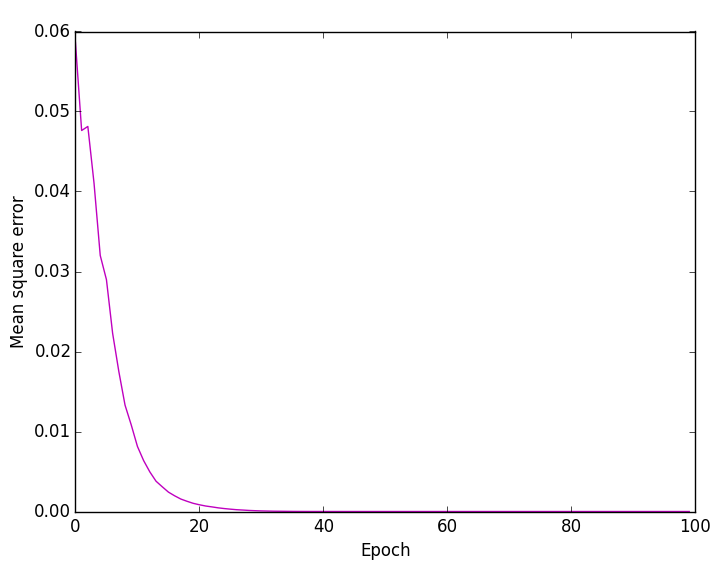}
        \caption{Mean Squared Error optimization.}
    \end{subfigure}
    \begin{subfigure}{.5\textwidth}
        \centering
        \includegraphics[width=1.\linewidth]{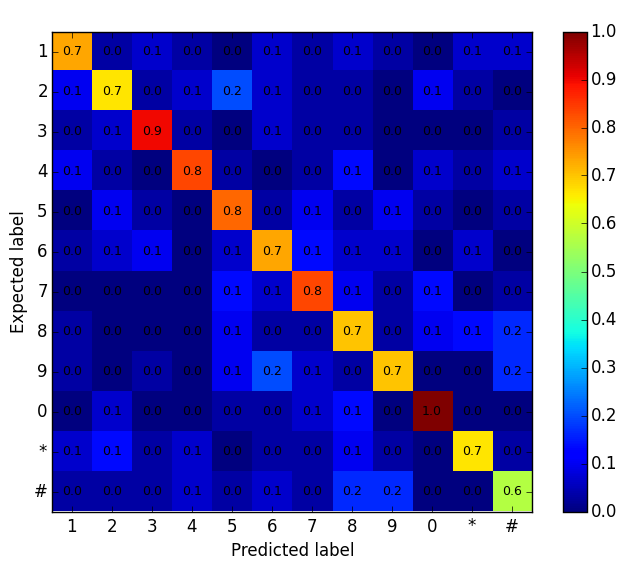}
        \caption{Confusion Matrix.}
    \end{subfigure}
    \caption{Keylogging with RNN-LSTM R-H}
    \label{fig:keylogginglstmrh}
\end{figure}

\section{Results for Experiment 3: from Touchlogging to Keylogging}

\begin{figure}[H]
    \begin{subfigure}{.5\textwidth}
        \centering
        \includegraphics[width=1.\linewidth]{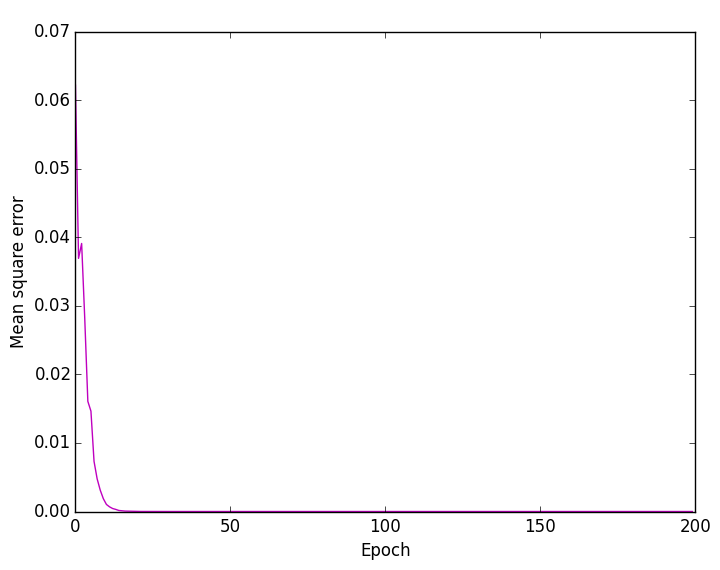}
        \caption{Mean Squared Error optimization.}
    \end{subfigure}
    \begin{subfigure}{.5\textwidth}
        \centering
        \includegraphics[width=1.\linewidth]{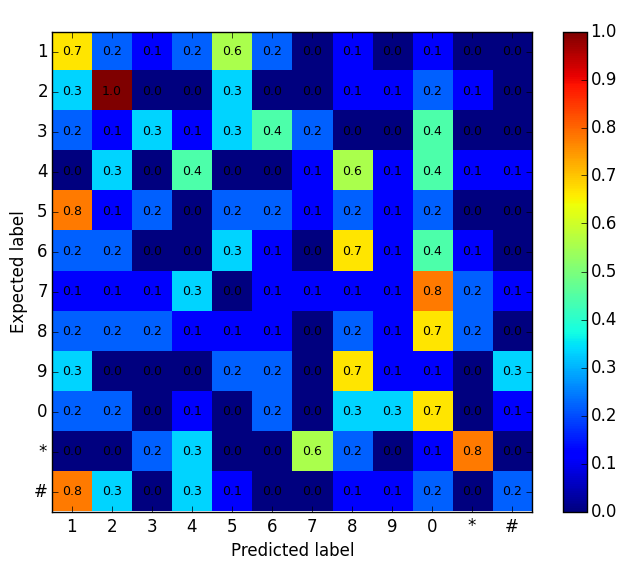}
        \caption{Confusion Matrix.}
    \end{subfigure}
    \caption{Touchlogging to Keylogging with RNN-LSTM P-T}
    \label{fig:exp3lstmpt}
\end{figure}

\begin{figure}[H]
    \begin{subfigure}{.5\textwidth}
        \centering
        \includegraphics[width=1.\linewidth]{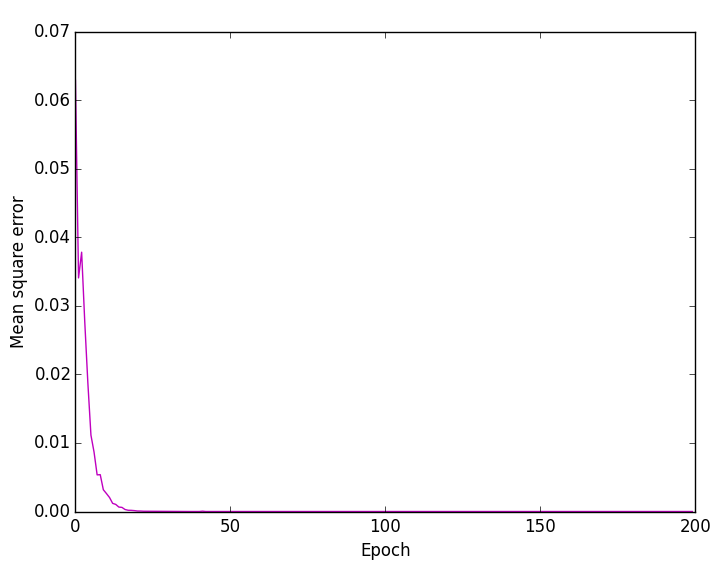}
        \caption{Mean Squared Error optimization.}
    \end{subfigure}
    \begin{subfigure}{.5\textwidth}
        \centering
        \includegraphics[width=1.\linewidth]{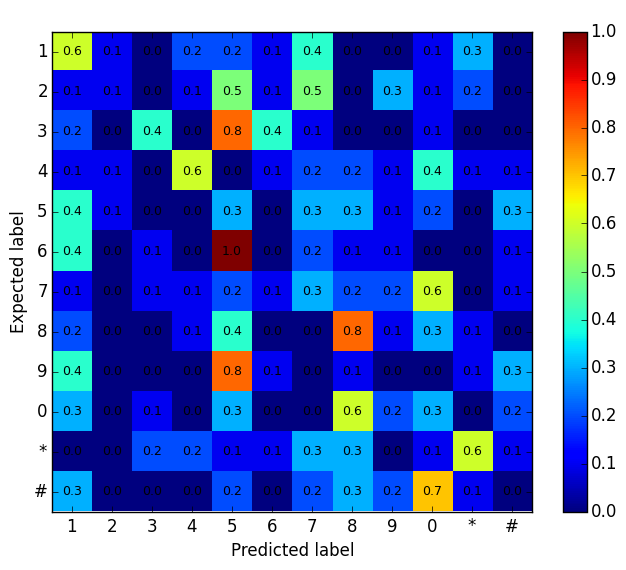}
        \caption{Confusion Matrix.}
    \end{subfigure}
    \caption{Touchlogging to Keylogging with RNN-LSTM P-H}
    \label{fig:exp3lstmph}
\end{figure}

\begin{figure}[H]
    \begin{subfigure}{.5\textwidth}
        \centering
        \includegraphics[width=1.\linewidth]{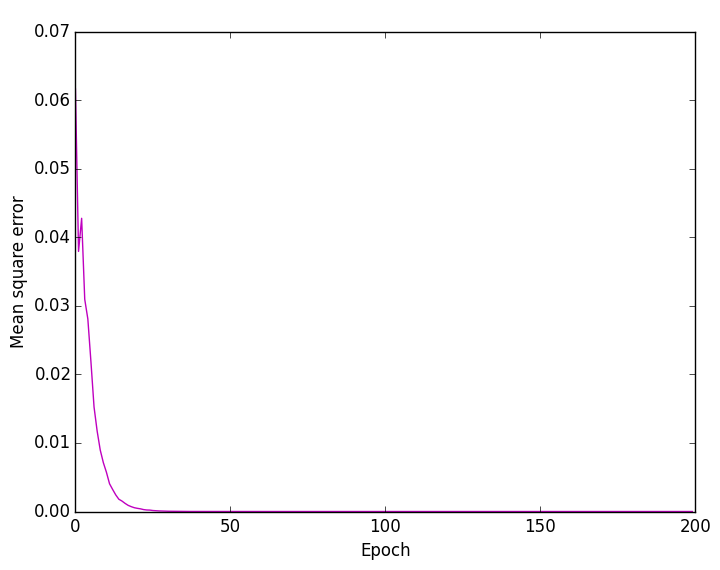}
        \caption{Mean Squared Error optimization.}
    \end{subfigure}
    \begin{subfigure}{.5\textwidth}
        \centering
        \includegraphics[width=1.\linewidth]{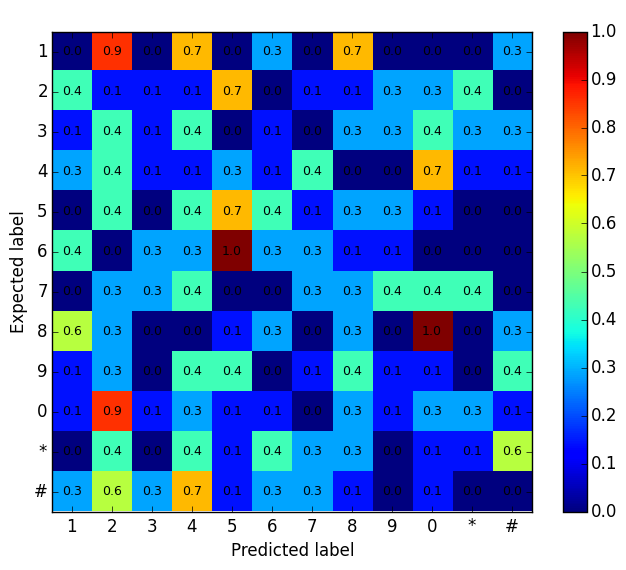}
        \caption{Confusion Matrix.}
    \end{subfigure}
    \caption{Touchlogging to Keylogging with RNN-LSTM R-T}
    \label{fig:exp3lstmrt}
\end{figure}

\begin{figure}[H]
    \begin{subfigure}{.5\textwidth}
        \centering
        \includegraphics[width=1.\linewidth]{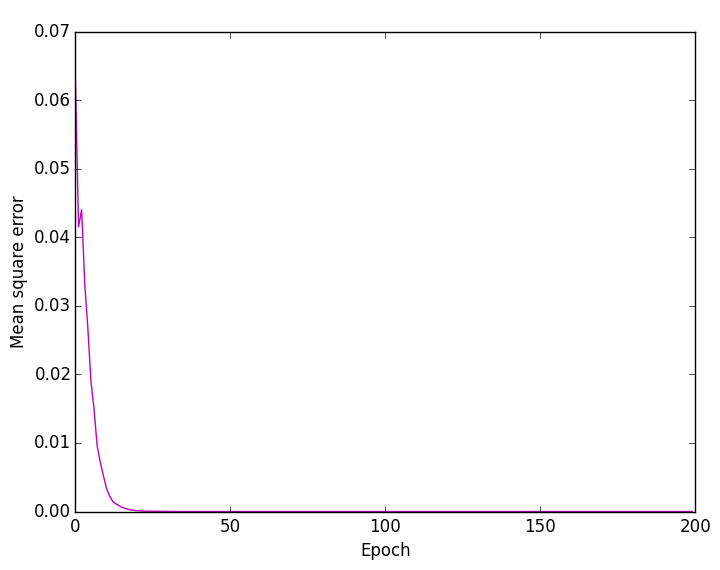}
        \caption{Mean Squared Error optimization.}
    \end{subfigure}
    \begin{subfigure}{.5\textwidth}
        \centering
        \includegraphics[width=1.\linewidth]{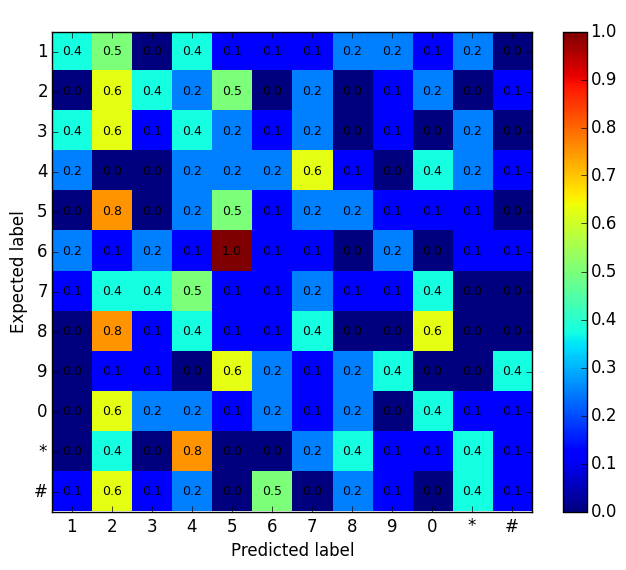}
        \caption{Confusion Matrix.}
    \end{subfigure}
    \caption{Touchlogging to Keylogging with RNN-LSTM R-H}
    \label{fig:exp3lstmrh}
\end{figure}
\end{appendices}

\end{document}